\documentclass[fleqn,usenatbib]{mnras}

\usepackage{newtxtext,newtxmath}

\usepackage[T1]{fontenc}
\usepackage{ae,aecompl}

\usepackage{graphicx}	
\usepackage{graphics}
\usepackage{amsmath}	
\usepackage{caption}
\usepackage{tikz}
\usetikzlibrary{shapes.geometric,arrows}
\usepackage{lipsum,adjustbox}
\usepackage{varwidth}
\usepackage{titlesec}
\usepackage{rotating}
\usepackage{hyperref}
\usepackage{float}

\newcommand{\XMM}{{\em XMM}}
\titleclass{\subsubsubsection}{straight}[\subsection]

\newcounter{subsubsubsection}[subsubsection]
\renewcommand\thesubsubsubsection{\thesubsubsection.\arabic{subsubsubsection}}
\renewcommand\theparagraph{\thesubsubsubsection.\arabic{paragraph}} 

\titleformat{\subsubsubsection}
{\normalfont\normalsize\itshape}{\thesubsubsubsection}{1em}{}
\titlespacing*{\subsubsubsection}
{0pt}{3.25ex plus 1ex minus .2ex}{1.5ex plus .2ex}
\definecolor{lime}{HTML}{A6CE39}
\DeclareRobustCommand{\orcidicon}{%
	\begin{tikzpicture}
	\draw[lime, fill=lime] (0,0) 
	circle [radius=0.16] 
	node[white] {{\fontfamily{qag}\selectfont \tiny ID}};
	\draw[white, fill=white] (-0.0625,0.095) 
	circle [radius=0.007];
	\end{tikzpicture}
	\hspace{-2mm}
}

\foreach \x in {A, ..., Z}{%
	\expandafter\xdef\csname orcid\x\endcsname{\noexpand\href{https://orcid.org/\csname orcidauthor\x\endcsname}{\noexpand\orcidicon}}
}

\title[SDSSRM-XCS Cluster Sample]{The XMM Cluster Survey analysis of the SDSS DR8 redMaPPer Catalogue: Implications for scatter, selection bias, and isotropy in cluster scaling relations}

\author[P. A. Giles et al.]
{P. A. Giles$^{1}$\thanks{p.a.giles@sussex.ac.uk}\orcidA{}, 
A. K. Romer$^{1}$\orcidB{},
R. Wilkinson$^{1}$\orcidC{}, 
A. Bermeo$^{1}$, 
D. J. Turner$^{1}$\orcidD{},
\newauthor
M. Hilton$^{2,3}$\orcidE{},
E. W. Upsdell$^{1}$\orcidH{},
P. J. Rooney$^{1}$,
S. Bhargava$^{1,4}$,
L. Ebrahimpour$^{5,6}$\orcidJ{},
\newauthor
A. Farahi$^{7}$\orcidI{},
R. G. Mann$^{8}$\orcidK{},
M. Manolopoulou$^{8}$,
J. Mayers$^{1}$,
C. Vergara$^{1}$,
\newauthor
P. T. P. Viana$^{5,6}$\orcidG{},
C. A. Collins$^{9}$,
D. Hollowood$^{10}$,
T. Jeltema$^{10}$,
C. J. Miller$^{11}$,
\newauthor
R. C. Nichol$^{12}$,
R. Noorali$^{10}$,
M. Splettstoesser$^{10}$,
J. P. Stott$^{13}$\orcidF{}
\\
$^{1}$Department of Physics and Astronomy, University of Sussex, Brighton, BN1 9QH, UK \\
$^{2}$Astrophysics Research Centre, University of KwaZulu-Natal, Westville Campus, Durban 4041, SA \\
$^{3}$School of Mathematics, Statistics, and Computer Science, University of KwaZulu-Natal, Westville Campus, Durban 4041, SA \\
$^{4}$AIM, CEA, CNRS, Université Paris-Saclay, Université Paris Diderot, Sorbonne Paris Cité, F-91191 Gif-sur-Yvette, France \\
$^{5}$Instituto de Astrof\'isica e Ci\^{e}ncias do Espa\c co, Universidade do Porto, CAUP, Rua das Estrelas, 4150-762 Porto, Portugal \\
$^{6}$Departamento de F\'isica e Astronomia, Faculdade de Ci\^{e}ncias, Universidade do Porto, Rua do Campo Alegre, 687, 4169-007 Porto, Portugal \\
$^{7}$Department of Statistics and Data Sciences, The University of Texas at Austin, Austin, TX 78712, USA \\
$^{8}$Institute for Astronomy, University of Edinburgh, Royal Observatory, Blackford Hill, Edinburgh EH9 3HJ, UK \\
$^{9}$Astrophysics Research Institute, Liverpool John Moores University, Liverpool Science Park, 146 Brownlow Hill, Liverpool L3 5RF, UK \\
$^{10}$Santa Cruz Institute for Particle Physics, University of California, Santa Cruz, 1156 High St, Santa Cruz, CA 95064, USA \\
$^{11}$Department of Astronomy, University of Michigan, Ann Arbor, MI, 48109, USA \\
$^{12}$Institute of Cosmology and Gravitation, Dennis Sciama Building, Burnaby Road, Portsmouth PO1 3FX, UK \\
$^{13}$Department of Physics, Lancaster University, Lancaster LA1 4YB, UK 
}

\date{Accepted 2022 August 17. Received 2022 August 12; in original form 2022 May 5}

\pubyear{2022}

\begin{document}
\label{firstpage}
\pagerange{\pageref{firstpage}--\pageref{lastpage}}
\maketitle

\begin{abstract}
In this paper, we present the X-ray analysis of SDSS DR8 redMaPPer (SDSSRM) clusters using data products from the XMM Cluster Survey (XCS).  In total, 1189 SDSSRM clusters fall within the {\em XMM}-Newton footprint. This has yielded 456 confirmed detections accompanied by X-ray luminosity ($L_{X}$) measurements. Of these clusters, 381 have an associated X-ray temperature measurement ($T_{X}$).  This represents one of the largest samples of coherently derived cluster $T_{X}$ values to date. 
Our analysis of the X-ray observable to richness scaling relations has demonstrated that scatter in the $T_{X}-\lambda$ relation is roughly a third of that in the $L_{X}-\lambda$ relation, and that the $L_{X}-\lambda$ scatter is intrinsic, i.e. will not be significantly reduced with larger sample sizes. 
Analysis of the scaling relation between $L_{X}$ and $T_{X}$ has shown that the fits are sensitive to the selection method of the sample, i.e. whether the sample is made up of clusters detected ``serendipitously'' compared to those deliberately targeted by {\em XMM}. These differences are also seen in the $L_{X}-\lambda$ relation and, to a lesser extent, in the $T_{X}-\lambda$ relation. Exclusion of the emission from the cluster core does not make a significant impact on the findings.  A combination of selection biases is a likely, but yet unproven, reason for these differences.
Finally, we have also used our data to probe recent claims of anisotropy in the $L_{X}-T_{X}$ relation across the sky. We find no evidence of anistropy, but stress this may be masked in our analysis by the incomplete declination coverage of the SDSS.
\end{abstract}




\renewcommand{\thefootnote}{\textsuperscript{\arabic{footnote}}}

\section{Introduction}

Clusters of galaxies are the largest gravitationally bound objects in the Universe, residing at the intersections of the dark matter filamentary structure.  Enabled by a new generation of imaging surveys, from across the electromagnetic spectrum, clusters are expected to play an important role in forthcoming attempts to measure cosmological parameters to percent level accuracy \citep[e.g. see Figure G2 in][]{2018arXiv180901669T}.  Several detection methods will be used to deliver cluster samples of sufficient size, quality, and redshift grasp to meet the requirements of Stage IV (and beyond) Dark Energy Experiments \citep[][]{2016arXiv160407626D}. These methods include detections of spectral distortions to the Cosmic Microwave Background (CMB), of extended X-ray emission, and of projected overdensities (in the optical/near-IR band) of member galaxies. Relevant, ongoing, or soon to begin, experiments include the South Pole Telescope, the Atacama Cosmology Telescope, the Simons Observatory (CMB\footnote[1]{\tt pole.uchicago.edu, act.princeton.edu, simonsobservatory.org}), the eROSITA telescope (X-ray\footnote{\tt mpe.mpg.de/eROSITA}), the Dark Energy Survey (DES), the Hyper Suprime-Cam Subaru Strategic Program, the Legacy Survey of Space and Time and the EUCLID mission (optical/near-IR\footnote{\tt darkenergysurvey.org, hsc.mtk.nao.ac.jp, www.lsst.org, sci.esa.int/euclid}).

Even after these new cluster samples become available, there will remain significant challenges to overcome before unbiased cosmological parameters can be reliably extracted. This has been illustrated by the surprising inconsistency between parameter estimates derived from clusters compared to those derived, using different techniques, from the same input data. For example, there is a 2.4$\sigma$ tension with the DES Y1 galaxy clustering and cosmic shear results \citep{2019PhRvD.100b3541A}. Similar tension was found by the Planck team when comparing their analysis of clusters with the CMB anisotropy spectrum \citep{2016A&A...594A..24P}. One way to address those challenges is to exploit synergies between data sets collected at different wavelengths \citep[e.g.][] {2010ApJ...713.1207W,2021MNRAS.504.1253G}. The work presented herein aims to provide X-ray support to the efforts of the DES and LSST-DESC\footnote{\tt lsstdesc.org} collaborations to realise the potential of optical/near-IR detected clusters for cosmological studies. For this, we use X-ray data collected by the {\em XMM-Newton} telescope and analysed by the XMM Cluster Survey team \citep{2001ApJ...547..594R}. We focus specifically on clusters identified using the red-sequence Matched-filter Probabilistic Percolation technique \citep[or redMaPPer,][hereafter RM]{2014ApJ...785..104R,2016ApJS..224....1R}. However, this work will also be analogous to other cluster samples generated from optical/near-IR surveys, e.g. those identified using the CAMIRA \citep{2014MNRAS.444..147O} or WaZP \citep{2021MNRAS.502.4435A} algorithms.

The extraction of cosmological parameters from RM samples relies on the use of a Mass Observable Relation (MOR), i.e. a description of how the dark mater halo mass scales with the detection observable. The latter is quantified in RM samples by the so-called {\em richness} measure, which describes the number of galaxies detected per cluster (see Section~\ref{sec:sdssrm} for more information). The halo mass is estimated from the weak lensing (WL) signal. However, the signal per cluster is so small that it is necessary to bin the sample, by richness and redshift, in order to measure the MOR \citep{2019MNRAS.482.1352M}. The main drawback of this binning, or ``stacking'' method is the loss of any information about the intrinsic scatter of the observable with mass. As shown in \cite{2009MNRAS.397..577S}, knowledge of the scatter and, its evolution with mass and redshift, is needed for accurate parameter estimation. The benefit of X-ray follow-up, such as that described herein, is that an X-ray observable to richness scaling relation will provide information about the scatter in the stacked MOR \citep[e.g.][]{2019MNRAS.490.3341F}.

Another drawback of using WL to calibrate the MOR for RM samples, is that the WL signal is diluted if there is an offset between the RM determined cluster centroid and the dark matter halo centre of mass. The impact of the offset needs to be modelled to mitigate the impact on derived cosmological parameters, for which X-ray follow-up is essential. This is because the X-ray surface brightness is a much better tracer of the underlying mass than the projected galaxy density. This type of mis-centering correction using X-ray data has been demonstrated in e.g.,  \cite{2019MNRAS.482.1352M}.

In summary, the work presented herein was motivated by the desire to support RM cluster cosmology in two ways: estimating intrinsic scatter on the MOR and  determining a mis-centering model. The first step required to meet both goals is to gather as much high quality X-ray data as possible, and in $\S$\ref{sec:sampleanalysis} we discuss the development of RM cluster samples with X-ray observations in the {\em XMM} public archive. The X-ray analysis of these clusters is described in $\S\ref{sec:specanalysis}$.  We go on to present scaling relations between X-ray and RM observables, and their associated scatter, in $\S$\ref{sec:relations}.  Analysis of the samples for the purposes of mis-centering modelling is the subject of a companion publication, \citealt{2019MNRAS.487.2578Z}.  In \S\ref{sec:discussion} we explore the impact of selection bias on scaling relation scatter measurements by splitting the sample into clusters detected serendipitously and those specifically targeted by {\em XMM}. We also investigate a recent claim of anisotropy in scaling relations across the sky.  Conclusions are summarised in \S\ref{sec:conc}. Throughout this paper we assume a cosmology of $\Omega_{M}$=0.3, $\Omega_{\Lambda}$=0.7 and $H_{0}$=70 km s$^{-1}$ Mpc$^{-1}$.     

\section{Development of the SDSSRM-XCS cluster samples}
\label{sec:sampleanalysis}

\begin{figure*}
\begin{center}
\includegraphics[width=13.0cm]{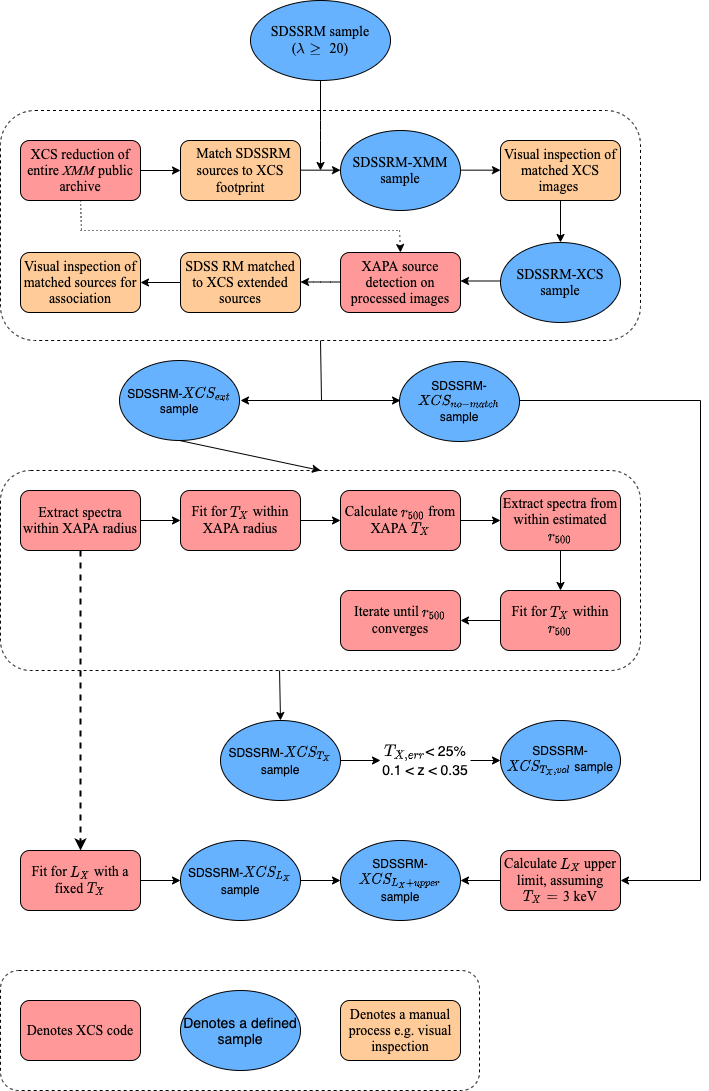}
\end{center}
\caption[]{A flowchart outlining the process used to generate a sample of RM clusters with measured X-ray properties in the SDSS DR8 RM footprint.}
\label{fig:flowchart}
\end{figure*}

In this section, we describe the construction of the X-ray cluster samples used throughout this work. The process starts with the parent SDSS optical cluster catalog described in \citep{2014ApJ...785..104R}.  A flowchart outlining the various steps involved is shown in Figure~\ref{fig:flowchart}.  

\subsection{The SDSS redMaPPer cluster catalogue}
\label{sec:sdssrm}

The red-sequence Matched-filter Probabilistic Percolation (again, denoted RM throughout), cluster finding algorithm \citep{2014ApJ...785..104R}, is a powerful tool for finding clusters from optical/near-IR photometric survey data and has already been successfully applied to SDSS \citep{2014ApJ...785..104R} and DES \citep{2016ApJS..224....1R}.  RM self-trains the red sequence model to any available spectroscopic redshifts, and then calculates, in an iterative fashion, photometric redshifts for each cluster identified.  The richness estimated by RM (hereafter, $\lambda_{\rm RM}$) of each cluster is calculated as the sum of membership probabilities over all galaxies within a scale radius, R$_{\lambda}$, where R$_{\lambda}=1.0h^{-1}{\rm Mpc}(\lambda / 100)^{0.2}$.  The specific RM cluster sample used throughout this work is based upon the 8th data release of the Sloan Digital Sky Survey\footnote{https://www.sdss.org/} \citep[or SDSS-DR8,][]{2011ApJS..193...29A}. The RM SDSS-DR8 catalog \citep{2014ApJ...785..104R} contains a total of 396,047 clusters.  The analysis was restricted to clusters with $\lambda_{\rm RM}>$20, because numerical simulations show that, at this threshold, 99\% of RM clusters can be unambiguously mapped to an individual dark matter halo \citep{2016MNRAS.460.3900F}.  Based upon this $\lambda_{\rm RM}$ cut, our initial sample contained 66,028 clusters (we denote this as the `SDSSRM' sample hereafter, see Table~\ref{tab:samples}).

\subsection{The XCS image database and source catalogue}
\label{sec:reduction}

The results presented in this paper were derived using X-ray data from all publicly available {\em XMM} observations\footnote{\href{http://nxsa.esac.esa.int/nxsa-web/}{{\em XMM} database}} (as of September 2018) with usable European Photon Imaging Camera (EPIC) science data.  The {\em XMM} observations were analysed as part of the {\em XMM} Cluster Survey \citep[][hereafter XCS]{1999astro.ph.11499R}.  The aim of XCS is to catalogue and analyse all X-ray clusters detected during the {\em XMM} mission. This includes both those that were the intended target of the respective observation, and those that were detected serendipitously (e.g. in the outskirts of an {\em XMM} observation targeting a quasar).  The XCS reduction process was fully described in \citet[][hereafter LD11]{2011MNRAS.418...14L}, but a brief outline is as follows. 

The data were processed using XMM-SAS version 14.0.0, and events lists generated using the {\sc EPCHAIN} and {\sc EMCHAIN} tools. In order to exclude periods of high background levels and particle contamination, we generated light curves in 50s time bins in both the soft (0.1 -- 1.0 keV) and hard (12 -- 15 keV) bands.  An iterative 3$\sigma$ clipping process was performed on the light curves; time bins falling outside this range were excluded.  

Single camera (i.e. PN, MOS1 and MOS2) images, along with the corresponding exposure maps, were then generated from the cleaned events files, spatially binned with a pixel size of 4.35$^{\prime\prime}$.  The images and exposure maps were extracted in the 0.5 -- 2.0 keV band, which is typical for soft band X-ray image analysis.  Individual camera images were merged to create a single image per observation, likewise the exposure maps.  The {\sc mos} cameras were scaled to the PN during the merging by the use of energy conversion factors (ECFs) derived using the {\sc xspec} \citep{1996ASPC..101...17A} package.  The ECFs were calculated based upon an absorbed power-law model.

Using the merged images and exposure maps, we applied a bespoke {\sc wavdetect} \citep{2002ApJS..138..185F} based source detection routine, the XCS Automated Pipeline Algorithm ({\sc xapa}). Once the source detection stage was complete, {\sc xapa} proceeded to classify the resulting sources as either point-like or extended. After removal of duplicates, a master source list (MSL) was generated.  The MSL used in this work contained a total of 326,294 X-ray sources, of which 35,575 were classified as extended detections.

\subsection{Identifying SDSSRM clusters in the XCS footprint}
\label{sec:sdss-xmm}

The SDSSRM cluster sample (Sect~\ref{sec:sdssrm}) was compared to the footprint of the XCS image archive (Sect~\ref{sec:reduction}). If a given RM centroid position fell within 15$^{\prime}$ of the aimpoint of one or more {\em XMM} observations, then that cluster was flagged as having a preliminary {\em XMM} match.  The matched list was then filtered based upon the total exposure time, where the total exposure time is a combination of the exposure times for each of the PN, MOS1 and MOS2 cameras, defined as 0.5$\times$PN$_{exp}$+0.5$\times$(MOS1$_{exp}$+MOS2$_{exp}$). Only those clusters with a total mean exposure (defined within a 5 pixel radius centered on the RM position) of greater than 3ks, and a median exposure of greater than 1.5ks, were retained in the match list.  The median exposure limit excluded RM clusters that had significant overlap with chip gaps or bad pixels.  Next, an additional exposure (mean and median) filter was carried out at a position 0.8R$_{\lambda}$ away from the RM defined centre (in the direction away from the {\em XMM} aimpoint).  This was done to encapsulate the expected range of mis-centering between RM and {\sc xapa} centroids \citep[see][]{2019MNRAS.487.2578Z}.  

Based on these matching criteria, 1,246 SDSSRM clusters fall within the active area of one or more XCS processed {\em XMM} observations. Hereafter, these 1,246 SDSSRM clusters are referred to as the `SDSSRM-{\em XMM}' sample (see Table~\ref{tab:samples}).  We then performed a visual inspection to remove clusters falling in observations with abnormally high background levels (e.g. Figure~\ref{fig:xmmbad}(a)), and those that were corrupted due to proximity to a very bright point source\footnote{Such sources produce artefacts in the {\em XMM} images including readout trails and ghost images of the telescope support structure.} (e.g. Figure~\ref{fig:xmmbad}(b)).   We removed 57 observations, therefore, after this filtering step, 1,189 clusters remained. We denote this set as the `SDSSRM-XCS' sample (see Table~\ref{tab:samples}).

\begin{figure*}
\begin{center}
\begin{tabular}{cc}
\includegraphics[clip, trim=4cm 0.25cm 6cm 0.25cm, width=8cm]{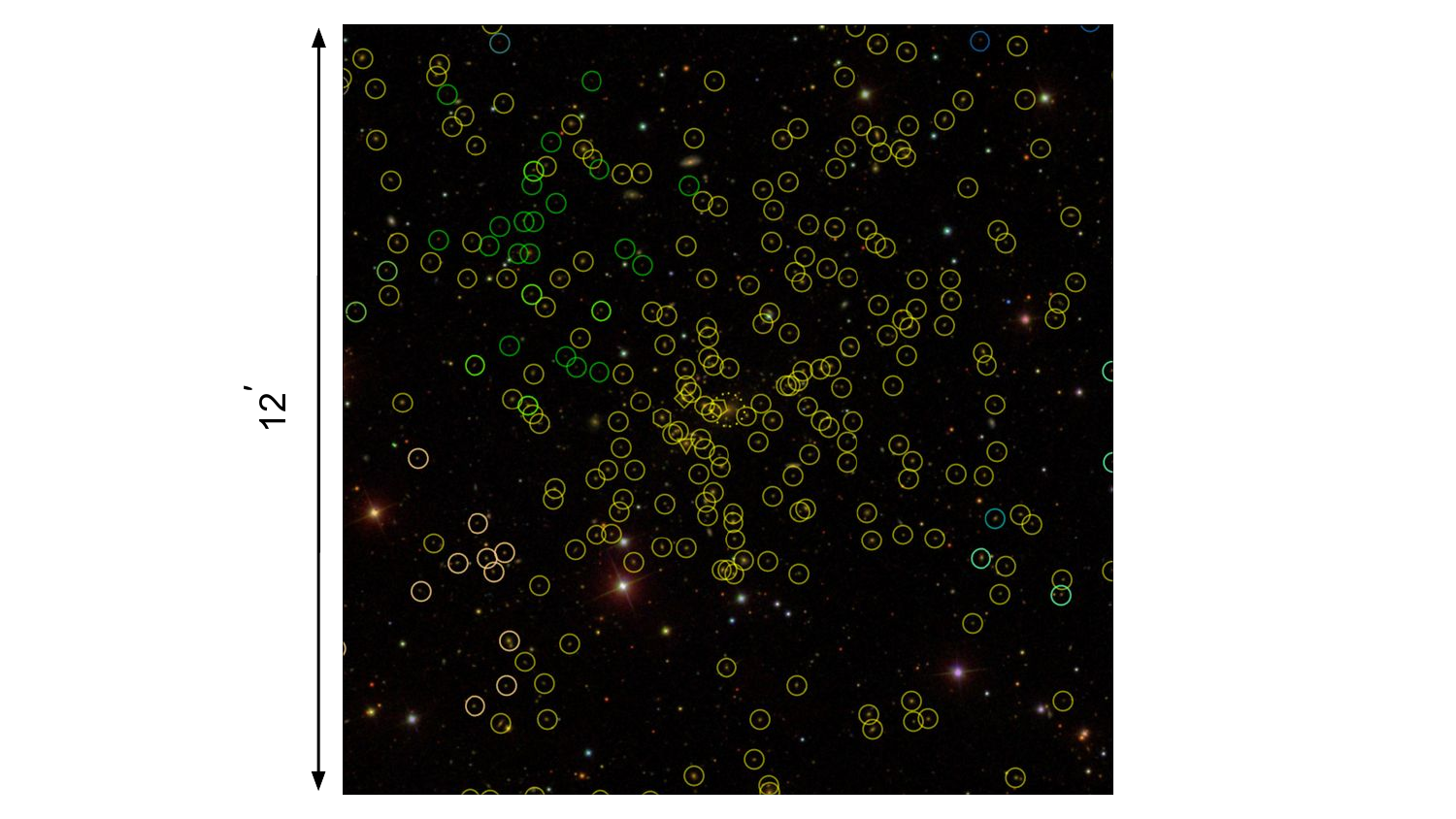} &
\hspace{-0.4cm}
\includegraphics[clip, trim=4cm 0.25cm 6cm 0.25cm, width=8cm]{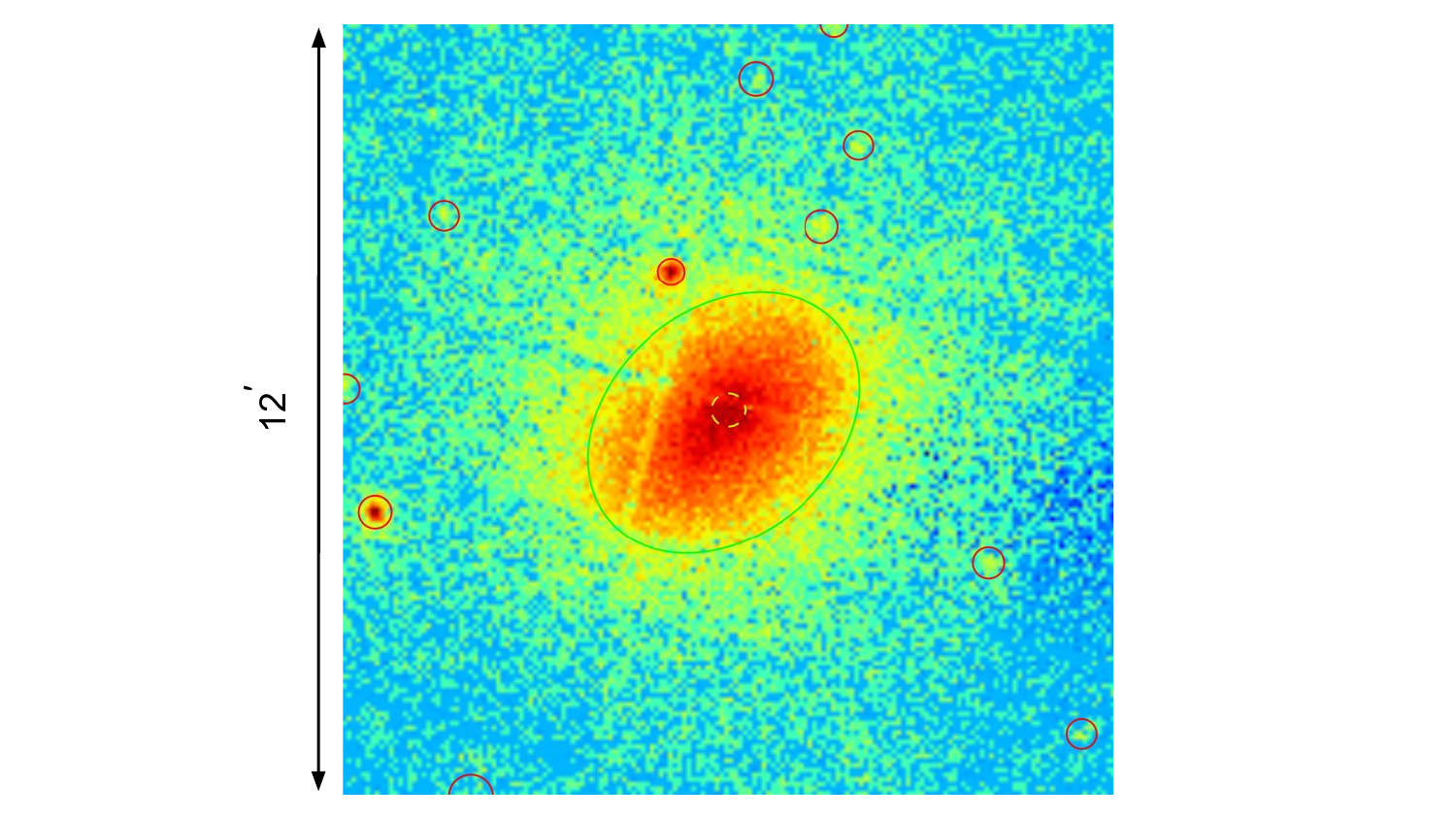} \\
\end{tabular}
\begin{tabular}{c}
\includegraphics[clip, trim=4cm 0.25cm 6cm 0.25cm, width=8cm]{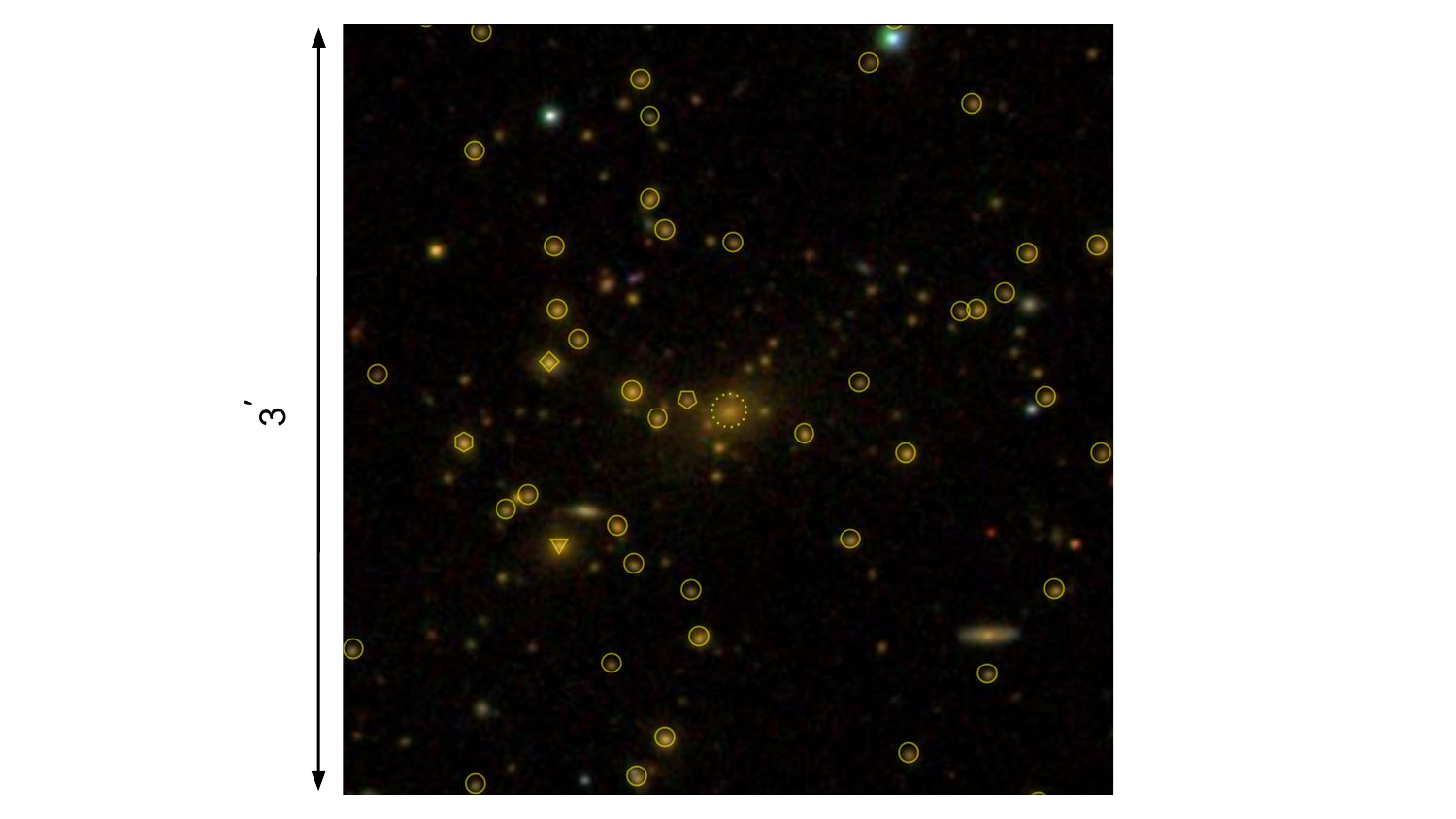} \\
\end{tabular}
\end{center}
\caption[]{An example of a cluster in the SDSSRM-{\em XMM}$_{\rm ext}$ sample (Table~\ref{tab:samples}). Top left: SDSS optical image of the cluster XMMXCS J164020.2+464227.1 (SDSS RMID=2), commonly referred to as Abell 2219, at a redshift of $z=0.23$ and richness of $\lambda_{\rm RM}=$199.  The dashed yellow circle and the other solid yellow shapes highlight the galaxies associated RMID=2.  Galaxy members of other nearby RM clusters nearby are circled in different colours (e.g. green, turquoise and cream); Top right: {\em XMM} X-ray observation of the matched XCS extended source.  Green (red) regions highlight extended (point) {\sc XAPA} sources detected in the observation. Bottom middle:  A zoom into the SDSS optical image.  The yellow circle (dashed), triangle, diamond, pentagon and hexagon represent, respectively, the 1st, 2nd, 3rd, 4th and 5th most probable, according to the RM algorithm, candidate for the central galaxy.}
\label{fig:cutouts}
\end{figure*}

\subsection{Cross-matching the SDSSRM-XCS sample with XCS extended sources}
\label{sec:sdss-xcs}

Although all 1,189 SDSSRM-XCS clusters (Sect.~\ref{sec:sdss-xmm}) fall within the XCS defined {\em XMM} footprint of SDSS, this doesn't guarantee they are matched to an extended {\sc xapa} source.  In this context, a match was defined to mean that the respective centroids were within $2$~$h^{-1}$~Mpc of each other, where the distance was calculated assuming the RM cluster redshift. If more than one extended {\sc xapa} source met this criterion, we made the assumption that the closest (on the sky) match was the correct association. By this definition, 782 -- of the input SDSSRM-XCS sample of 1,189 -- were initially matched to an extended {\sc xapa} source (the remaining, 407 SDSSRM-XCS entries are discussed further in Sect.~\ref{sec:undet}).   

The 782 XCS extended sources matched to SDSSRM clusters were then examined by eye to exclude cases where the X-ray emission was unlikely to be physically associated with the RM cluster in question. An example of a cluster that passed this test is shown in Figure~\ref{fig:cutouts} (one that did not is shown in Figure~\ref{fig:unassociated}). The top left panel of Figure~\ref{fig:cutouts} shows, with yellow circles, all the galaxies associated, by RM, with the cluster in question (other coloured circles depict the galaxies associated with other RM clusters in the field).  In the bottom panel, the dashed circle highlights the position of the galaxy defined by RM as the most likely central galaxy.  The 2nd, 3rd, 4th and 5th most likely candidates, are highlighted by the yellow triangle, diamond, pentagon and hexagon respectively.  The top right panel shows the {\em XMM} image of the matched XCS extended source. Following the visual inspection process, only 456, of the 782 checked, clusters were retained. These 456 are referred to hereafter as the `SDSSRM-XCS$_{\rm ext}$' sample, see Table~\ref{tab:samples} (the remaining 326 entries are discussed further in Sect.~\ref{sec:undet}).  

\begin{table*}
\centering
\caption[]{{\small Summary of the SDSSRM cluster sub-samples produced during the matching process described throughout Section~\ref{sec:sampleanalysis}}\label{tab:samples}}
\vspace{1mm}
\begin{tabular}{l|lcc}
\hline
\hline
 Sample & Brief description & \# clusters & Relevant section \\
\hline
SDSSRM & SDSSRM DR8 clusters with a richness $\lambda_{\rm RM}>$20 & 66,028 & $\S$~\ref{sec:sdssrm} \\
\hline
SDSSRM-{\em XMM} & \hspace{-3.5mm}\begin{tabular}{@{}l@{}}SDSSRM clusters fall within the active area of one or more \\ XCS processed {\em XMM} observations\end{tabular} & 1246 & $\S$~\ref{sec:sdss-xmm} \\
\hline
SDSSRM-XCS & \hspace{-3.5mm}\begin{tabular}{@{}l@{}}As above, but after visual inspection to remove matches to \\ problematic {\em XMM} images \end{tabular} & 1189 & $\S$~\ref{sec:sdss-xcs} \\
\hline
SDSSRM-XCS$_{\rm ext}$ & \hspace{-3.5mm}\begin{tabular}{@{}l@{}}SDSSRM-XCS clusters that are matched to an {\it extended} \\ XCS source \end{tabular} & 456 & $\S$~\ref{sec:sdss-xcs} \\
\hline
SDSSRM-XCS$_{\rm unm}$ & \hspace{-3.5mm}\begin{tabular}{@{}l@{}}SDSSRM-XCS clusters that are unmatched to an {\it extended} \\ XCS source \end{tabular} & 733 & $\S$~\ref{sec:undet} \\
\hline
\end{tabular}
\end{table*}


\subsubsection{Accounting for incidences of redMaPPer mispercolations}
\label{sec:misperc}

The RM algorithm employs a process known a ``percolation'' that aims to assign galaxies to the correct system when there are two or more RM clusters in close proximity on the sky \citep[][$\S$~9.3]{2014ApJ...785..104R}.  However, sometimes this process fails, with the result that RM assigns a low value of $\lambda_{\rm RM}$ to a genuinely rich cluster when it is close (in projection) to a less rich system, and vice versa.  This RM failure mode is known as ``mispercolation'' \citep[see][]{2019ApJS..244...22H}.  An example is shown in Figure~\ref{fig:misperc}.  The yellow circles in Figure~\ref{fig:misperc} (a) highlight the galaxies associated with a $\lambda_{\rm RM}=166$ RM cluster.  From the distribution of the X-ray emission of the system (Figure~\ref{fig:misperc} (b)), it is clear that the large richness has been incorrectly assigned to the low flux sub-halo of a nearby massive cluster (incorrectly assigned a richness of $\lambda_{\rm RM}=20$).  

During the visual inspection process that generated the SDSSRM-XCS$_{\rm ext}$ sample (see Sect.~\ref{sec:sdss-xcs}), we identified three pairs of clusters affected by mispercolation.  In order to correct their $\lambda_{\rm RM}$ values, we followed the method outlined in \cite{2019ApJS..244...22H}, i.e. the originally assigned $\lambda_{\rm RM}$ value for the main halo was manually switched with that of the sub-halo. However, unlike \cite{2019ApJS..244...22H}, we did not remove the lower flux system from further analysis if  $\lambda_{\rm RM}\geq$20.  Table~\ref{tab:misperc_data} provides properties of the clusters effected by mispercolation.  Of the 6 clusters effected by mispercolation, one is not included in the final SDSSRM-XCS$_{\rm ext}$ sample, as its richness has a value of $\lambda_{\rm RM}<$20.  

\subsubsection{SDSSRM-{\em XMM} entries not associated with XCS extended sources}
\label{sec:undet}

A total of 733 members of the SDSSRM-XCS sample are not included in the SDSSRM-XCS$_{\rm ext}$ sample.  This is because they have not been matched to an XCS extended source.  Of these, 407 are not close to any XCS extended source, whereas the remaining 326 were close in projection, but were deemed, after the eye-balling step, unlikely to be physically associated with it. Combined, these 733 ``unmatched'' clusters are denoted as the SDSSRM-XCS$_{\rm unm}$ sub-set. For these clusters, we determine luminosity upper limits in Section~\ref{sec:calcupperlim} so that they can be included in the scaling relation analysis presented in Section~\ref{sec:alldatarelations}. 

\begin{figure*}
\begin{center}
\begin{tabular}{cc}
\includegraphics[width=.5\textwidth]{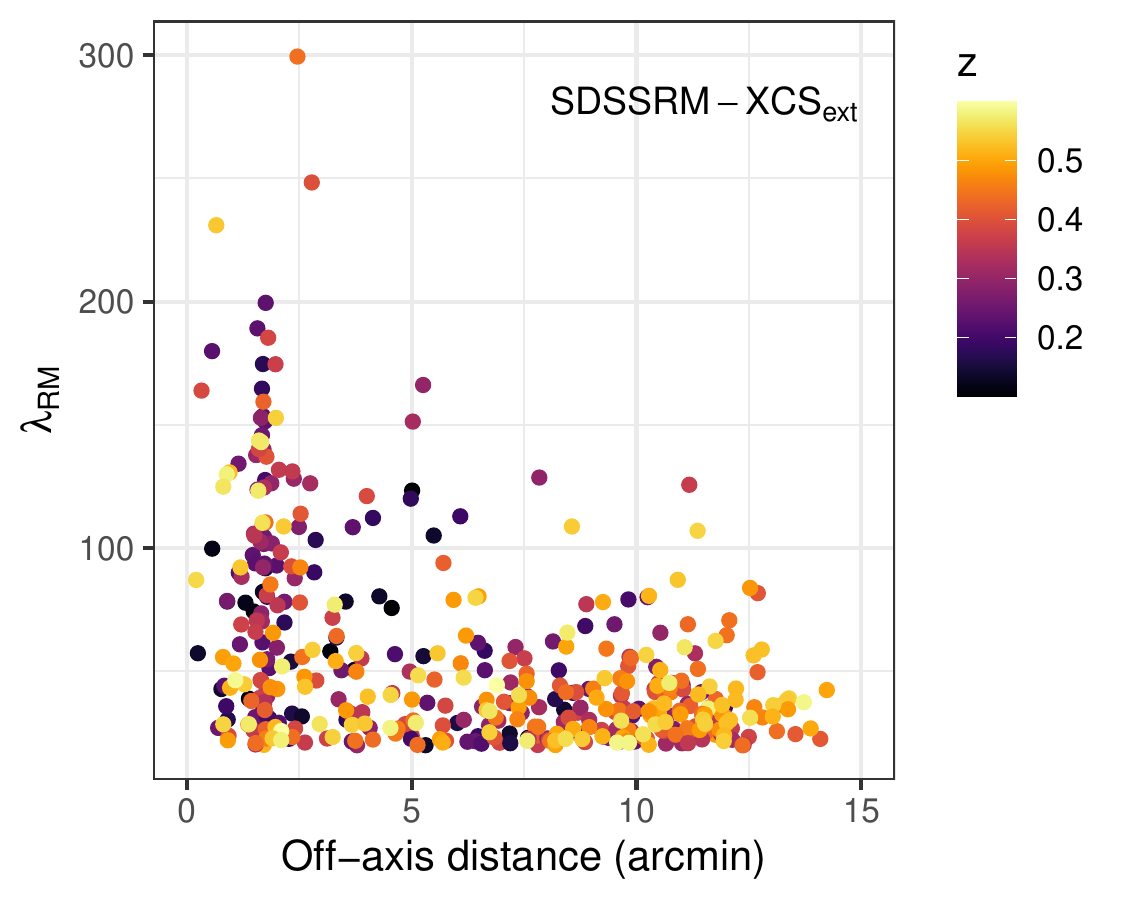} &
\hspace{-0.7cm} \includegraphics[width=.5\textwidth]{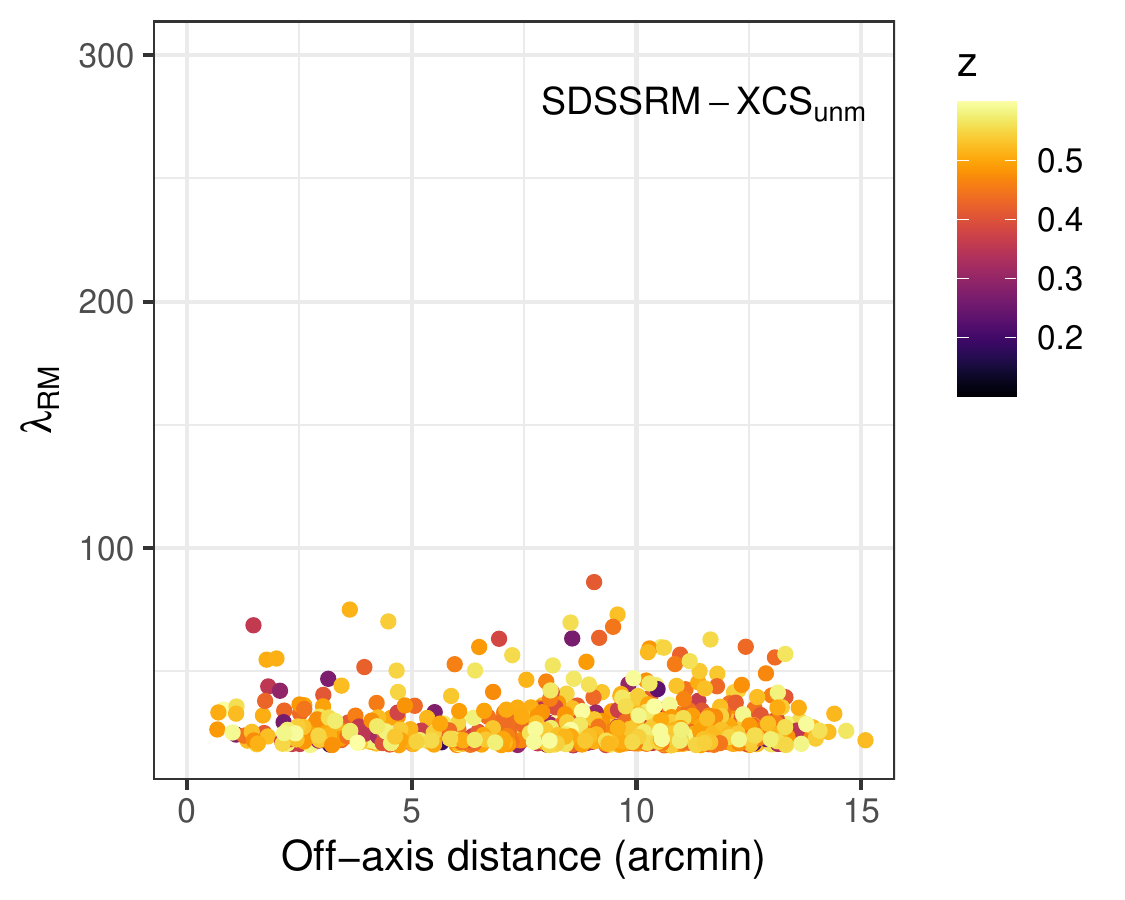} \\
(a) & (b) \\
\end{tabular}
\end{center}
\caption{Distributions of richness ($\lambda_{\rm RM}$) and off-axis distance for the SDSSRM-XCS$_{\rm ext}$ (a) and SDSSRM-XCS$_{\rm unm}$ (b) subsets.  The off-axis distance is defined as the distance from the RM defined central galaxy to the centre of the {\em XMM} observation. In each case, the points are colour-coded by redshift, given by the inset colour-bar.}
\label{fig:3d_distribution}
\end{figure*}

To better understand why certain clusters were not detected in their respective {\em XMM} observation(s), we compared the distributions of their richness, off-axis distance and redshift with those of the detected SDSSRM-${\rm XCS}_{\rm ext}$ sample (see Figure~\ref{fig:3d_distribution}). Here, we defined the off-axis distance as the angular separation of the observation aimpoint to the RM defined central galaxy: both the effective exposure time and the point spread function (PSF) degrade significantly with off-axis distance. To emphasise the redshift difference between the two samples, the points are colour-coded by redshift. As expected, we find that the majority of SDSSRM-XCS$_{\rm unm}$ clusters fall at larger off-axis positions, higher redshifts, and lower richnesses, than SDSSRM-XCS$_{\rm ext}$ clusters.   

\subsection{False-Positive Rate}
\label{sec:falsepos}

\begin{figure}
	\begin{centering}
    \includegraphics[width=.5\textwidth]{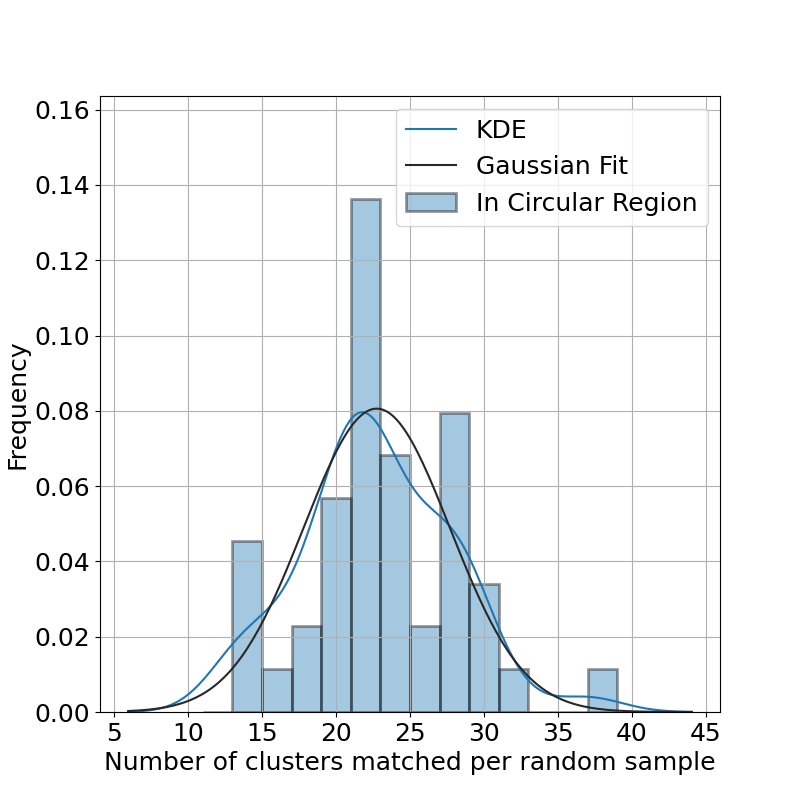}
	\caption[]{Distribution of the number of RM$^{rd}$ clusters matched to an extended XCS source, repeated using 43 samples of random positions created to match the size of the SDSS DR8 RM cluster sample (see~\ref{sec:falsepos}).\label{fig:falsepos}}
	\end{centering}
\end{figure}

In order to determine the false-positive rate of matches between the SDSS DR8 catalogue and the XCS MSL, we make use of the SDSSRM random catalogue\footnote{http://risa.stanford.edu/redMaPPer/}.  Full details of the construction of the random catalog can be found in \citet[][$\S$11]{2014ApJ...785..104R}.  The random catalog is constructed such as to map the detectability of clusters as a function of redshift and richness, taking into account the large-scale structure that is already imprinted on the galaxy catalog.  The random catalogue contains $\approx$3$\times$10$^{6}$ clusters (which we denote as RM$^{rd}$), a factor $\sim$100 larger than the SDSS DR8 RM catalogue.  We draw at random from the RM$^{rd}$ clusters and create samples of equal size to the SDSS DR8 RM catalogue (i.e. 66,028 clusters), resulting in 43 separate catalogues of RM$^{rd}$ clusters.  For each separate catalogue, we first determined the number of RM$^{rd}$ positions falling on an {\em XMM} observation using the method described in Sect.~\ref{sec:sdss-xmm} (i.e. a mean and median exposure cut of 3~ks and 1.5~ks respectively).  We note that the RM$^{rd}$ clusters do not contain a $R_{\lambda}$ estimate, therefore, we do not employ the additional exposure cut at a position 0.8$R_{\lambda}$ away from the RM position (see Section~\ref{sec:sdss-xmm}).  From the 43 mock catalogues, we determined that, on average, 1548$\pm$33 RM$^{rd}$ clusters fell inside the {\em XMM} footprint.  Next, we matched the RM$^{rd}$ clusters to XCS extended sources.

We defined a RM$^{rd}$ cluster to be associated with an extended source when the centroid fell within the {\sc xapa} detection region.  Note that {\sc xapa} provides elliptical regions but, for this matching, we circularised the {\sc xapa} region by making the radius equal to the semi-major axis of the {\sc xapa} source. Figure~\ref{fig:falsepos} shows the distribution of these associations for all 43 random catalogues.  Based upon a Gaussian fit to the distribution, we find we would, on average, randomly match to an extended {\sc xapa} source 22.8$\pm$5.0 times. We thus estimate a contamination rate in the SDSSRM-XCS sample of $\simeq1.5\%$ ($23/1548$).  We note that since we made the simplifying assumption of a RM$^{rd}$ match when falling within a {\sc xapa} (circularised) region, and no eyeballing performed, this estimate is likely an upper limit. 

\section{X-ray Analysis of the SDSSRM-XCS sample}
\label{sec:specanalysis}

We used the XCS Post Processing Pipeline ({\tt XCS3P}) to derive the X-ray properties of the SDSSRM-XCS$_{\rm ext}$ clusters, i.e. their temperature ($T_{\rm X}$) and luminosity ($L_{\rm X}$). {\tt XCS3P} can be run in batch mode and applied to hundreds of clusters at a time. 

A detailed description of {\tt XCS3P} can be found in LD11, but a brief overview is as follows. Cluster spectra were extracted using the SAS tool {\sc evselect} and fit using {\sc xspec} \citep{1996ASPC..101...17A}.  The fits were performed in the 0.3-7.9 keV band with an absorbed {\tt APEC} model \citep{2001ApJ...556L..91S} using the $c$-statistic \citep{1979ApJ...228..939C}. The {\tt APEC} component accounts for the emission from a hot diffuse gas enriched with various elements.  Relative abundances of these elements are defined as their ratio to Solar abundances ($Z_{\odot}$).  The absorption due to the interstellar medium was taken into account using a multiplicative {\tt Tbabs} model \citep{2000ApJ...542..914W} in the fit, with the value of the absorption ($n_{H}$) taken from \cite{2016A&A...594A.116H} and frozen during the fitting process.  The abundance was fixed at 0.3~$Z_{\odot}$, a value typical for X-ray clusters \citep{2012ARA&A..50..353K}.  The redshift was fixed to the value as determined by RM.  We note that redshift uncertainties are not taken into account in the fit since the typical photometric redshift uncertainty for SDSS RM clusters is very small \citep[$\frac{\sigma_{z}}{1 + z}\leq 0.04$ out to a redshift of $z_{\lambda}$=0.6, see Fig.~9 in][]{2014ApJ...785..104R}.  The {\tt APEC} temperature and normalisation were free to vary during the fitting process.  Temperature errors were estimated using the {\tt XSPEC} {\tt ERROR} command, and quoted within 1-$\sigma$. Finally, luminosities (and associated 1-$\sigma$ errors) were estimated from the best-fit spectra using the {\tt XSPEC} {\tt LUMIN} command (in both the bolometric and 0.5--2.0~keV, rest-frame, bands).   

\subsection{Updates to XCS3P since LD11}
\label{sec:updateLD11}

Improvements have been made to {\tt XCS3P} since LD11 was published, and these are described in the subsections below.  

\subsubsection{Spectral extraction region} 
\label{sec:densityradius}

In LD11, the spectral extraction region was based on the {\sc xapa} (see Section~\ref{sec:reduction}) characterized detection region i.e. an elliptical aperture defined using the lengths of the {\sc xapa} determined major and minor axes.  The extraction region has since been updated to be within a circular overdensity radius ($r_{\Delta}$).  Overdensity radii are defined as the radius at which the density is $\Delta$ times the critical density of the Universe at the cluster redshift.  We used two radii common in the X-ray cluster literature i.e. $r_{\rm 500c}$ and $r_{\rm 2500c}$, where the radii were estimated using the relation given in \cite{2005A&A...441..893A}:
\begin{align}
\label{equ:rdelta}
E(z)r_{\Delta} &= B_{\delta}\left(\frac{T_{\rm X}}{5~{\rm keV}} \right)^{\beta},
\end{align}
where $E(z)$=$\sqrt{\Omega_{M}(1 + z)^3 + \Omega_{\Lambda}}$.  In the case of $r_{500c}$, $B_{\delta}$=1104~kpc and $\beta$=0.57.  The process is iterative because we do not know a priori what $T_{\rm X}$ is; an initial temperature was calculated within the {\sc xapa} defined elliptical source region, which is then used to estimate $r_{\rm 500c}$ (using Equ.~\ref{equ:rdelta}).  A new $T_{\rm X}$ value was then measured from a spectrum extracted from a circular region with $r_{\rm 500c}$ radius. The new $T_{\rm X}$ was then used to define a new $r_{\rm 500c}$ value. The process was repeated until $r_{\rm 500c}$ converged (the ratio of the new to old $0.9> r_{\rm 500c,new}/r_{\rm 500c,old} <1.1$).  We employed the condition that at least three iterations were performed, regardless of the convergence.  To account for the background in the spectral analysis, we made use of a local background annulus centered on the cluster, with an inner and outer radii of 1.05$r_{\rm 500c}$ and 1.5$r_{\rm 500c}$ respectively (see blue edged outer annulus in Fig~\ref{fig:sourceexclude}).  It is also beneficial to compute core excluded properties for analysis \citep[e.g. the use of core-excluded luminosities reduces the scatter in the luminosity-mass relation, see][]{2018MNRAS.473.3072M}.  Therefore, we repeat the process described above, but exclude the inner 0.15$r_{\rm 500c}$ region \citep[as used in many studies in the literature e.g.][]{2009A&A...498..361P,2012MNRAS.421.1583M,2020ApJ...892..102L}.  

In the $r_{\rm 2500c}$ case, Equation~\ref{equ:rdelta} was used, with $B_{\delta}$=491~kpc and $\beta$=0.56.  The local background was taken into account using an annulus centered on the cluster with an inner and outer radius of 2$r_{\rm 2500c}$ and 3$r_{\rm 2500c}$ respectively. In all other respects, the derivation of $T_{\rm X,2500}$ values followed that used for the $T_{\rm X,500}$ values.

\subsubsection{Selection of spectra}  
\label{sec:selectclustspec}

\begin{figure}
	\begin{centering}
	\includegraphics[width=.47\textwidth]{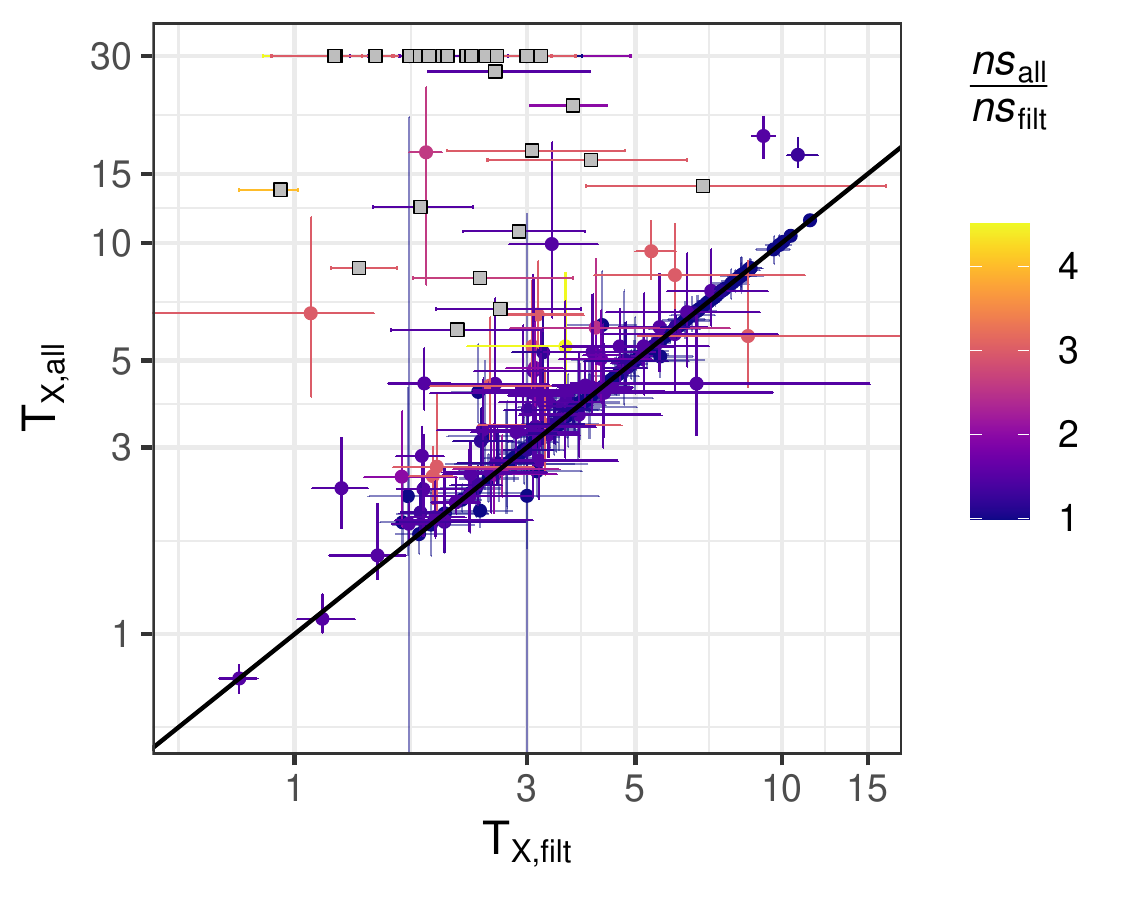}
	\caption{\small{Comparison of the measured temperature when only including spectra in the simultaneous fit that pass quality controls ($T_{X,filt}$, Sect.~\ref{sec:selectclustspec}), and the temperature when using all available spectra in the simultaneous fit ($T_{X,all}$, Sect.~\ref{sec:selectclustspec}).  Grey squares highlight clusters failing to converge during the iteration process (see Sect.~\ref{sec:densityradius}) when determining $T_{X,all}$.  The remaining clusters are colour-coded by the ratio of the number of spectra when including all available spectra to the number passing the quality controls ($\frac{ns_{\rm all}}{ns_{\rm filt}}$).  The black dashed line represents the 1:1 relation.}}
\label{fig:temperrorcomp}
	\end{centering}
\end{figure}

In the LD11 version of {\tt XCS3P}, all available spectra were used in a simultaneous {\sc xspec} fit ($ns_{\rm all}$). This included spectra derived from each of the three (PN, MOS1 and MOS2) {\em XMM} cameras and, where available, multiple {\em XMM} observations (up to 25 per cluster in some  cases). However, we have subsequently discovered that using all available spectra, irrespective of data quality, can increase the measured temperature.  This is demonstrated in Figure~\ref{fig:temperrorcomp} which compares the temperature estimated using all available spectra ($T_{X,all}$) to those determined by filtering out spectra that did not ($T_{X,filt}$), individually, produce a fitted temperature (complete with $1\sigma$ upper and lower limit values) in the range 0.08 $<$ $T_{X}$ $<$ 20 keV. The number of available spectra after filtering is defined as $ns_{\rm filt}$. In Figure~\ref{fig:temperrorcomp} we plot $T_{X,all}$ against $T_{X,filt}$ , with each point representing a cluster and colour coded by the ratio of the number of spectra used when determining $T_{X,all}$ and $T_{X,filt}$ (defined as $\frac{ns_{\rm all}}{ns_{\rm filt}}$).  Grey squares indicate clusters that do not fulfil the criteria of a converged $r_{\rm 500c}$ temperature for the $T_{X,all}$ analysis. Therefore, by using a filtered sample, we were able to extract more $T_{X}$ values. Moreover, where $T_{X,all}$ and $T_{X,filt}$ differ, the former are typically higher. This suggests that there is residual background flaring in low signal-to-noise observations, because the particle background has a hard spectrum. For these reasons, {\tt XCS3P} now only uses filtered spectra sets during the simultaneous fitting.

\subsubsection{Measurement of luminosity uncertainties}  
\label{sec:lxerror}

When estimating the luminosity in {\sc xspec}, the absorption component ($n_{H}$) must be set to zero in order to represent conditions at the cluster (i.e. unabsorbed).  However, the luminosity uncertainties will be in error if they are also determined while $n_{H}$ is set to zero, since the uncertainties are determined from the spectral fit to the absorbed data. This error was present in the LD11 version of {\tt XCS3P}, and has now been corrected. In the latest version of {\tt XCS3P}, the uncertainties are determined using an initial luminosity ($L_{ini}$) calculation, before $n_{H}$ has been set to zero.  Then, $n_{H}$ is set to zero and the luminosity extracted ($L_{0}$).  The uncertainties are then scaled by the ratio of $L_{0}$ to $L_{ini}$. 

\begin{figure*}
	\begin{centering}
	\includegraphics[width=.8\textwidth]{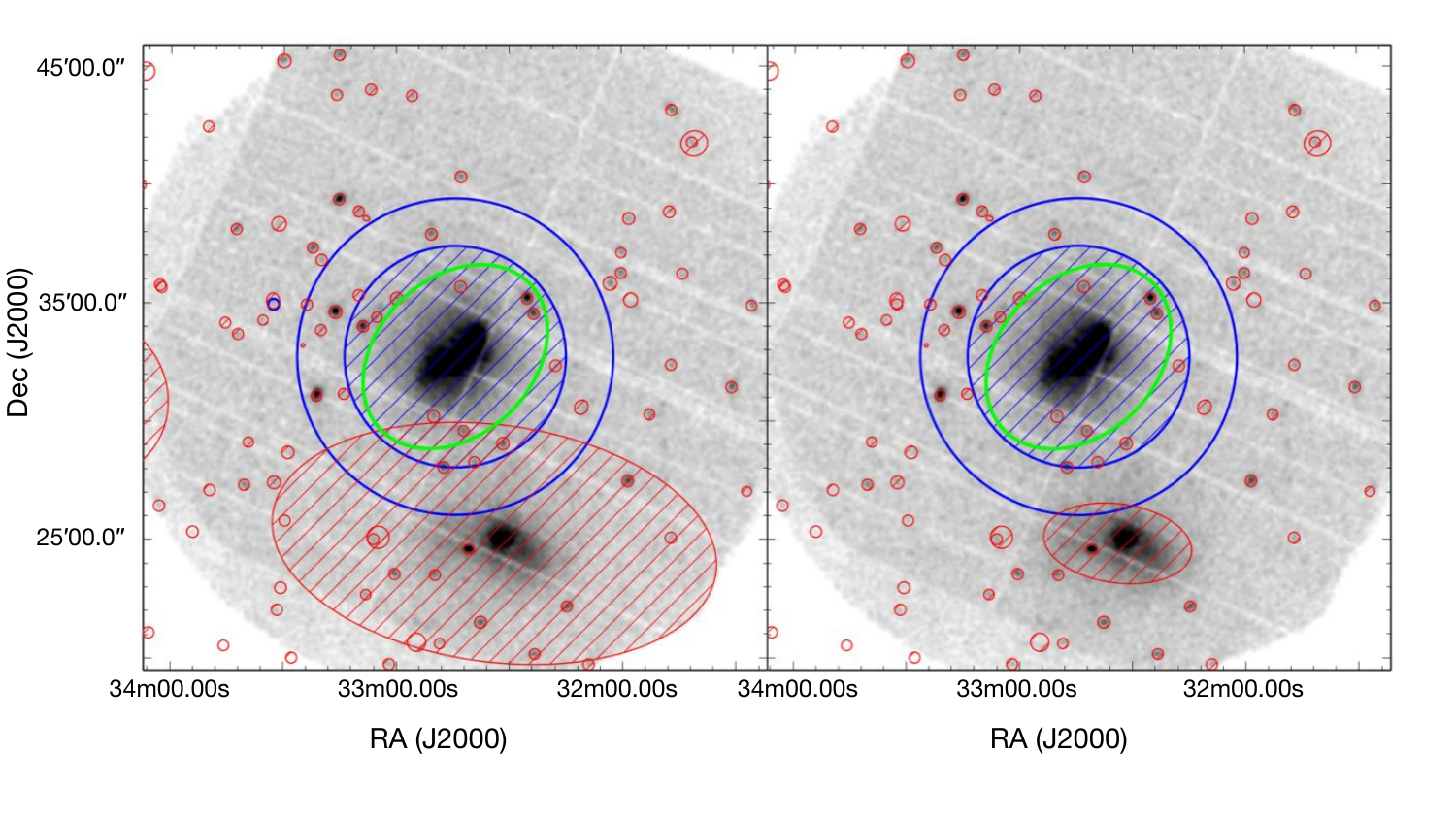}
	\caption{Updated approach to source masking in {\tt XCS3P} compared to LD11. In each image the green ellipse represents the cluster extraction region, the blue edged annulus represents the background region and the red small circles are excluded point sources. The nearby extended source is excluded using the red hashed ellipse. In the LD11 analysis, the exclusion region was too large (left image). This has been corrected in the current version of {\tt XCS3P} (Right image, see Section~\ref{sec:extsources}).} 
\label{fig:sourceexclude}
	\end{centering}
\end{figure*}

\subsubsection{Exclusion of extended sources}  
\label{sec:extsources}

The method used in LD11 to exclude nearby extended sources (NES) sometimes overestimated the area to `drill out' around the NES, because the exclusion area was scaled by number of NES counts.  Figure~\ref{fig:sourceexclude} (left image) highlights the region used to exclude a NES in the LD11 analysis (red hashed ellipse).  In this case, the excluded region overlaps with the source extraction region (green circle), removing a fraction of the source flux.   Therefore, the scaling factor used in LD11 has been deprecated, see Figure~\ref{fig:sourceexclude} (right image).

\subsection{Luminosity estimates when $T_{X}$ is fixed}
\label{sec:lxnotx}

\begin{figure}
	\begin{center}
    \includegraphics[width=.47\textwidth]{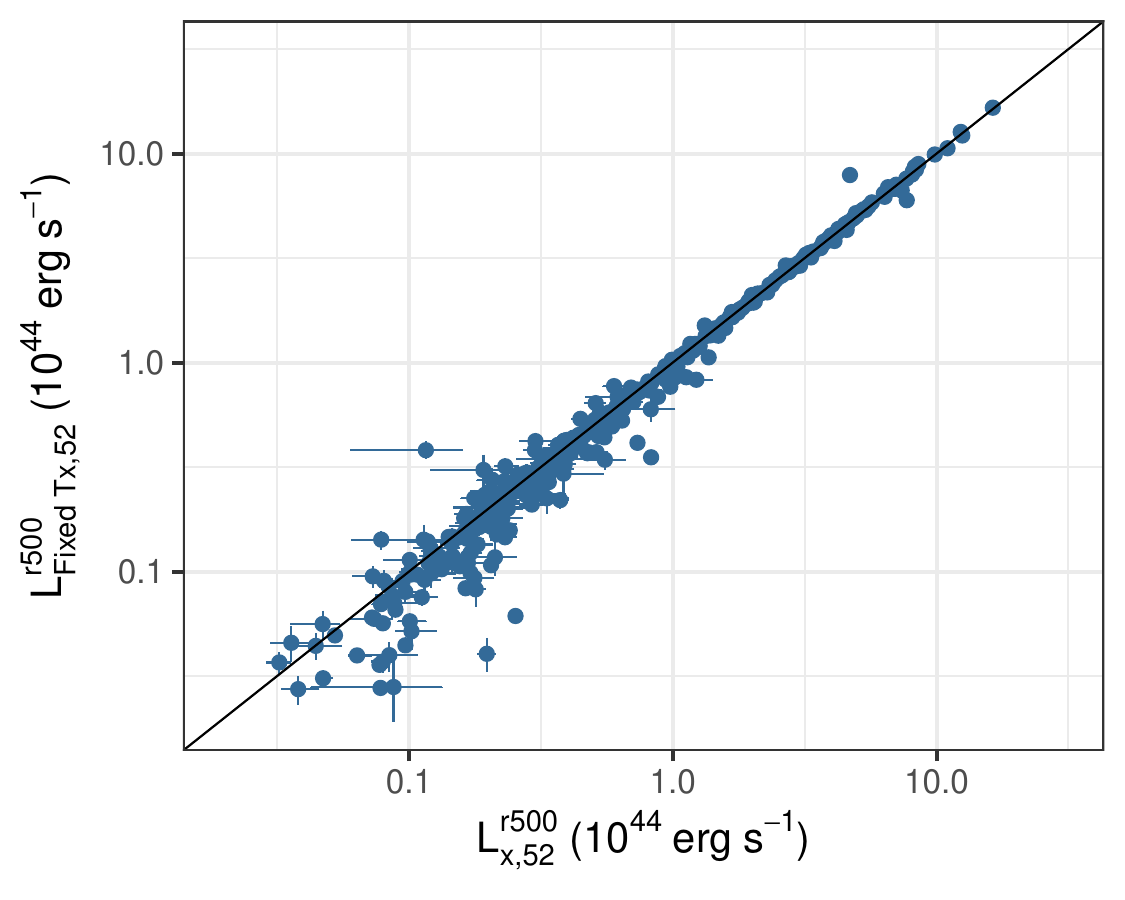}
	\caption[]{Comparison of luminosity determined using a fixed temperature (L$_{\rm Fixed~Tx,52}^{r500}$) estimated from the luminosity temperature relation (see Sect.~\ref{sec:lxnotx} for details), to that determined from spectra extracted within $r_{500}$ (see Sect.~\ref{sec:specanalysis}).  The solid black line indicates the 1:1 relation.}\label{fig:lx_no_tx_comp}
	\end{center}
\end{figure}

Not all 456 clusters in the SDSSRM-XCS$_{\rm ext}$ sample yielded a reliable temperature measurement. However, it was still possible to estimate a luminosity value for them from the extracted spectra using an adapted version of the iterative procedure outlined in Sect.~\ref{sec:specanalysis}. In this adaptation, the temperature was fixed in the spectral fit.  Initially, spectra were extracted within the {\sc xapa} defined region, and an {\sc XSPEC} fit was performed with the $T_{X}$ in the model fixed at 3~keV. This produced an initial luminosity value, which was fed into the luminosity-temperature relation presented in Sect.~\ref{sec:ltrelation} (with parameters given in Table~\ref{tab:bestfit}) to derive a more appropriate $T_{X}$ value. An $r_{\rm 500c}$ was estimated using Equation~\ref{equ:rdelta} using this $T_{X}$ value and a new spectrum was extracted and fit. The process was repeated until the change in the $r_{\rm 500c}$ radius was less than within 10\%. Luminosities estimated in this way are denoted $L_{\rm Fixed~Tx,52}^{r500}$.  To test the validity of this method, we applied it to all clusters in the SDSSRM-XCS$_{\rm ext}$ sample, of which 351 have a measured $L_{X}$ from the spectral extraction method described above (throughout Sect.~\ref{sec:specanalysis}).  In Figure~\ref{fig:lx_no_tx_comp}, these luminosities (estimated using a fixed $T_{X}$) are compared to the luminosities estimated for the 381 clusters in the SDSSRM-XCS$_{T_{X}}$ sample (see Sect~\ref{sec:cosmosample}, i.e., clusters where the luminosities were estimated from the spectral analysis with $T_{X}$ free).  The 1:1 relation is highlighted by the solid black line.  This comparison shows there is a good agreement between the two luminosity estimates.

\subsection{Upper limit estimates in the absence of an XCS detection}
\label{sec:calcupperlim}

There are 733 SDSSRM-XCS clusters that have no corresponding confirmed match to an XCS extended source (the SDSSRM-XCS$_{\rm unm}$ sample, see Table~\ref{tab:samples}).  For these systems, we calculated upper limit luminosities in the following way. First, we assumed each RM cluster has a temperature of 3~keV and calculated $r_{\rm 500c}$ using Equation~\ref{equ:rdelta} (note, we only estimated upper limits within $r_{\rm 500c}$).  We used a fixed temperature of 3~keV for the upper limit analysis to avoid bias coming from the correlation between the richness and luminosity (as would happen if one were to estimate the temperature from $\lambda_{\rm RM}$).  The choice of 3~keV was motivated by previous studies \citep[e.g.,][who use 3~keV]{2019ApJS..244...22H}, and that the mean temperature of $\lambda_{\rm RM}\approx$20-30 clusters in our SDSSRM-XCS$_{T_{X},vol}$ sample is 2.5~keV, close to our assumed value.  The majority of the SDSSRM-XCS$_{\rm unm}$ clusters have $\lambda_{\rm RM} \approx20-30$.  We then measured a 3$\sigma$ upper limit on the count-rate within those apertures, using the SAS tool {\sc eregionanalyse}, which implements the method of \cite{1991ApJ...374..344K}.  Point and extended sources are masked out from the analysis.  The background region had radii with inner and outer values of 1.05$r_{\rm 500c}$ \& 1.5$r_{\rm 500c}$ respectively.

In order to convert the count-rate upper limit into a luminosity upper limit, we used an energy conversion factor (ECF).  First, an Auxiliary Response File (ARF) and Redistribution Matrix File (RMF) were produced at the position of the RM cluster, assuming the relevant overdensity radius.  The ARF and RMF were then used to generate a fake spectrum in {\tt xspec} using the {\sc fakeit} tool.  The process requires the use of a model with which to produce the fake spectrum, for which we assumed a {\sc Tbabs}$\times${\sc APEC} model (the same one used to estimate cluster properties as in Sect.~\ref{sec:specanalysis}).  We assumed an $nH$ calculated at the RM position, the redshift as determined by RM, and the abundance fixed at 0.3~Z$_{\odot}$.  The temperature was assumed to be 3~keV.  An arbitrarily high exposure time, of 100~ks, was used to generate the spectrum. The ECF was then calculated as the ratio between the count-rate and the measured flux from the fake spectrum.  Using this ECF, the count-rate upper limit is converted to a flux, and finally converted to a luminosity upper limit.  Using this method, we measure upper limit luminosities for 599 of the SDSSRM-XCS$_{\rm unm}$ sample (representing $\approx$80\% of the input sample)\footnote{The remaining clusters could not have an upper limit measured since they fall on or near a chip gap, or the region was masked due to the presence of a point source or un-associated extended source}.

\subsection{Introducing the various SDSSRM-XCS sub-samples}
\label{sec:cosmosample}

In Table~\ref{tab:analysissamples} we overview the various sub-samples of SDSSRM clusters that have been analysed in this work. The cluster sample that we use most (e.g., Sections~\ref{sec:ltrelation},~\ref{sec:xray_richness},~\ref{sec:targets}) is known as `SDSSRM-XCS$_{T_{X},vol}$'. It contains 150 clusters that have accurate temperature estimates, defined as having an average percentage temperature error of $T_{X,err}<$25\%, and falling in the redshift range corresponding to the SDSSRM volume-limited sample \citep[as estimated in][0.1$\le$ z $\le$0.35]{2014ApJ...785..104R}. The SDSSRM-XCS$_{T_{X},vol}$ sample is a subset of the SDSSRM-XCS$_{T_{X}}$ sample, which contains 381 clusters with $T_{X,err}<$100\% and no redshift limits imposed (see Sect.~\ref{sec:alldatarelations}).

The largest sub-sample of SDSSRM clusters with measured luminosities is known as  SDSSRM-XCS$_{L_{X}}$ and contains 456 clusters (no $z$ limits imposed).  In this case, the  $L_{X}$ values were estimated with a fixed (not fitted) $T_{X}$ parameter (see Sect.~\ref{sec:lxnotx}).  A subset of SDSSRM-XCS$_{L_{X}}$ clusters in the 0.1$\le$ z $\le$0.35 range has 178 entries and is known as SDSSRM-XCS$_{L_{X},vol}$.

Finally, the SDSSRM-XCS$_{L_{X}}$ sample is supplemented with upper limit luminosities determined for the SDSSRM-XCS$_{\rm unm}$ sample.  These luminosities are added to the SDSSRM-XCS$_{L_{X}}$ sample to create a sample of 1055 clusters, which we denote as the SDSSRM-XCS$_{L_{X}+upper}$ subset (no $z$ limits).  A subset of SDSSRM-XCS$_{L_{X}+upper}$ clusters in the 0.1$\le$ z $\le$0.35 range has 222 entries and is known as SDSSRM-XCS$_{L_{X}+upper,vol}$.  This subset is used in the analyses presented in Section~\ref{sec:alldatarelations}.

A data table containing properties for the cluster sample outlined in this work can be found at \href{http://users.sussex.ac.uk/pag22/SDSSRM-XCS/sdssrm-xcs-sample-data.csv}{data table}, along with a \href{http://users.sussex.ac.uk/pag22/SDSSRM-XCS/column_names.txt}{table description}.

\subsubsection{Comparison to the literature}
\label{sec:litcompv2}

\subsubsubsection{Sample size}

\begin{figure*}
\begin{center}
\begin{tabular}{cc}
\includegraphics[width=8.0cm]{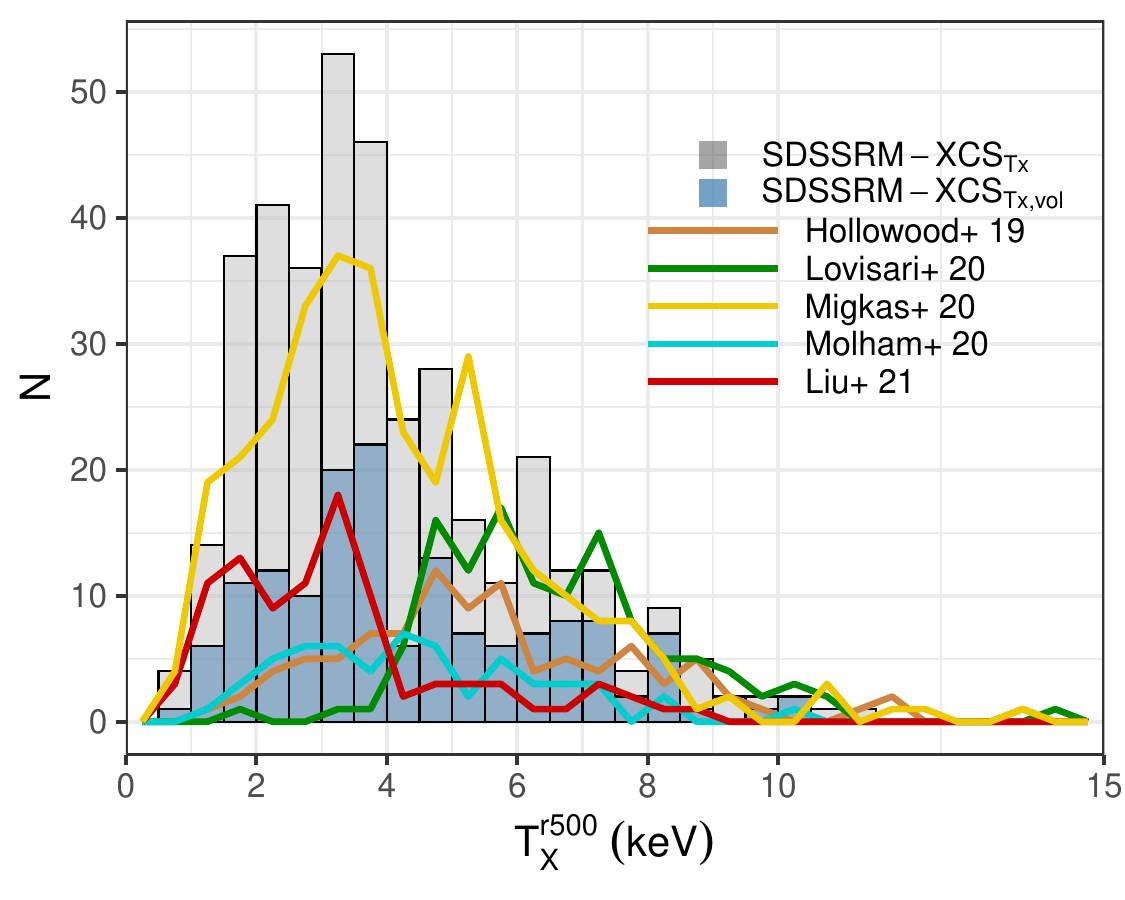} &
\includegraphics[width=8.0cm]{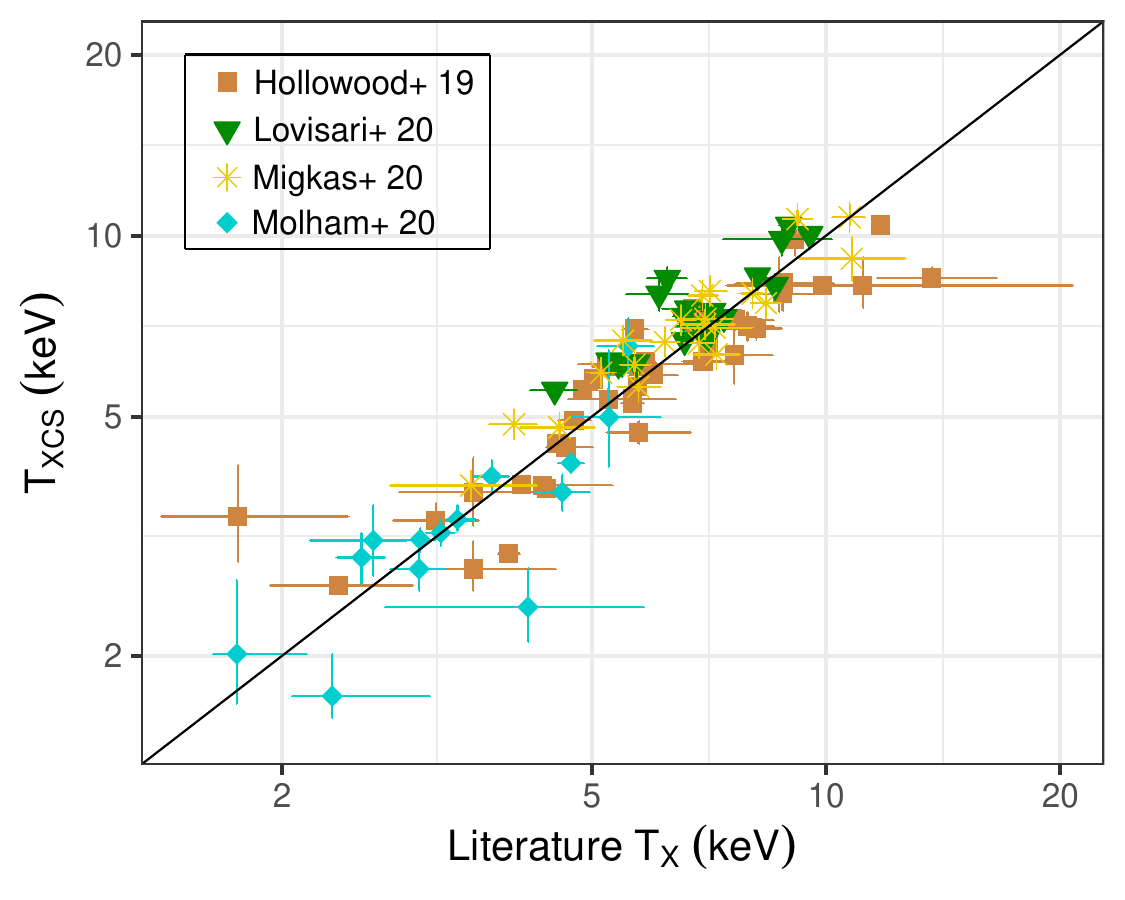} \\
(a) & (b) \\
\end{tabular}
\end{center}
\caption[]{(a) Temperature distribution of the SDSSRM-XCS$_{T_{X}}$ (grey histogram) and SDSSRM-XCS$_{T_{X},vol}$ (lightblue histogram) samples.  Distributions from \citet[][brown line]{2019ApJS..244...22H}, \citet[][green line]{2020ApJ...892..102L}, \citet[][yellow line]{2020A&A...636A..15M}, \citet[][cyan line]{2020MNRAS.494..161M} and \citet[][red line]{2021arXiv210614518L} are highlighted for comparison.  Note that \cite{2019ApJS..244...22H} and \cite{2020A&A...636A..15M} temperatures are scaled from {\em Chandra} to {\em XMM} using \cite{2016ApJS..224....1R} and  \cite{2015A&A...575A..30S} respectively, and the \cite{2021arXiv210614518L} temperatures are scaled by the offset between {\em eROSITA} and {\em XMM} found in \cite{2021arXiv210911807T}.  (b) Comparison of the measured temperatures in the SDSSRM-XCS$_{T_{X},vol}$ sample to \citet[][brown squares]{2019ApJS..244...22H}, \citet[][green downward triangles]{2020ApJ...892..102L}, \citet[][yellow stars]{2020A&A...636A..15M} and \citet[][cyan diamonds]{2020MNRAS.494..161M}, with 43, 20, 20 and 15 clusters in common respectively.  In most cases $T_{X}^{r500}$ is compared, however, in the case of \cite{2020A&A...636A..15M} we compare our $T_{X}^{(0.15-1)r_{500}}$ to their $T_{X}^{(0.2-0.5)r_{500}}$.  The 1:1 relation is given by the black solid line.} \label{fig:littempcomp}
\end{figure*}

We have delivered one of the largest cluster samples with coherently measured $T_{X}$ values to date. The only equivalent sample is the XCS First Data release \citep[XCSDR1]{2012MNRAS.423.1024M}. XCSDR1 included 401 clusters with measured temperatures distributed across the entire extragalactic sky (i.e. extending beyond the SDSSDR8 footprint). In Figure~\ref{fig:littempcomp}(a), we show the number of SDSSRM-XCS$_{T_{X}}$ clusters per $T_{X}$ bin (the subset of 150 in the SDSSRM-XCS$_{T_{X},vol}$ subsample is highlighted in blue).  The $T_{X}$ value distributions from a non-comprehensive list of other recently published samples are over-plotted as spline curves, described by the following:

\begin{itemize}

\item{The cyan curve shows the 57 {\em XMM} derived $T_{X}$ values for SDSS DR8 RM clusters (spanning a redshift range of 0.14$< z <$0.59) in the \citet[][hereafter Mol20]{2020MNRAS.494..161M} sample. This sample is a subset of the X-CLASS catalog \citep{2012MNRAS.423.3545C,2021A&A...652A..12K}.
}

\item{The red curve shows the 95 {\em eROSITA} derived $T_{X}$ values X-ray selected clusters (spanning a redshift range 0.049$< z <$0.708) in the \citet[][hereafter L21]{2021arXiv210614518L}\footnote{\href{https://erosita.mpe.mpg.de/edr/eROSITAObservations/Catalogues/}{Liu+21 sample}}. This sample is a subset of 542 clusters extracted from the 140 deg$^{2}$ contiguous {\em eROSITA} Final Equatorial-Depth Survey (eFEDS). We note that the eFEDS $T_{X}$ values were derived from spectra extracted from a circular $r<$ 500~kpc region, and that they have been scaled by a factor of 1.25 to account for the measured $T_{X}$ offset between {\em eROSITA} and {\em XMM} \citep[][]{2021arXiv210911807T}.}

\item{The brown curve shows the 97 {\em Chandra} derived $T_{X}$ values for SDSSRM clusters ($0.1 \leq z \leq 0.35$) in the \citet[][hereafter H19]{2019ApJS..244...22H}\footnote{\href{ftp://cdsarc.u-strasbg.fr/pub/cats/J/ApJS/244/22}{Hollowood+ 19 sample}} sample. For the purposes of illustration, the {\em Chandra} $T_{X}$ values are scaled to {\em XMM} using the calibration found in \cite{2016ApJS..224....1R}.}  

\item{The green curve shows the 120 {\em XMM} derived $T_{X}$ values for {\em Planck} clusters (spanning a redshift range of 0.059$< z <$0.546) in the \citet[hereafter L20]{2020ApJ...892..102L}\footnote{\href{https://vizier.cds.unistra.fr/viz-bin/VizieR?-source=J/ApJ/892/102}{Lovisari+ 20 sample}}. This sample is a subset of the {\em Planck} Early Sunyaev-Zeldovich \citep{2011A&A...536A...8P} cluster catalog.}

\item{The yellow curve shows the 313 {\em XMM} and {\em Chandra} derived $T_{X}$ values for X-ray selected clusters (spanning a redshift range 0.004$< z <$0.447, with 70\% at $z<$0.1) in the \citet[][hereafter Mig20]{2020A&A...636A..15M} sample\footnote{\href{https://vizier.cds.unistra.fr/viz-bin/VizieR?-source=J/A+A/636/A15}{Migkas+ 20 sample}}.  This sample is a subset of the Meta-Catalogue of X-ray detected Clusters of galaxies \citep[MCXC][]{2011A&A...534A.109P}. We note that the {\em Chandra} values in Mig20 are scaled to {\em XMM} using \citet[][as used in Mig20]{2015A&A...575A..30S}, and that all 313 $T_{X}$ values were derived from spectra extracted from a (0.2-0.5)$r_{500}$ region.}

\end{itemize}

\subsubsubsection{Temperature estimates}

To demonstrate the reliability of the $T_{X}$ values estimated in this work, we have compared our values (using the SDSSRM-XCS$_{T_{X},vol}$ sample) to those for clusters in common with the H19, L20, Mig20 and Mol20 samples mentioned above. There are 43, 20, 20 and 15 examples respectively. For the purposes of this comparison, a {\em Chandra}-to-{\em XMM} scaling as been applied to the H19 and Mig20 $T_{X}$ values as described above. Figure~\ref{fig:littempcomp}(b) plots the comparison of the temperature for these three literature samples.  The black line shows the 1:1 relation, highlighting both that the various $T_{X}$ measurements are broadly consistent, and that the XCS values generally have smaller errors.

\begin{table*}
\centering
\caption[]{{\small Summary of the SDSSRM cluster sub-samples used for scaling relation analysis in this work.}\label{tab:analysissamples}}
\vspace{1mm}
\begin{tabular}{l|lcc}
\hline
\hline
 Sample & Brief description & \# clusters & Relevant sections  \\
\hline
SDSSRM-XCS$_{T_{X}}$ & \hspace{-3.5mm}\begin{tabular}{@{}l@{}}SDSSRM-XCS$_{\rm ext}$ clusters with a measured temperature value \\ (with $T_{X,err}<$100\%) \end{tabular} & 381 & $\S$\ref{sec:cosmosample} \\
\hline
SDSSRM-XCS$_{T_{X},vol}$ & As above, but limited to systems with 0.1$<$z$<$0.35 and $T_{X,err}<$25\% & 150 & $\S$\ref{sec:cosmosample} \\
\hline
SDSSRM-XCS$_{L_{X}}$ & \hspace{-3.5mm}\begin{tabular}{@{}l@{}}SDSSRM-XCS$_{\rm ext}$ clusters where the luminosity was measured \\ assuming a fixed temperature \end{tabular} & 456 & $\S$\ref{sec:lxnotx} \\
\hline
SDSSRM-XCS$_{L_{X},vol}$ & As above, but limited to systems with 0.1$<$z$<$0.35 & 178 & $\S$\ref{sec:lxnotx} \\
\hline
SDSSRM-XCS$_{L_{X}+upper}$ & \hspace{-3.5mm}\begin{tabular}{@{}l@{}}SDSSRM-XCS$_{L_{X}}$ sample, supplemented with upper limit luminosities \end{tabular} & 1055 & $\S$\ref{sec:lxnotx}, \ref{sec:calcupperlim} \\
\hline
SDSSRM-XCS$_{L_{X}+upper,vol}$ & As above, but limited to systems with 0.1$<$z$<$0.35 & 222 & $\S$\ref{sec:lxnotx}, \ref{sec:calcupperlim} \\
\hline
\end{tabular}
\end{table*}

\section{Scaling relations derived from the SDSSRM-XCS samples}
\label{sec:relations}

In this section, we present the scaling relations derived from some of the SDSSRM-XCS samples described in Section~\ref{sec:cosmosample}  and Table~\ref{tab:analysissamples}.
In Sections~\ref{sec:ltrelation} and~\ref{sec:xray_richness}, we focus on the sample with the most robustly measured X-ray properties and that is restricted to the RM volume limited redshift range i.e., the SDSSRM-XCS$_{T_{X},vol}$ cluster sample (see Sect.~\ref{sec:cosmosample}).  
In Section~\ref{sec:alldatarelations} we present fits to samples with less conservative cuts, to explore the relative importance of sample size over measurement accuracy.
Fits to the scaling relations were performed in log space using the R package LInear Regression in Astronomy ({\sc lira}\footnote{{\sc lira} is available as an R package from https://cran.r-project.org/web/packages/lira/index.html}), fully described in \cite{LIRA}.  Formally, scaling relations are fitted with a power-law of the form
\begin{align}
Y = A + B\cdot Z \pm \epsilon
\end{align}
where ${\rm var}(\epsilon)=\sigma^{2}_{Y|Z}$ and $Z$ is the intrinsic cluster property.  For simplicity, throughout, the scaling relations are denoted by the cluster properties in question and the scatter given by $\sigma$ (for example, see Equ~\ref{equ:lt}).  For these analyses we used core-included temperatures and soft band luminosities within $r_{\rm 500c}$ unless otherwise stated.  Temperature and luminosities estimated in this way are denoted $T_{X}^{r500}$ and $L_{X,52}^{r500}$ respectively.

\begin{table*}           
\begin{center}
\caption[]{{\small Best-fit parameters of the cluster scaling relations (see $\S$\ref{sec:ltrelation} and $\S$\ref{sec:xray_richness} for details).  For each relation, parameters are given for the SDSSRM-XCS$_{T_{X},vol}$ ($T_{X,err}<25\%$ and 0.1$\le$z$\le$0.35) cluster sample.  Best-fit parameters are given for the $L_{X}-T_{X}$, $L_{X}-\lambda_{\rm RM}$ and $T_{X}-\lambda_{\rm RM}$ relations, given by equations~\ref{equ:lt},~\ref{equ:l-lambda} and~\ref{equ:t-lambda} respectively.}\label{tab:bestfit}}
\vspace{1mm}
\begin{tabular}{cccccc}
\hline
\hline
 Relation & Fit & Normalisation & Slope & Scatter & Figure \\
(sample) & & & & & \\
\hline
$L^{r500}_{X,52}-T^{r500}_{X}$ &  & $A_{LT}$ & $B_{LT}$ & $\sigma_{LT}$ & \\ 
SDSSRM-XCS$_{T_{X},vol}$ & {\sc lira} & 0.97$\pm$0.06 & 2.63$\pm$0.12 & 0.68$\pm$0.04 & \ref{fig:ltrelation} \\
 & {\sc linmix} & 0.98$\pm$0.06 & 2.63$\pm$0.12 & 0.69$\pm$0.03 & -- \\
SDSSRM-XCS$_{T_{X}}$ & {\sc lira} & 0.94$\pm$0.04 & 2.49$\pm$0.08 & 0.64$\pm$0.03 & \ref{fig:all_data_relations}(a) \\
\hline
$L^{r500}_{X,bol}-T^{r500}_{X}$ & & $A_{LbT}$ & $B_{LbT}$ & $\sigma_{LbT}$  & \\ 
SDSSRM-XCS$_{T_{X},vol}$ & {\sc lira} & 3.05$\pm$0.18 & 3.07$\pm$0.12 & 0.68$\pm$0.04 & -- \\
\hline
$L^{r500}_{X,52}-\lambda_{\rm RM}$ & & $A_{L\lambda}$ & $B_{L\lambda}$ & $\sigma_{L\lambda}$ & \\ 
SDSSRM-XCS$_{T_{X},vol}$ & {\sc lira} & 0.98$\pm$0.09 & 1.61$\pm$0.14 & 1.07$\pm$0.06  & \ref{fig:x-ray_richness}(a) \\
 & {\sc linmix} & 0.98$\pm$0.09 & 1.62$\pm$0.14 & 1.08$\pm$0.06 & -- \\
SDSSRM-XCS$_{L_{X+upper},vol}$ & {\sc lira} & 1.08$\pm$0.10 & 1.84$\pm$0.12 & 1.09$\pm$0.06 & \ref{fig:all_data_relations}(b) \\
\hline
$T^{r500}_{X}-\lambda_{\rm RM}$ & & $A_{T\lambda}$ & $B_{T\lambda}$ & $\sigma_{T\lambda}$ \\ 
SDSSRM-XCS$_{T_{X},vol}$ & {\sc lira} & 1.01$\pm$0.03 & 0.59$\pm$0.04 & 0.33$\pm$0.02  & \ref{fig:x-ray_richness}(b) \\
 & {\sc linmix} & 1.01$\pm$0.03 & 0.59$\pm$0.05 & 0.33$\pm$0.01 & -- \\
\hline
\end{tabular}
\end{center}
\end{table*}

\subsection{The Luminosity-Temperature relation derived from the SDSSRM-XCS$_{T_{X},vol}$ sample}
\label{sec:ltrelation}

\begin{figure}
	\begin{centering}
    \includegraphics[width=.46\textwidth]{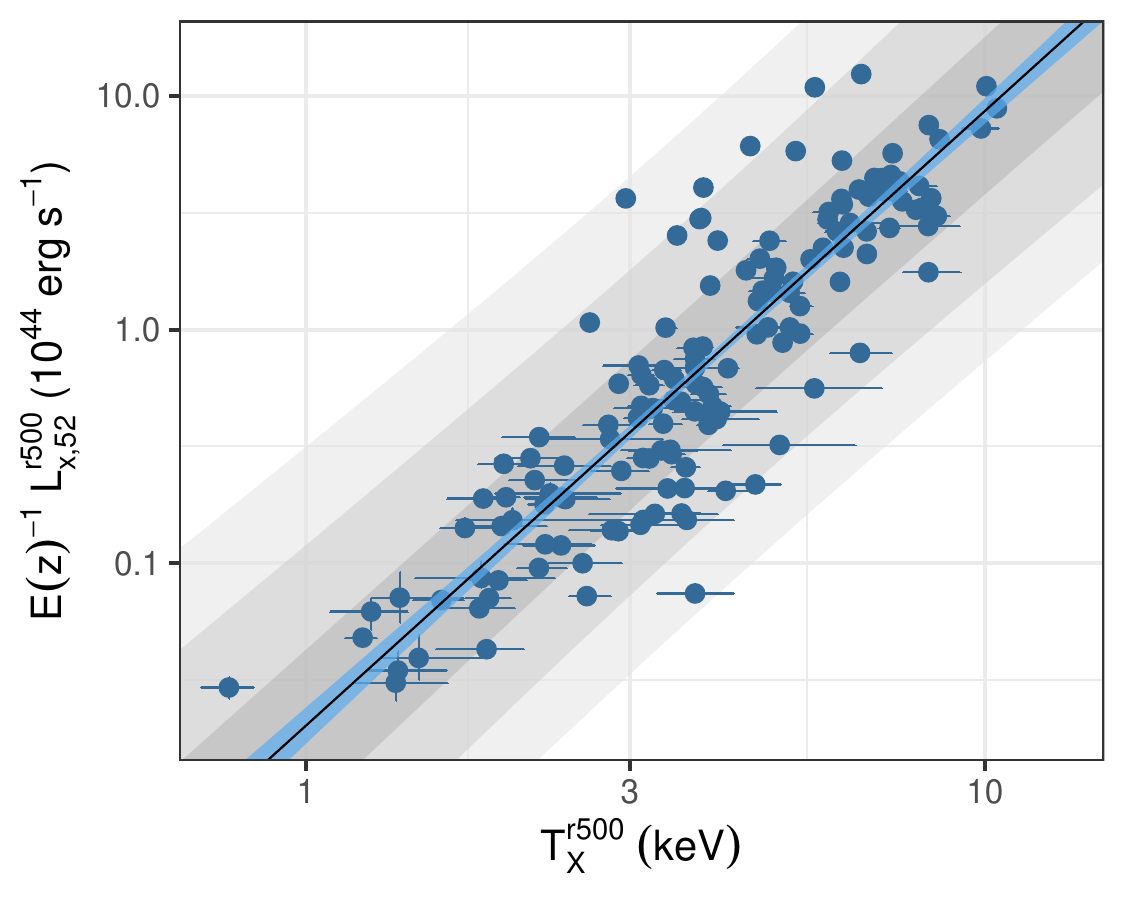}
	\caption[]{Luminosity-Temperature relation of the SDSSRM-XCS$_{T_{X},vol}$ subset (blue circles).  The best-fit to the data (see Sect.~\ref{sec:ltrelation}) is represented by the black solid line, and the lightblue shaded region represents the 68\% confidence interval.  The grey bands represent the 1, 2, and 3$\sigma$ intrinsic scatter. \label{fig:ltrelation}}
	\end{centering}
\end{figure}

The $L^{r500}_{X,52}-T^{r500}_{X}$ relation is shown in Figure~\ref{fig:ltrelation}.  The SDSSRM-XCS$_{T_{X},vol}$ data points are shown as blue circles.  A power law relation between $L^{r500}_{X,52}$ and $T^{r500}_{x}$ is fit to the data, which we express as
\begin{align}
{\rm log}\left(\frac{L^{r500}_{X,52}}{E(z)^{\gamma_{LT}}L_{0}}\right) &= {\rm log}(A_{LT}) + B_{LT}{\rm log}\left(\frac{T^{r500}_{X}}{T_{0}}\right) \pm \sigma_{L|T},
\label{equ:lt}
\end{align}
where $A_{LT}$ denotes the normalisation, $B_{LT}$ the slope, $\gamma_{LT}$ the evolution with redshift and $\sigma_{L|T}$ the intrinsic scatter.  Note that the intrinsic scatter is given in natural log space and can be interpreted as the fractional scatter.  We assumed $T_{0}=4$ keV, $L_{0}=0.8\times10^{44}$ erg s$^{-1}$ (roughly the median values for the SDSSRM-XCS$_{T_{X},vol}$ sample) and a self-similar evolution of the relation where $\gamma_{LT}=1$.  The fit to the SDSSRM-XCS$_{T_{X},vol}$ sample is highlighted by the blue solid line in Figure~\ref{fig:ltrelation}, with the lightblue shaded region representing the 68\% uncertainty. The grey bands represent the 1, 2 and 3$\sigma$ intrinsic scatter.  This scaling relation was used to estimate luminosities when the $T_{X}$ was fixed, rather than fitted (see Sect.~\ref{sec:lxnotx}).  The best-fit {\sc lira} parameters of the $L^{r500}_{X,52}-T^{r500}_{X}$ relations are given in Table~\ref{tab:bestfit}.  For comparison, we performed a fit using the {\sc linmix} routine \citep{2007ApJ...665.1489K}, with best-fit parameters also given in Table~\ref{tab:bestfit}.  Many literature studies using X-ray luminosities determine relations using the bolometric luminosity.  Therefore, we also fitted the bolometric luminosity - temperature ($L^{r500}_{X,bol}-T^{r500}_{X}$) relation, with the best-fit parameters from given in Table~\ref{tab:bestfit}.  

\subsection{The X-ray observable-Richness relations derived from the SDSSRM-XCS$_{T_{X},vol}$ sample}
\label{sec:xray_richness}

\begin{figure*}
\begin{center}
\begin{tabular}{cc}
\includegraphics[width=8.0cm]{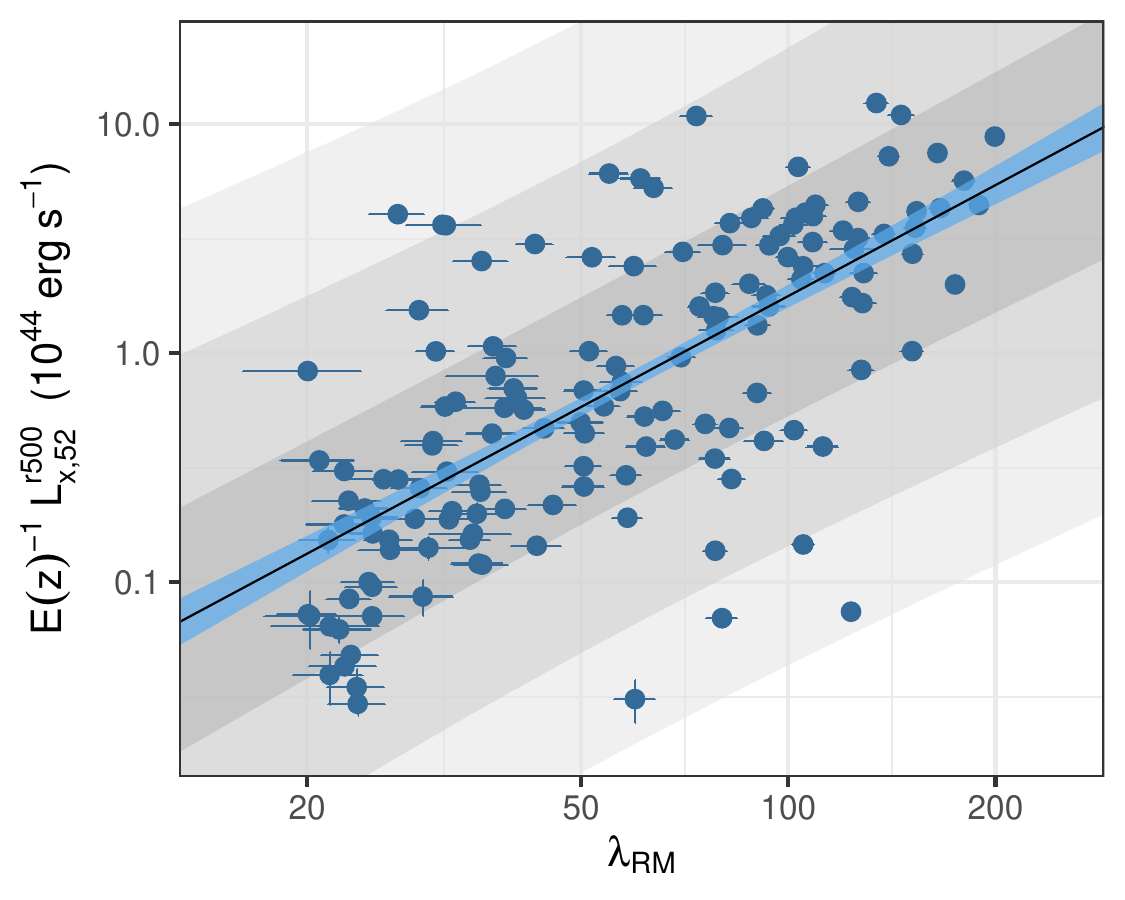} &
\includegraphics[width=8.0cm]{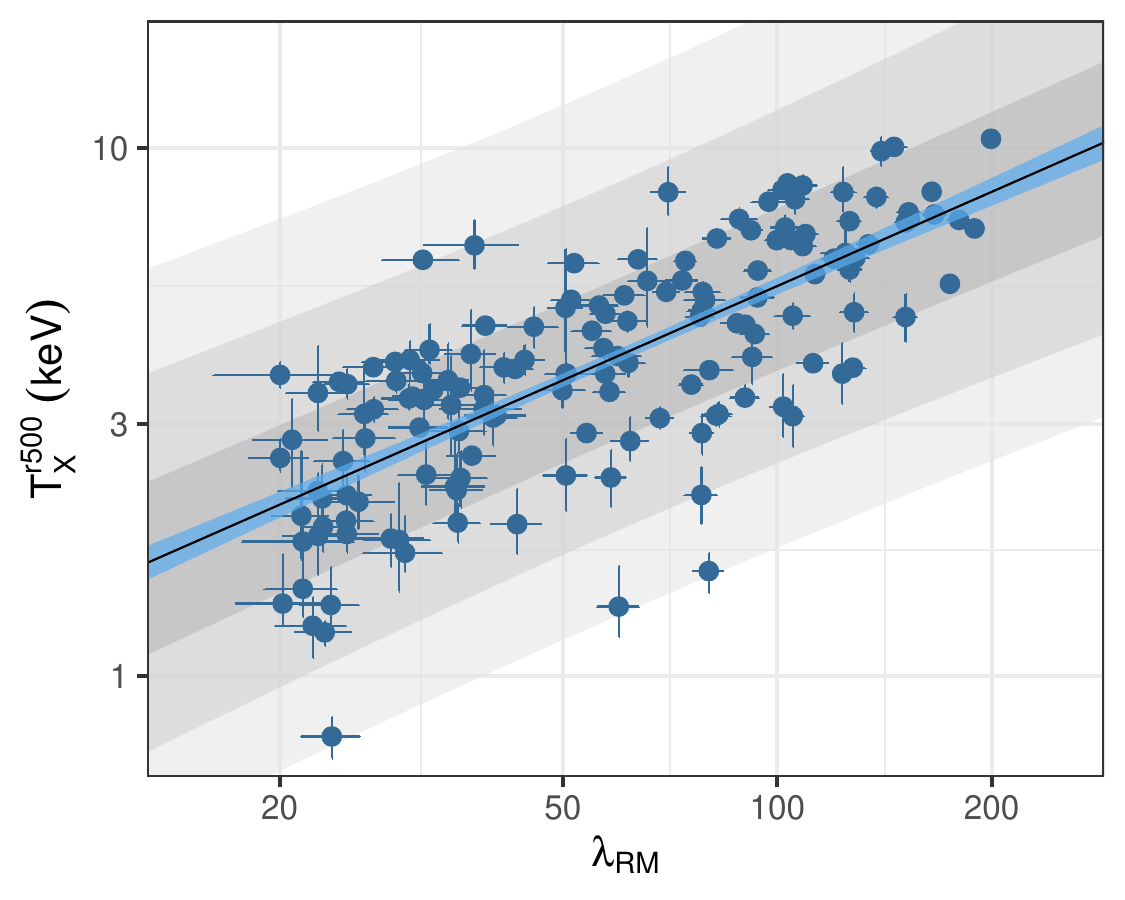} \\
(a) & (b) \\
\end{tabular}
\end{center}
\caption[]{Luminosity-richness (a) and temperature-richness (b) relation of the SDSSRM-XCS$_{T_{X},vol}$.  For each relation, the best-fit to the data (see Sect.~\ref{sec:xray_richness}) is represented by the blue solid line, and the light blue shaded region represents the 68\% confidence interval of the mean logarithmic relation.  The grey bands represent the 1, 2, and 3$\sigma$ intrinsic scatter. \label{fig:x-ray_richness}} 
\end{figure*}

The $L_{X,52}^{r500}-\lambda_{\rm RM}$ and $T_{X}^{r500}-\lambda_{\rm RM}$ relations for the SDSSRM-XCS$_{T_{X},vol}$ (steel blue circles) are shown in Figures~\ref{fig:x-ray_richness}(a) and (b) respectively.  We fit for the $L_{X,52}^{r500}-\lambda_{\rm RM}$ and $T_{X}^{r500}-\lambda_{\rm RM}$ relations, again, which we express as:

\begin{align}
{\rm log}\left(\frac{L_{X,52}^{r500}}{E(z)^{\gamma_{L\lambda}} L_{0}}\right) &=  {\rm log}(A_{L\lambda}) + B_{L\lambda}{\rm log}\left(\frac{\lambda_{\rm RM}}{\lambda_{0}}\right) \pm \sigma_{L|\lambda},
\label{equ:l-lambda}
\end{align}

\begin{align}
{\rm log}\left(\frac{T_{X}^{r500}}{T_{0}}\right) &= {\rm log}(A_{T\lambda}) + B_{T\lambda}{\rm log}\left(\frac{\lambda_{\rm RM}}{\lambda_{0}}\right) \pm \sigma_{T|\lambda},
\label{equ:t-lambda}
\end{align}
where $A_{L\lambda}$ and $A_{T\lambda}$ denote the normalisations, $B_{L\lambda}$ and $B_{T\lambda}$ represent the slopes and $\sigma_{T|\lambda}$ and $\sigma_{L|\lambda}$ denote the intrinsic scatters (once again the values are given in natural log space).  We assumed $L_{0}=0.8\times10^{44}$ erg s$^{-1}$ in equation~\ref{equ:l-lambda} and $T_{0}=4$ keV in equation~\ref{equ:t-lambda}, and in both relations assumed $\lambda_{0}=60$ (again, all roughly corresponding to the median values for the SDSSRM-XCS$_{T_{X},vol}$ sample).  Self-similar evolution for each relation is assumed such that $\gamma_{L\lambda}=1$ in equation~\ref{equ:l-lambda}. We note that the $E(z)$ correction cancels out in the $T_{X}-\lambda_{\rm RM}$ relation (hence the absence of the $E(z)$ parameter in Equ.~\ref{equ:t-lambda}).  The best-fit {\sc lira} parameters for each relation are given in Table~\ref{tab:bestfit} (again, {\sc linmix} parameters are also provided for comparison) and the best-fit relations are given by the blue solid lines in Figures~\ref{fig:x-ray_richness}(a) and (b), with the 68\% uncertainty given by the light blue shaded region.  The grey bands represent the 1, 2 and 3$\sigma$ intrinsic scatter.  A comparison of these results to those in the literature are presented in Section~\ref{sec:litcomp}.

In summary, we find that the measured scatter of the $L_{X}-\lambda_{\rm RM}$ relation is roughly three times that of the $T_{X}-\lambda_{\rm RM}$. This is not due to measurement error (indeed the percentage errors on the $L_{X}$ values are much smaller than those on the $T_{X}$ values) but likely because non-gravitational physics impacts the luminosity to a much greater extent than it does the temperature. Even expanding the sample of $L_{X}$ values by a large factor (as will be possible with the eROSITA All Sky Survey \citealt[][]{2021A&A...647A...1P}) will not bring the scatter down below that shown in Figure~\ref{fig:x-ray_richness}(a).

\subsection{Scaling relations with all available X-ray data}
\label{sec:alldatarelations}

The SDSSRM-XCS$_{T_{X},vol}$ sample only contains a fraction of the X-ray information available for SDSSRM-XCS clusters. In this section, we investigate whether there is a benefit to including additional clusters with less precise individual measurements.  

\begin{figure*}
\begin{center}
\begin{tabular}{cc}
\includegraphics[width=8.3cm]{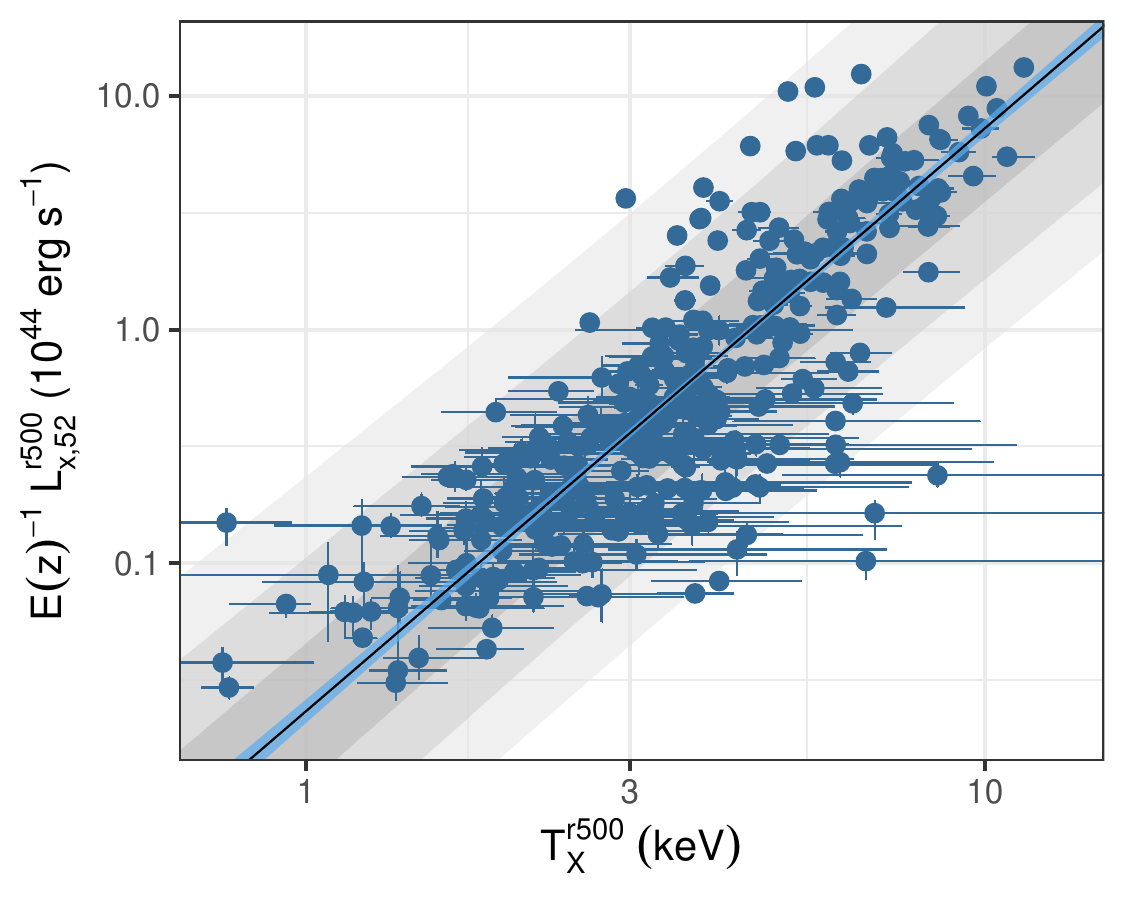} &
\includegraphics[width=8.3cm]{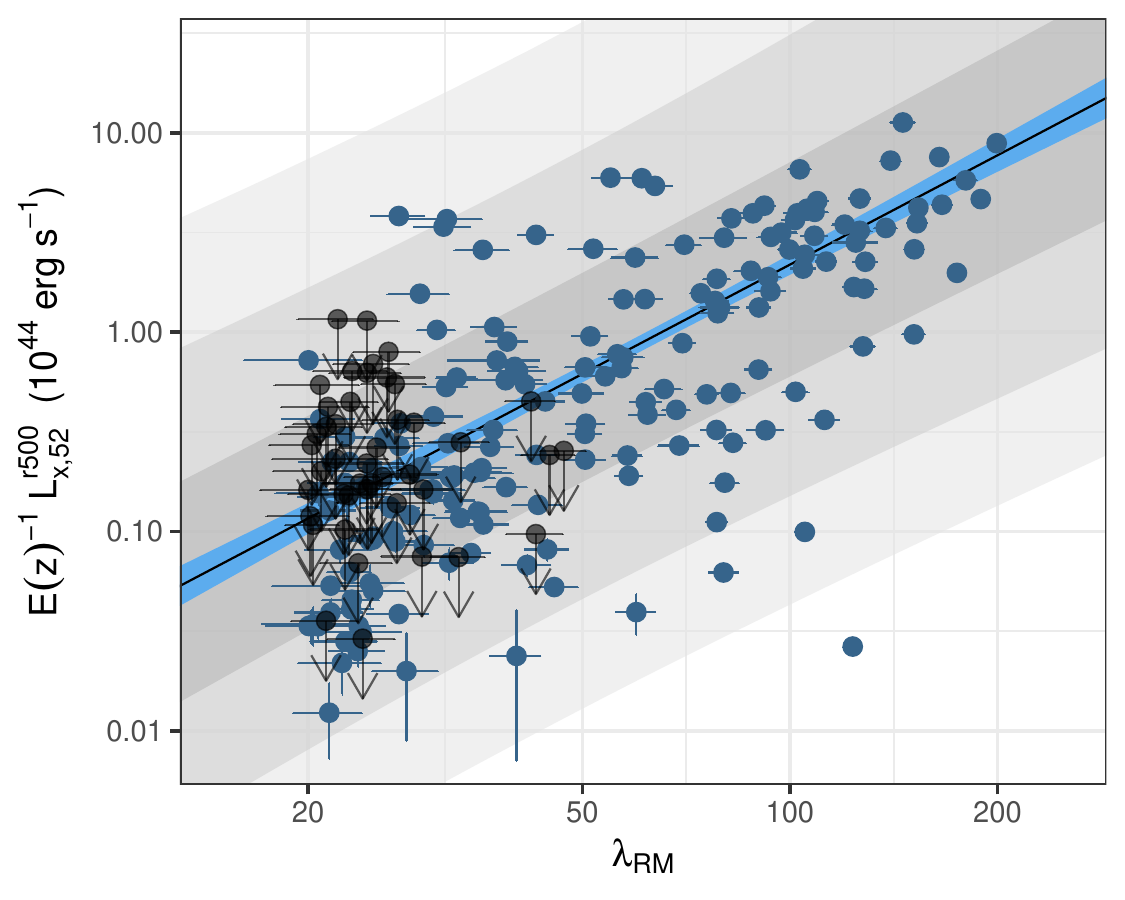} \\
(a) & (b) \\
\end{tabular}
\end{center}
\caption[]{(a) Luminosity-temperature relation of the SDSSRM-XCS$_{T_{X}}$ sample (blue circles).  The best-fit to the data is represented by the black solid line, and the lightblue shaded region represents the 68\% uncertainty. (b) Luminosity-richness relation of the SDSSRM-XCS$_{L_{X}+upper,vol}$ subset.  Clusters are given by the blue points and luminosity upper limits for SDSSRM-XCS without an XCS detection given by the black circles (and downward arrows).  The black solid line represents a fit to the data (including upper limits) with the lightblue shaded region highlighting the 68\% uncertainty.  In each plot, the grey bands represent the 1, 2, and 3$\sigma$ intrinsic scatter.} 
\label{fig:all_data_relations}
\end{figure*}

To explore the impact of $T_{X}$ measurement errors on the derived $L_{X}-T_{X}$ relation (see Sect.~\ref{sec:ltrelation}), we have added all 381 clusters with a measured $T_{X}$ value in the SDSS-XCS$_{T_{X}}$ sample. The results are shown in Figure~\ref{fig:all_data_relations}(a) and best-fit parameters given in Table~\ref{tab:bestfit}. It is clear that there is no significant change in the fitted relation when less accurate $T_{X}$ values are included. There is some marginal benefit to including more clusters in the fit (e.g. the scatter drops a little, although not significantly).

To explore the impact of $L_{X}$ measurement errors, and, to some extent, sample incompleteness, on the derived luminosity-richness relation, we make use of luminosities estimated with a fixed temperature (see Sect.~\ref{sec:lxnotx}) for all 456 clusters in the SDSSRM-XCS$_{\rm ext}$ sample, combined with luminosity upper limits (Section~\ref{sec:calcupperlim}) where available. 
The results are shown in Figure~\ref{fig:all_data_relations}(b) and best-fit parameters given in Table~\ref{tab:bestfit}. It is clear that when less accurate $L_{X}$ values, and upper limits, are included that the measured scatter goes up a little, but does not change significantly. However, there are perceptible changes to the slope and normalisation, which are likely a result of a combination of the change in $L_{X}$ measurement method, and in the selection function. 

In summary, it is probably worthwhile including all available $T_{X}$ values when assessing $T_{X}-\lambda$ scatter for cosmological studies, i.e. the fitted parameters are robust to both measurement errors and selection effects. However, one should exercise more caution when using $L_{X}-\lambda$ relations. The impact of selection on the $L_{X}-\lambda$ relation will be explored in \cite{upsdelldes}, which explores completeness and contamination in the low $\lambda$ regime using an XCS analysis of contiguous {\em XMM} survey regions (totalling $\sim$57 deg$^{2}$) that overlap with the DES Year 3 data release \citep{2018ApJS..239...18A}.

\subsection{Comparison to the literature}
\label{sec:litcomp}

\begin{figure*}
\begin{center}
\begin{tabular}{cc}
\includegraphics[width=8.0cm]{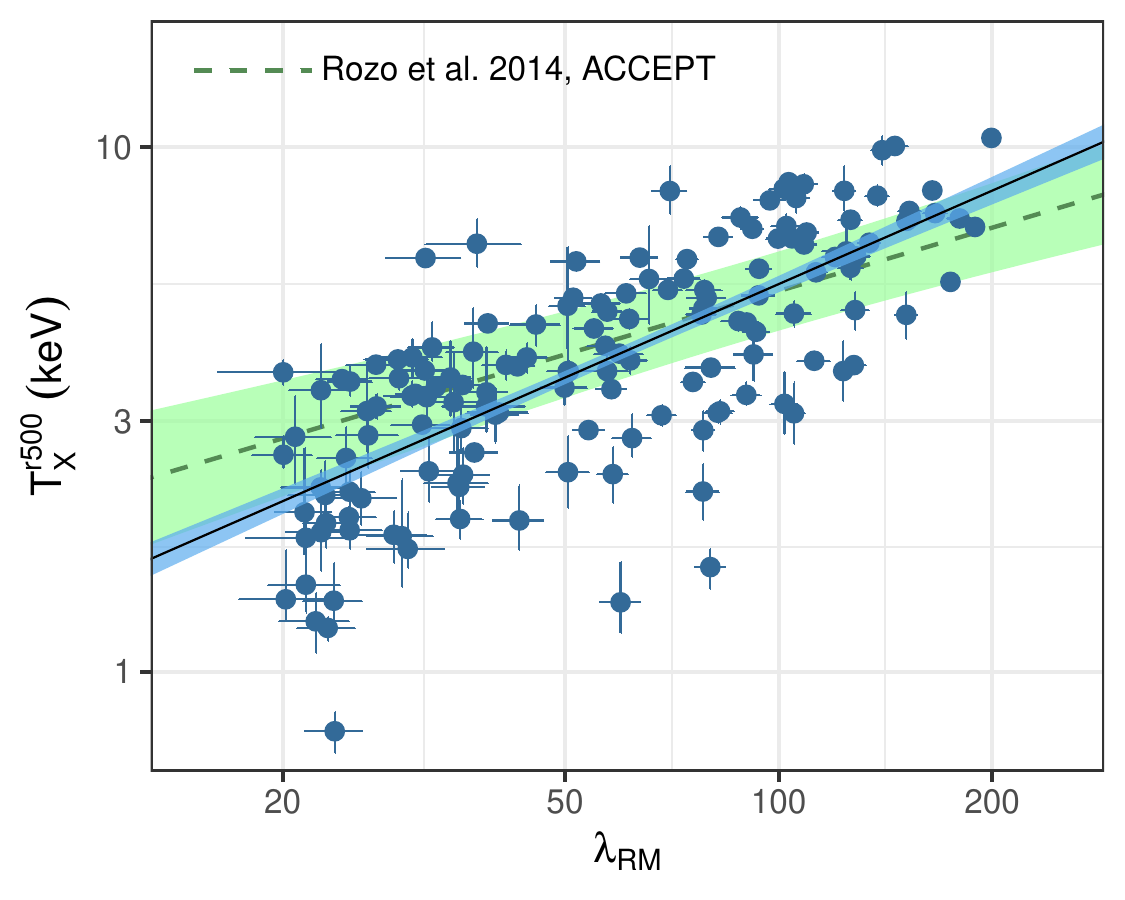} &
\includegraphics[width=8.0cm]{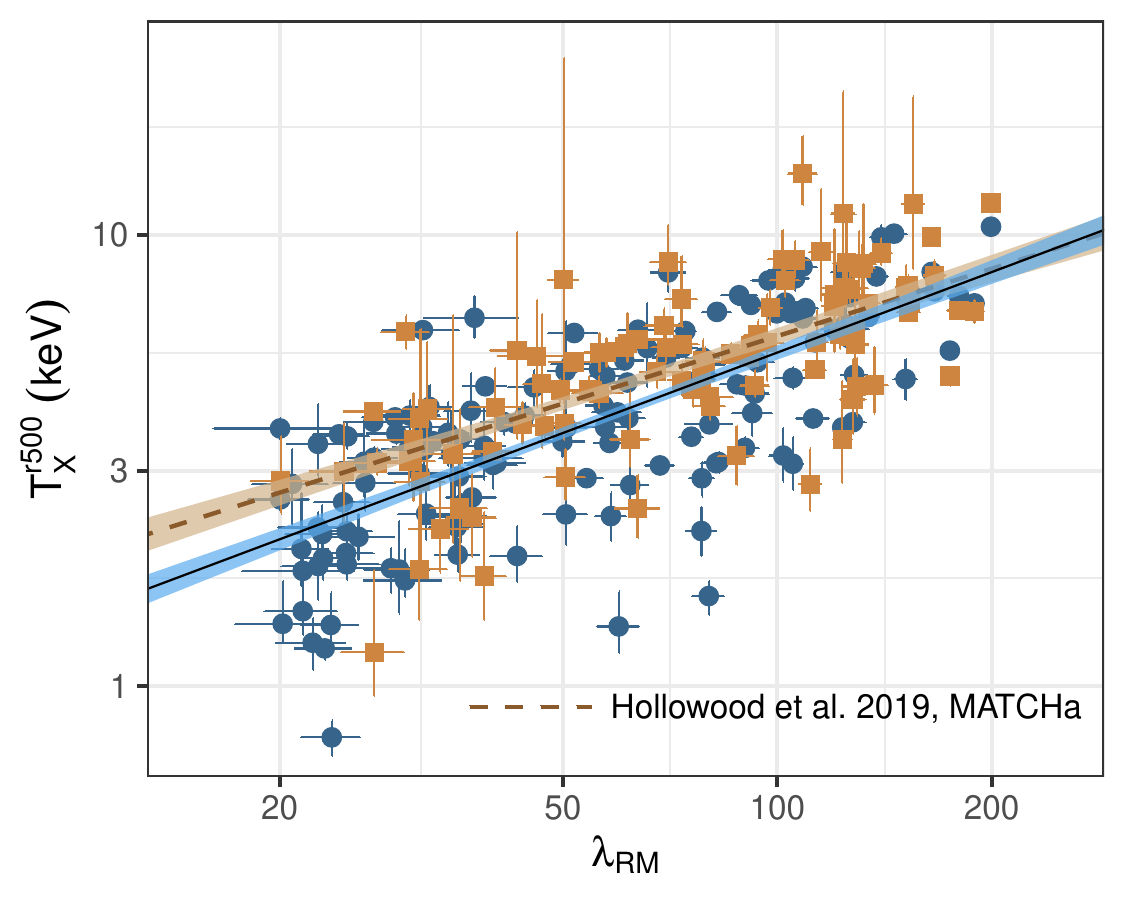} \\
(a) & (b) \\
\includegraphics[width=8.0cm]{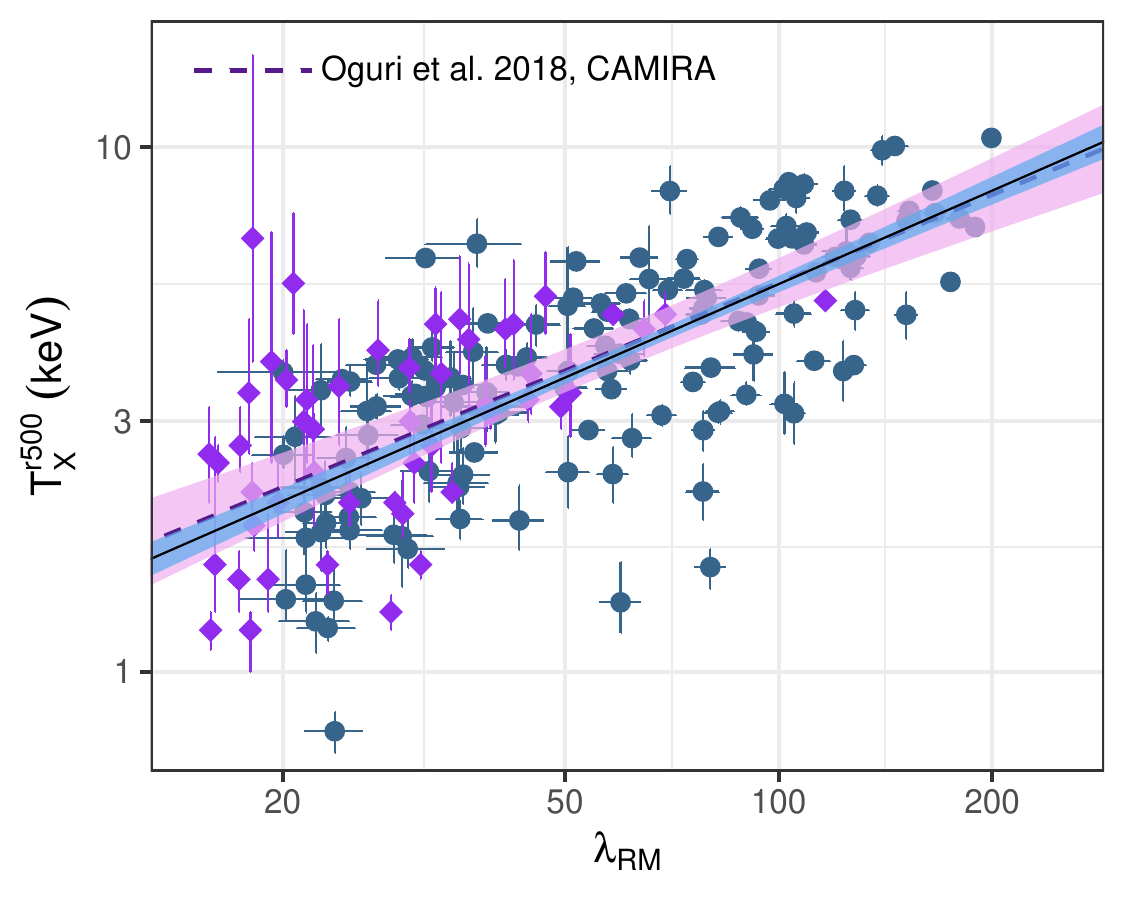} &
\includegraphics[width=8.0cm]{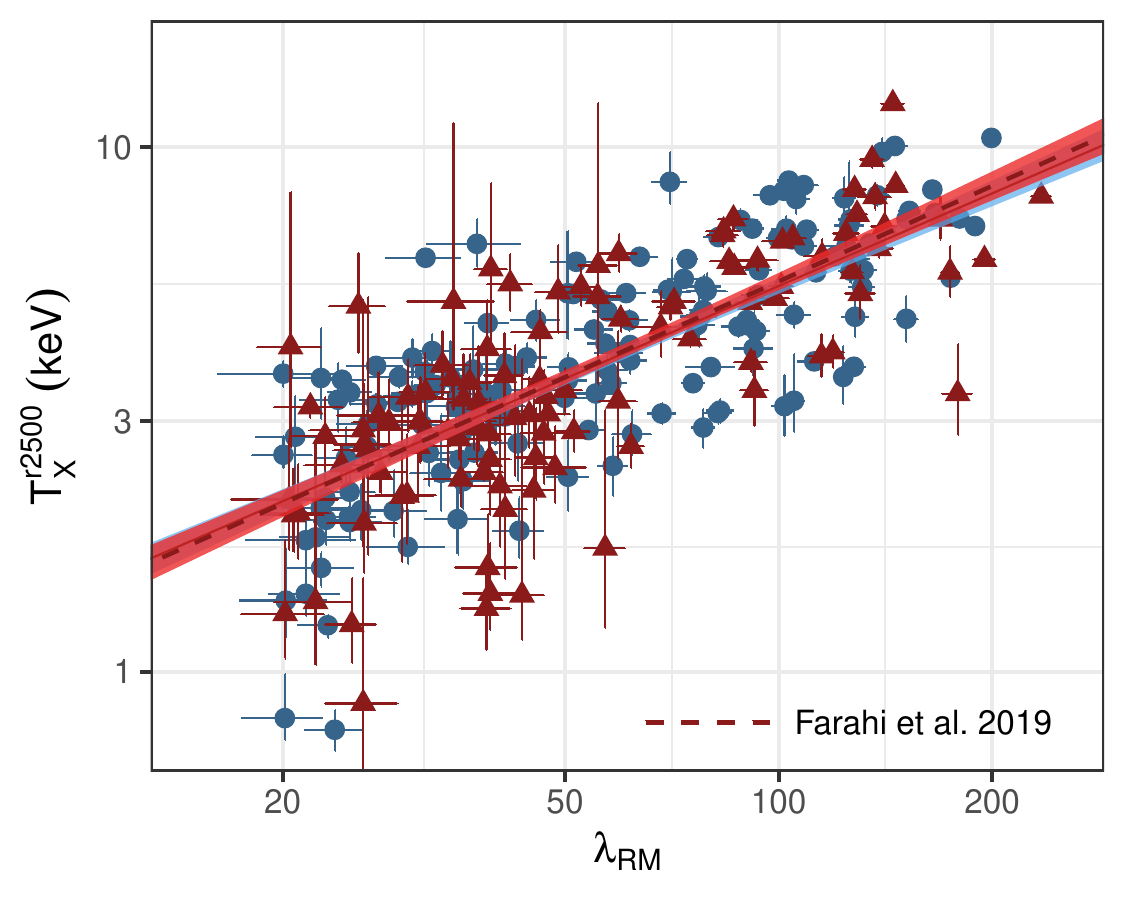} \\
(c) & (d) 
\end{tabular}
\end{center}
\caption[]{Comparison of the SDSSRM-XCS$_{T_{X},vol}$ temperature-richness relation to previously published relations.  In each case, the black solid line represents a fit to the SDSSRM-XCS$_{T_{X},vol}$ data (as given in Table~\ref{tab:bestfit}) and the light-blue shaded region the corresponding 68\% uncertainty.  Comparison to the fit provided in \citet[][green dashed line and light-green shaded region highlighting the 68\% uncertainty]{2014ApJ...783...80R} using the R14$_{\rm ACCEPT}$ sample \citep[note the relation has been scaled to {\em XMM} temperatures using][]{2016ApJS..224....1R} is shown in (a).  Comparison to the data given in \cite{2019ApJS..244...22H} is show in plot (b), scaled to {\em XMM} temperatures using \cite{2016ApJS..224....1R}, with the {\sc lira} fit to the data given by the brown dash line and the 68\% uncertainty given by the brown shaded region.  Comparison to the data given in \cite{2018PASJ...70S..20O} is shown in plot (c), with the {\sc lira} fit to the data given by the purple dash line and 68\% uncertainty given by the light-purple shaded region.  Note, the richness is estimated from CAMIRA, denoted $N_{\rm mem (CAMIRA)}$. Comparison to the data given in \citet[][dark-red triangles]{2019MNRAS.490.3341F} is shown in plot (d), with the {\sc lira} fit given by the dark-red dashed line and the 68\% uncertainty given by the red shaded region (note the agreement between the fits obscures much of the comparison, and in this case the $T_{\rm X}$ values were measured in a $r_{2500}$ apertures).}
\label{fig:tlambda_comp}
\end{figure*}

Figure~\ref{fig:lumin_temp_comps}(b) and (d) demonstrate that $T_{X}$ estimates values are insensitive to the details of the measurement process, be that the extraction aperture, or the inclusion of the cluster core.  Furthermore, the comparisons in Figure~\ref{fig:littempcomp}(b) show consistency of our measured $T_{X}$ to those in the literature.  Therefore, we can have confidence that comparisons of the $T_{X}$-$\lambda$ scaling relations presented in Section~\ref{sec:xray_richness} with those available in the literature will be meaningful. We do not make similar comparisons to relation involving $L_{X}$ since $L_{X}$ estimates can vary significantly for a given cluster depending on the adopted methodology, see Figures~\ref{fig:lumin_temp_comps}(a) and (c).  Furthermore, relations involving $L_{X}$ are more dependent on sample selection than $T_{X}$.

We first compare to the $T_{X}$-$\lambda_{\rm RM}$ scaling relations presented in \citet[][hereafter RR14]{2014ApJ...783...80R}. These are based on SDSSDR8 RM clusters ($0.1<z<0.3$), and so the $\lambda$ values are consistent with those used herein. Two samples are presented in RR14, one contains 25 {\em XMM} derived $T_{X}$ values taken from the first XCS data release \citep{2012MNRAS.423.1024M}, hereafter the RR14$_{\rm XCS}$ sample. The other contains 54 {\em Chandra} derived $T_{X}$ values, hereafter the RR14$_{\rm ACCEPT}$ sample.  These 54 are a subsample of the 329 clusters in the ACCEPT database \citep{2009ApJS..182...12C}. The input data vectors used in the RR14 are not available, therefore the comparison here is limited to the fitted relations (taken from Table 2 of that paper). It is important to note that, for the ACCEPT sample, the RR14 fit was scaled to account for the offset in {\em Chandra} and {\em XMM} temperature measurements \citep[using][]{2016ApJS..224....1R}. As expected (given that the $T_{X}$ methodology was very similar to that used herein), there is excellent agreement in the case of the RR14$_{\rm XCS}$ sample. The fit to the RR14$_{\rm ACCEPT}$ sample is also consistent with that to the SDSSRM-XCS$_{T_{X},vol}$ sample.  However, as can be seen in Figure~\ref{fig:tlambda_comp}(a), the slope is steeper (although at $<$3$\sigma$). There is consistency in normalisation at the pivot point ($\lambda_{0}$=60).  This contrary to that found in RR14, who found a $\approx$40\% difference between their fits to RR14$_{\rm ACCEPT}$ and to R14$_{\rm XCS}$ because RR14 did not carry out any {\em Chandra} to {\em XMM} $T_{X}$ scaling.

In Figure~\ref{fig:tlambda_comp}(b) we compare our $T_{X}$-$\lambda$ scaling relation  to that derived from the H19 sample of 97 SDSSRM clusters (brown curve/points in Figure~\ref{fig:littempcomp}). In this case, the input data vector was available, so we were able to perform a new fit following the approach in Section~\ref{sec:xray_richness}, i.e. with $\lambda_{0}$=60 and $T_{0}$=4~keV, to maximise uniformity in the method.  The comparison of the data and fits are given in Figure~\ref{fig:tlambda_comp}(c, with the appropriate {\em Chandra} and {\em XMM} $T_{X}$ scaling applied). The H19 data are given by the brown squares, with the {\sc lira} fit given by the brown dashed line (and brown shaded region highlighting the 68\% uncertainty).  We obtain fit parameters of the normalisation and slope of $A_{T\lambda,H19}=1.16\pm0.04$ and $B_{T\lambda,H19}=0.50\pm0.05$ respectively.  There is a small (14\%) offset in normalisation at the pivot point ($\lambda_{0}$=60) significant at the 2.9$\sigma$ level.  While not significant, we assess the impact of the choice of {\em Chandra}-to-{\em XMM} temperature scaling on the above comparison.  Therefore, we rescaled the H19 temperatures to {\em XMM} using the scaling found in \cite{2015A&A...575A..30S} and re-fit the H19 $T_{X}$-$\lambda$ relation.  We obtain fit parameters of $A_{T\lambda,H19}=1.23\pm0.03$ and $B_{T\lambda,H19}=0.44\pm0.05$.  The offset in normalisation increases to $\approx$20\%, significant at the 5.1$\sigma$ level.  This highlights the potential difficulty of combining {\em Chandra} and {\em XMM} data.  However, we note, one cannot exclude the effects of differences in selection between the two archival samples.      

In Figure~\ref{fig:tlambda_comp}(c) we compare our  $T_{X}$-$\lambda$ scaling relation to that based on the CAMIRA analysis of Hyper Suprime-Cam (HSC) observations \cite{2018PASJ...70S..20O}.  The CAMIRA algorithm is similar to RM, in that it identifies clusters using the red-sequence, but the estimated richness values will differ. The $T_{X}$-$\lambda$ scaling relation analysis based on 50 CAMIRA clusters is presented in \cite{2018PASJ...70S..20O}, where the input $T_{X}$ values were derived from {\em XMM} observations.  For these 50 clusters, 34 $T_{X}$ values were taken from \cite{2016A&A...592A...3G} and 16 $T_{X}$ values taken from \cite{2014MNRAS.444.2723C}.  Again, we were able to refit the input data using the approach in Section~\ref{sec:xray_richness}, as they were kindly made available to us via priv. comm. by the authors. We obtain fit parameters of the normalisation and slope of $A_{T\lambda,O18}=1.04\pm0.10$ and $B_{T\lambda,O18}=0.56\pm0.11$ respectively.  Figure~\ref{fig:tlambda_comp}(d) compares the SDSSRM-XCS$_{T_{X},vol}$ $T_{X}-\lambda_{\rm RM}$ relation and the fit to the CAMIRA data (given by the purple diamonds with the best-fit relation given by the purple dashed line and light purple shaded region the 68\% uncertainty). We note that richness is defined as $N_{\rm mem}$ in \cite{2018PASJ...70S..20O}, but we keep the notation of $\lambda_{RM}$ in Figure~\ref{fig:tlambda_comp}(c) for clarity in the comparisons.  As seen in Figure~\ref{fig:tlambda_comp}(c), the two relation are fully consistent, albeit with the caveat that $\lambda_{RM} \neq N_{mem}$.

Finally, we compare to the $T_{X}$-$\lambda_{\rm RM}$ scaling relations presented in \citet[][hereafter F19]{2019MNRAS.490.3341F}.  The relations are constrained using RM clusters detected within 1500 deg$^{2}$ of the DES \citep[using the 1st year of DES observations][]{2018ApJS..235...33D}.  DES RM clusters were matched to {\em XMM} detected clusters using the same processes outlined in this work, resulting in a sample of 110 clusters used for the $T_{X}$-$\lambda_{\rm RM}$ scaling analysis.  The clusters fall within 0.2$<$z$<$0.7 and do not contain a temperature error cut (unlike in the SDSSRM-XCS$_{T_{X},vol}$ sample).  Furthermore, the temperatures are determined within r$_{2500}$.  The input data vector was obtained, and the $T_{X}^{2500}$-$\lambda_{\rm RM}$ relation fit following Section~\ref{sec:xray_richness} i.e., with $\lambda_{0}=$60 and $T_{0}=$4 keV.  The comparison of the data and fits are given in Figure~\ref{fig:tlambda_comp}(d).  The F19 data are given by the dark-red triangles, with the {\sc lira} fit given by the dark-red dashed line (and red shaded region the 68\% uncertainty).  The SDSSRM-XCS$_{T_{X},vol}$ relation is mostly obscured by the F19 fit, because the results are so consistent.

In summary, the results presented here (and in Sect.~\ref{sec:litcompv2}) are consistent with those in the literature and based on the largest compilation of $T_{X}$ and $\lambda_{\rm RM}$ data to date.  Furthermore, the extremely consistent comparison between this work and the results in F19 (Fig~\ref{fig:tlambda_comp}(d)), highlights that our sample can be combined with clusters from the DES for further analysis (see further discussion in Sect.~\ref{sec:ltisotropy}).  

\section{Discussion}
\label{sec:discussion}

As mentioned above (see Sect.~\ref{sec:cosmosample}), we have compiled one of the largest samples of consistently derived $T_{X}$ values to date. This allows us to explore factors that might influence measured (as opposed to intrinsic) scaling relations. For example, in Sect.~\ref{sec:targets}, we explore the impact of selection on the relations, specifically the difference between targeted and serendipitous detections.  And, in Sect.~\ref{sec:ltisotropy}, we investigate the recent claims of an anisotropy across the sky in the measured $L_{X}-T_{X}$ relation \citep{2020A&A...636A..15M}.

\subsection{The dependence of scaling relations on detection type (targeted or serendipitous)}
\label{sec:targets}

\begin{table*}           
\begin{center}
\caption[]{{\small Best-fit parameters of the cluster scaling relations for the SDSSRM-XCS$_{T_{X},vol}$ cluster sample, split between the targeted and serendipitous sub-samples (as defined in Sect.~\ref{sec:targets}).  Best-fit parameters are given for the $L_{X}-T_{X}$, $L_{X}-\lambda_{\rm RM}$ and $T_{X}-\lambda_{\rm RM}$ relations, given by equations~\ref{equ:lt},~\ref{equ:l-lambda} and~\ref{equ:t-lambda} respectively.}\label{tab:bestfit_targ_serend}}
\vspace{1mm}
\begin{tabular}{cccccc}
\hline
\hline
 Relation & Fit & Normalisation & Slope & Scatter & Figure \\
(sample) & & & & \\
\hline
$L^{r500}_{X,52}-T^{r500}_{X}$ & & $A_{LT}$ & $B_{LT}$ & $\sigma_{LT}$ & \\ 
SDSSRM-XCS$_{T_{X},vol}$ & & & & & \\
Targets & {\sc lira} & 1.04$\pm$0.09 & 2.63$\pm$0.20 & 0.74$\pm$0.06 & \ref{fig:lx_kt_targ_serend}(a) \\
Serendipitous & {\sc lira} & 0.66$\pm$0.09 & 2.00$\pm$0.22 & 0.52$\pm$0.06 & \ref{fig:lx_kt_targ_serend}(a) \\
\hline
$L^{r500}_{X,52}-\lambda_{\rm RM}$ & & $A_{L\lambda}$ & $B_{L\lambda}$ & $\sigma_{L\lambda}$ & \\ 
SDSSRM-XCS$_{T_{X},vol}$ & & & & & \\
Targets & {\sc lira} & 1.42$\pm$0.16 & 1.13$\pm$0.19 & 1.06$\pm$0.07 & \ref{fig:x-ray_lambda_targ_serend}(a) \\
Serendipitous & {\sc lira} & 0.48$\pm$0.09 & 1.23$\pm$0.27 & 0.79$\pm$0.08 & \ref{fig:x-ray_lambda_targ_serend}(a) \\
\hline
$T^{r500}_{X}-\lambda_{\rm RM}$ & & $A_{T\lambda}$ & $B_{T\lambda}$ & $\sigma_{T\lambda}$ \\ 
SDSSRM-XCS$_{T_{X},vol}$ & & & & & \\
Targets & {\sc lira} & 1.14$\pm$0.03 & 0.45$\pm$0.05 & 0.27$\pm$0.02 & \ref{fig:x-ray_lambda_targ_serend}(b) \\
Serendipitous & {\sc lira} & 0.81$\pm$0.06 & 0.50$\pm$0.12 & 0.34$\pm$0.04 & \ref{fig:x-ray_lambda_targ_serend}(b) \\
\hline
\end{tabular}
\end{center}
\end{table*}

\begin{figure}
\begin{center}
\begin{tabular}{c}
\includegraphics[width=8.0cm]{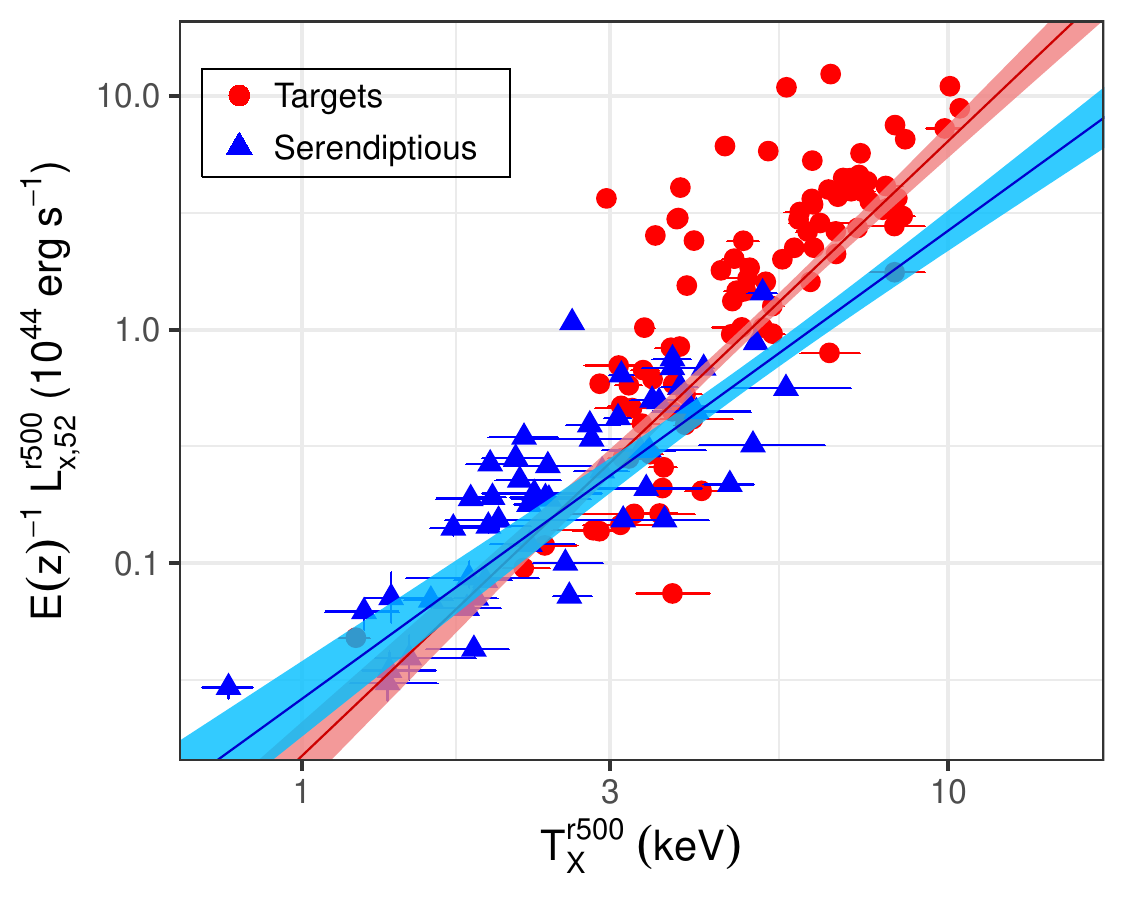} \\
(a) \\
\includegraphics[width=8.0cm]{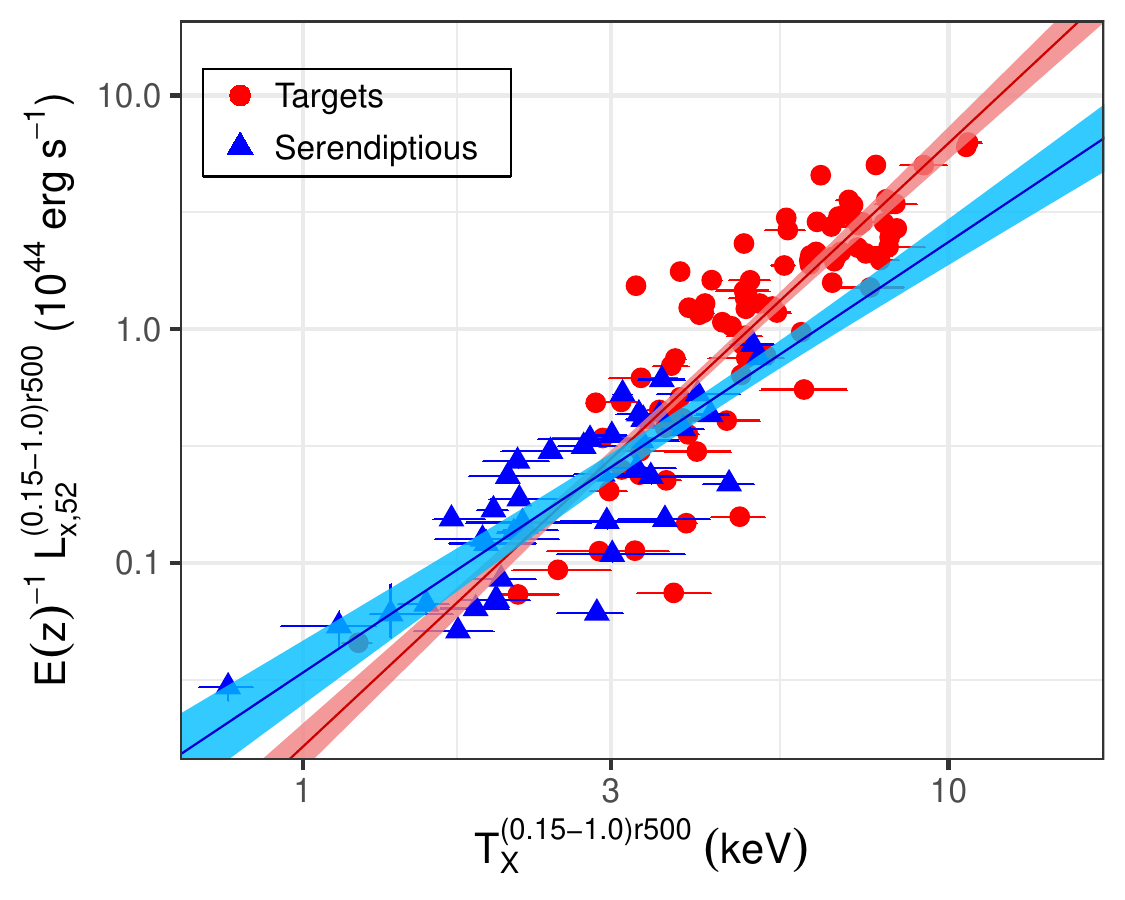} \\
(b)
\end{tabular}
\end{center}
\caption{Luminosity-temperature relation of the SDSSRM-XCS$_{T_{X},vol}$ sample, split between targeted (red circles) and serendipitous (blue triangles) clusters.  The fiducial core-included $r_{500}$ relations are shown in Plot (a, $L_{X,52}^{r500}-T_{X}^{r500}$), and the core-excluded relation in Plot (b, $L_{X,52}^{(0.15-1)r500}-T_{X}^{(0.15-1)r500}$).  The red and blue lines represent a fit to the targeted and serendipitous clusters respectively (fit using equation~\ref{equ:lt}).  The shaded regions around the respective lines represents the 68\% confidence interval of the mean logarithmic relation.} \label{fig:lx_kt_targ_serend}
\end{figure}

\begin{figure*}
\begin{center}
\begin{tabular}{cc}
\includegraphics[width=8.0cm]{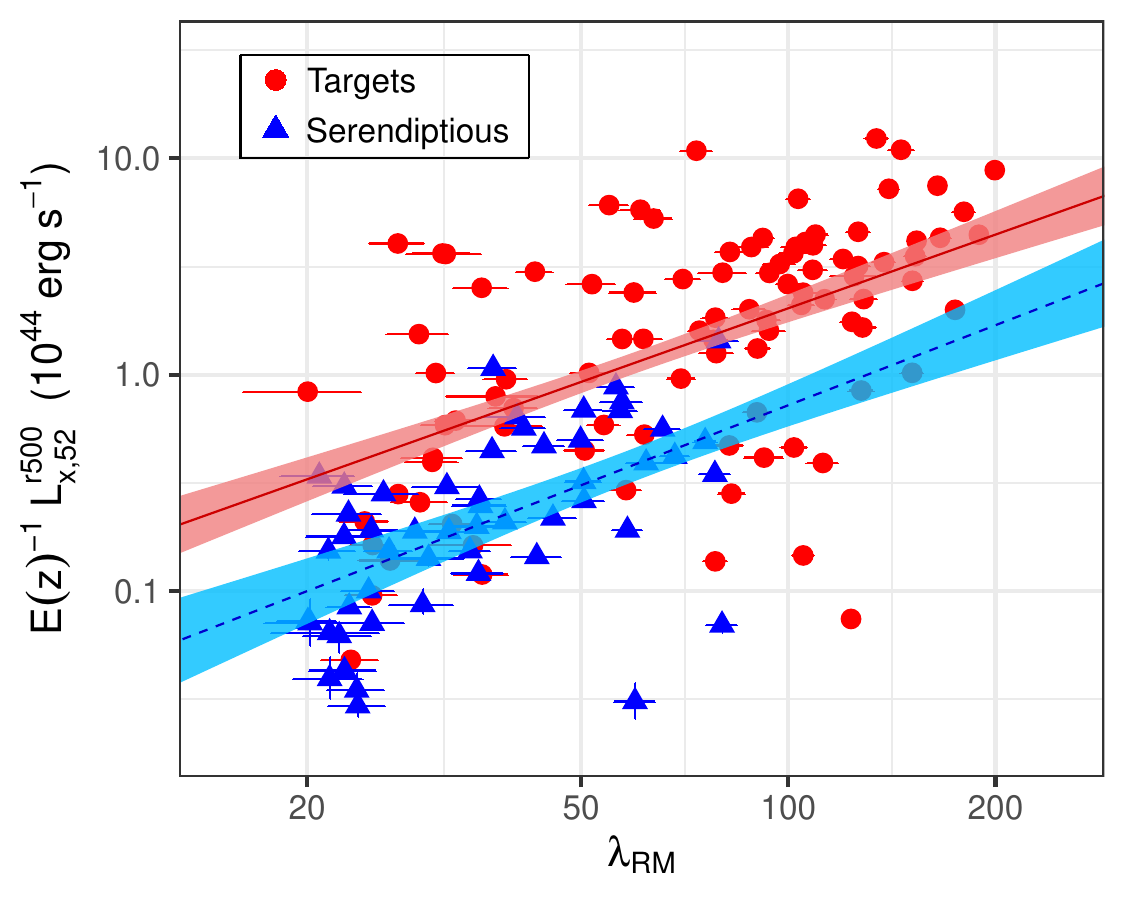} &
\includegraphics[width=8.0cm]{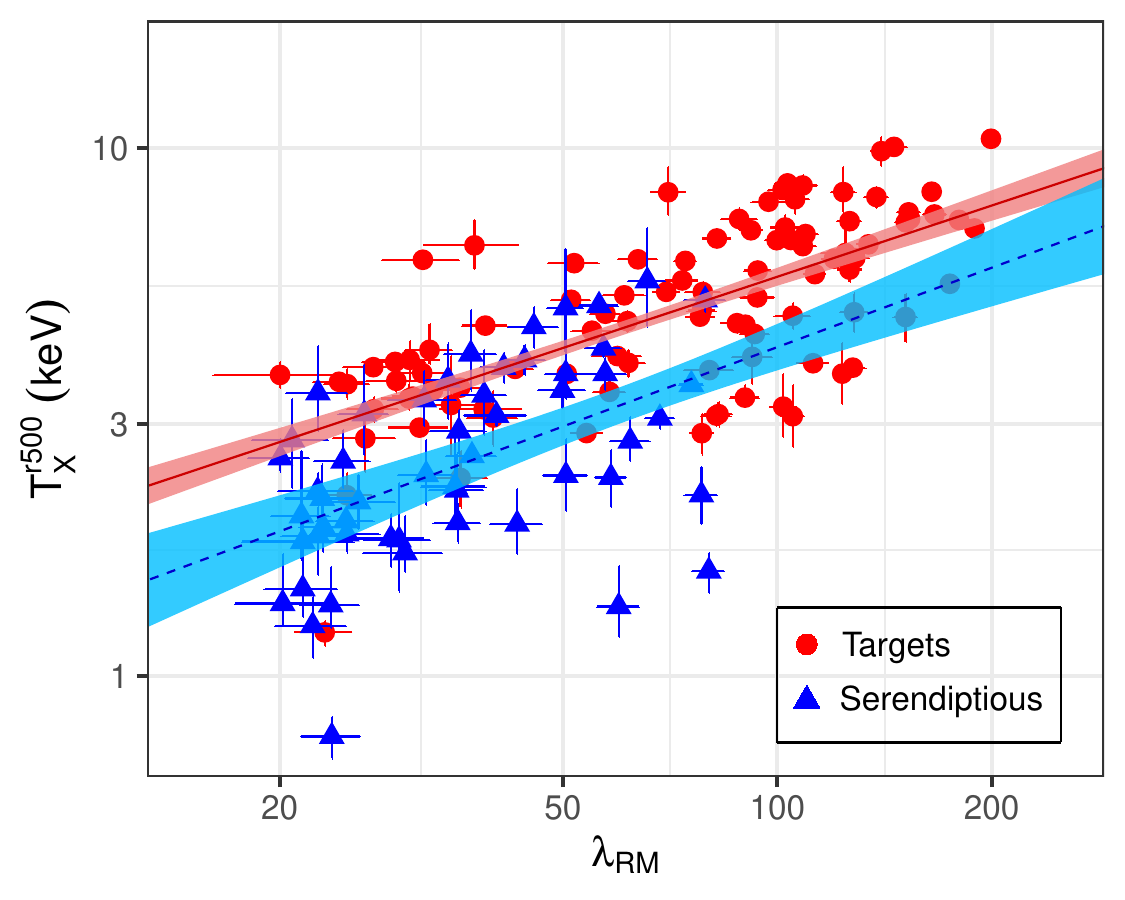} \\
(a) & (b) \\
\includegraphics[width=8.0cm]{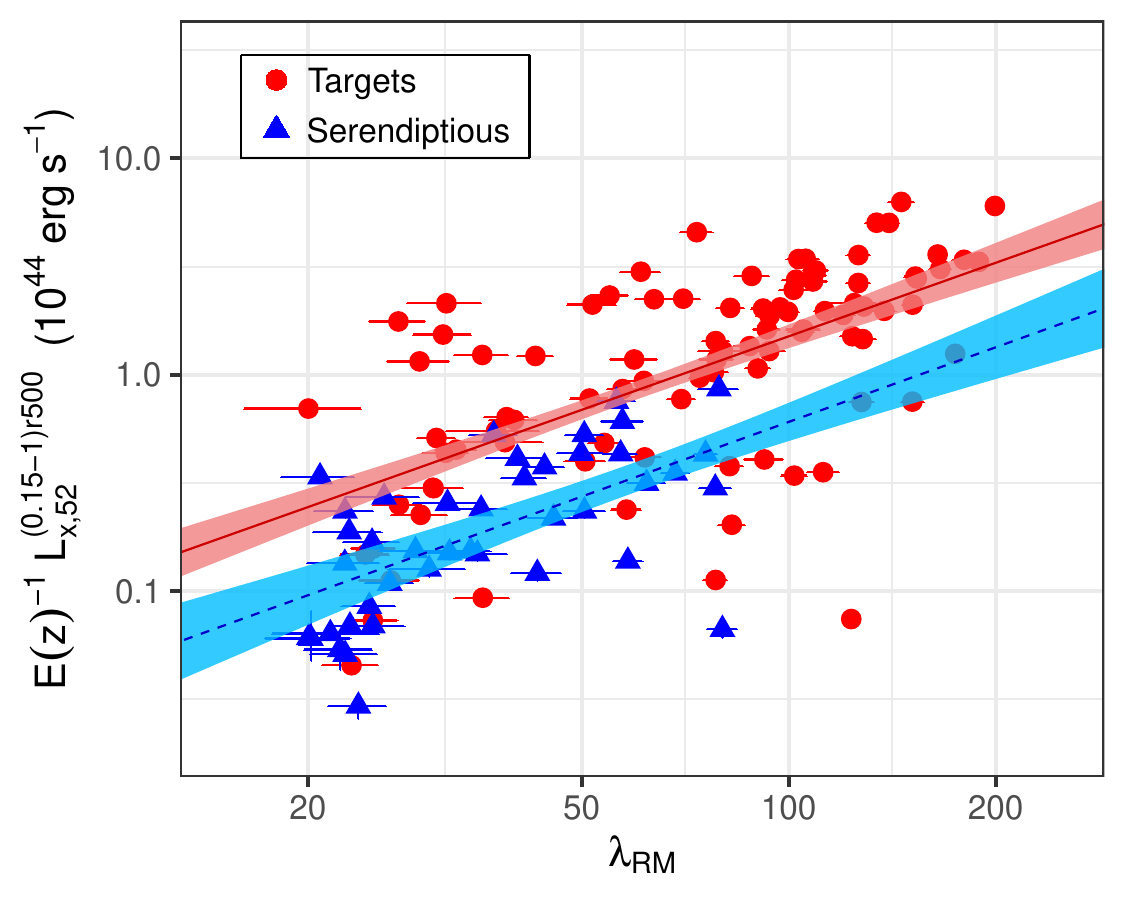} &
\includegraphics[width=8.0cm]{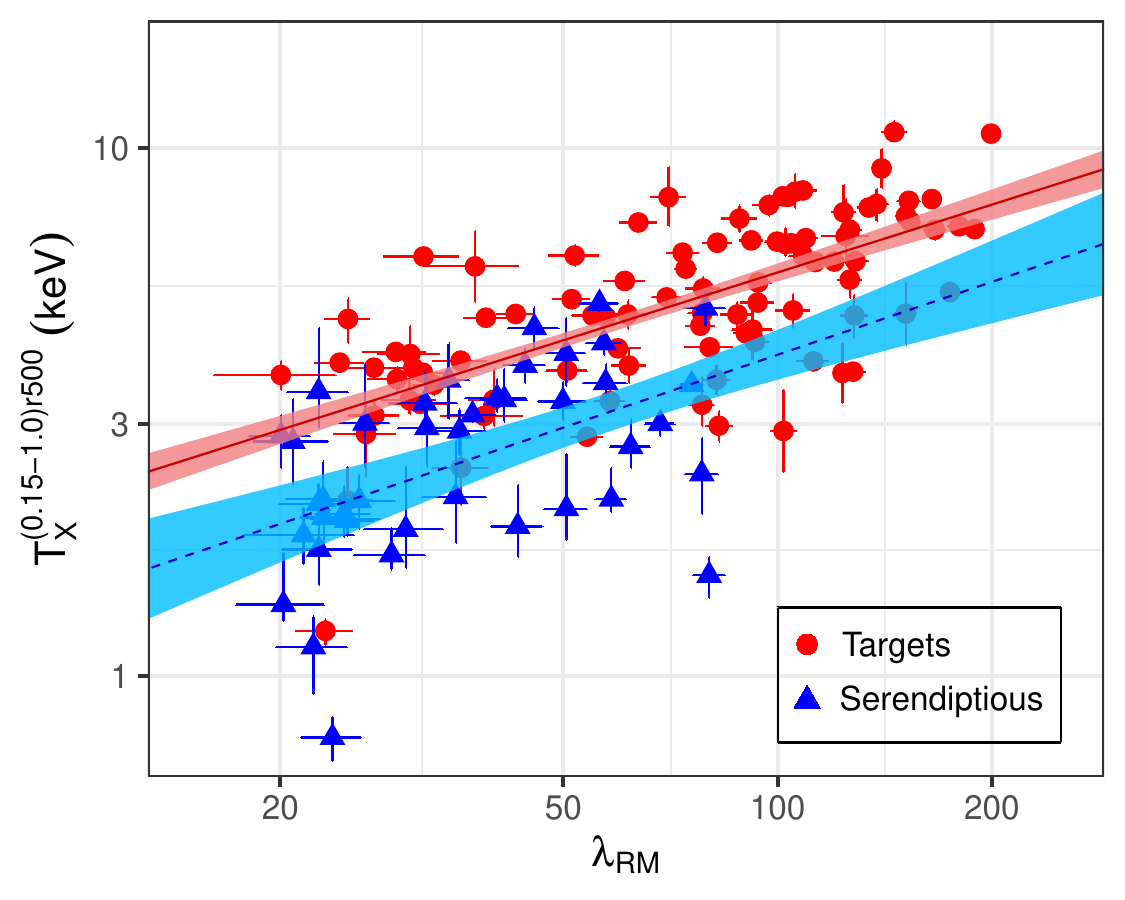} \\
(c) & (d) \\
\end{tabular}
\end{center}
\caption{X-ray observable-richness relations of the SDSSRM-XCS$_{T_{X},vol}$ sample, split between targeted (red circles) and serendipitous (blue triangles) clusters.  Figure (a) plots the $L_{X,52}^{r500}-\lambda_{\rm RM}$ relation and Figure (b) plots the $T_{X}^{r500}-\lambda_{\rm RM}$ relation, using the fiducial core included properties.  Figure (c) plots the $L_{X,52}^{(0.15-1)r500}-\lambda_{\rm RM}$ relation and Figure (d) plots the $T_{X}^{(0.15-1)r500}-\lambda_{\rm RM}$ relation, using core excluded properties.  In each case, the red line and blue line represent a fit to the targeted and serendipitous clusters respectively.  Shaded regions around each fit highlights the 1$\sigma$ uncertainty.} 
\label{fig:x-ray_lambda_targ_serend}
\end{figure*}

We have separated the SDSSRM-XCS$_{T_{X},vol}$ clusters into those that were the target of their respective {\XMM}-Newton and those that were detected ``serendipitously''.  The classification was done based upon a visual inspection of the X-ray images and information from the {\XMM}-Newton Science Archive (namely the target name and target type). Of the 150 SDSSRM-XCS$_{T_{X},vol}$ clusters, 97 were flagged as being {\XMM}-Newton targets, and 53 as serendipitous detections.  We then revisited the scaling relations presented in Table~\ref{tab:bestfit}.  The results are presented in Table~\ref{tab:bestfit_targ_serend}, plotted in Figures~\ref{fig:lx_kt_targ_serend}(a),~\ref{fig:x-ray_lambda_targ_serend}(a) and \ref{fig:x-ray_lambda_targ_serend}(b).  In all cases, the measured normalisation of the targeted sub-sample is higher than that of the serendipitous sub-sample (ranging between 2.9 to 5.1-$\sigma$). This remains true even when the emission from the cluster cores is excluded, see Table~\ref{tab:exccorebestfit} and Figures~\ref{fig:lx_kt_targ_serend}(b),~\ref{fig:x-ray_lambda_targ_serend}(c) and \ref{fig:x-ray_lambda_targ_serend}(d). While the measured slope of the $L_{X,52}^{r500}-T_{X}^{r500}$ differs, it is only significant at the $\approx$2$\sigma$ level. There is very little change in the richness scaling relations. 

The current data are not sufficient to draw a firm conclusion as to the cause of these differences. However, they are unlikely to be due to a systematic in the XCS analysis methods, i.e. whereby biases in measured $L_{X}$ or $T_{X}$ values are correlated with location on the detector: LD11 studied the effect of measuring temperatures for the same clusters that were detected at a high off-axis position and then re-observed at the on-axis aimpoint.  LD11 found a 1:1 relationship between the measured temperatures, finding no systematic offset (see Fig 18 in LD11).

Instead, we suggest the cause is due to incompleteness in the sub-samples. There is a dearth of X-ray bright objects in the serendipitous sub-sample because these clusters are intrinsically very rare and so have a low projected sky density: a small area serendipitous survey is unlikely to come across them by accident. In the targeted sample, many {\em XMM} (and {\em Chandra}) targets were historically drawn from samples detected by the ROSAT All-Sky Survey (which had a relatively bright flux limit) and followed-up clusters with a high luminosity.  Figure A1 in \citealt{2010MNRAS.406.1773M} demonstrates how biases (specifically a luminosity limit) can significantly flatten the measured slope of a scaling relation.  In addition, both sub-samples are incomplete at the low flux end due to biases in selection. It is possible to model the impact of incompleteness (as was done in \citealt{2010MNRAS.406.1773M}) but is beyond the scope of this work. The true normalisation and slope of the $L_{X}-\lambda$ relation should be uncovered by the X-ray selected samples from the eRASS project, but in the meantime it would be prudent to use only $T_{X}-\lambda$ relations for cosmological studies (as these are the least impacted by the sub-sample choice, see Figure~\ref{fig:x-ray_lambda_targ_serend},(b) \& (d)).

\subsection{Investigating $L_{X}-T_{X}$ isotropy with the SDSS-XCS$_{\rm ext}$ sample}
\label{sec:ltisotropy}

Recently, Mig20 made a claim relating to a possible anisotropy across the sky in the luminosity-temperature relation \citep{2020A&A...636A..15M}. This claim, if true, would add additional systematics and uncertainty when using cluster number counts as a cosmological probe. The main Mig20 result was based on 313 clusters with measured $T_{X}$ values (the yellow curve in Figure~\ref{fig:littempcomp}(a)). These 313 are made up of a compilation of both {\em XMM} and re-scaled {\em Chandra} $T_X$ values. So we felt it was worthwhile to re-explore the Mig20 result using the larger (381) SDSSRM-XCS$_{T_X}$ sample of clusters, with $T_X$ values drawn only from one telescope. Additional motivation comes from the results presented in Section~\ref{sec:targets}, the difference in normalisation seen in Figure~\ref{fig:lx_kt_targ_serend} is larger than that presented in Mig20.  Note that while we focus on the results of Mig20 using the 313 clusters, the conclusions of Mig20 were enhanced by using this main sample and a combination of clusters from the ACC \citep{2001PhDT........88H} and XCS-DR1 \citep{2012MNRAS.423.1024M}.

To demonstrate the robustness of our technique, we first repeated the analysis presented in Mig20, using the same input data vectors. In brief, the method is as follows: The sky is binned into regions over the full range of galactic longitude ($l$) and latitude ($b$), using a bin width of $\Delta l$=1$^{\circ}$ and $\Delta b$=1$^{\circ}$ (creating 65,160 bins on the sky).  At the centre of each bin, a cone with a radius $\theta$ is used to find a subset of all clusters within an angular separation of $\theta_{d}$ from the coordinates of the bin.  Using this subset, the `local' $L_{X}-T_{X}$ relation is fit using {\sc lira}, following the same method as described in Sect.~\ref{sec:ltrelation}.  However, as per Mig20, the slope of the local relation is fixed at all-sky value.  A statistical weighting is applied to each cluster in the subset by increasing the size of the uncertainties by a factor 
\begin{align}
\cos\left(\frac{\theta_{d}}{\theta}\times90^{\circ}\right).
\label{equ:anisotropy_err_scale}
\end{align}
At each position on the sky, the local $L_{X}-T_{X}$ normalisation, $A$, is divided by the normalisation of the all sky $L_{X}-T_{X}$ relation ($A_{\rm all}$), with sky maps plotted based upon $A$/$A_{\rm all}$. In Figure~\ref{fig:ltmigkas}(a), we replicate the results presented in Mig20 Figure 8 for the $\theta$=60$^{\circ}$ cone (thus confirming that our method is robust).  This test also shows that the dipole feature is present irrespective of the linear regression fitting method used.  Whereas we used {\sc lira}, the Mig20 analysis used a fitting method equivalent to the BCES Y|X fitting method \citep{1996ApJ...470..706A}.  

\begin{figure*}
\begin{center}
\begin{tabular}{cc}
\includegraphics[width=8.5cm]{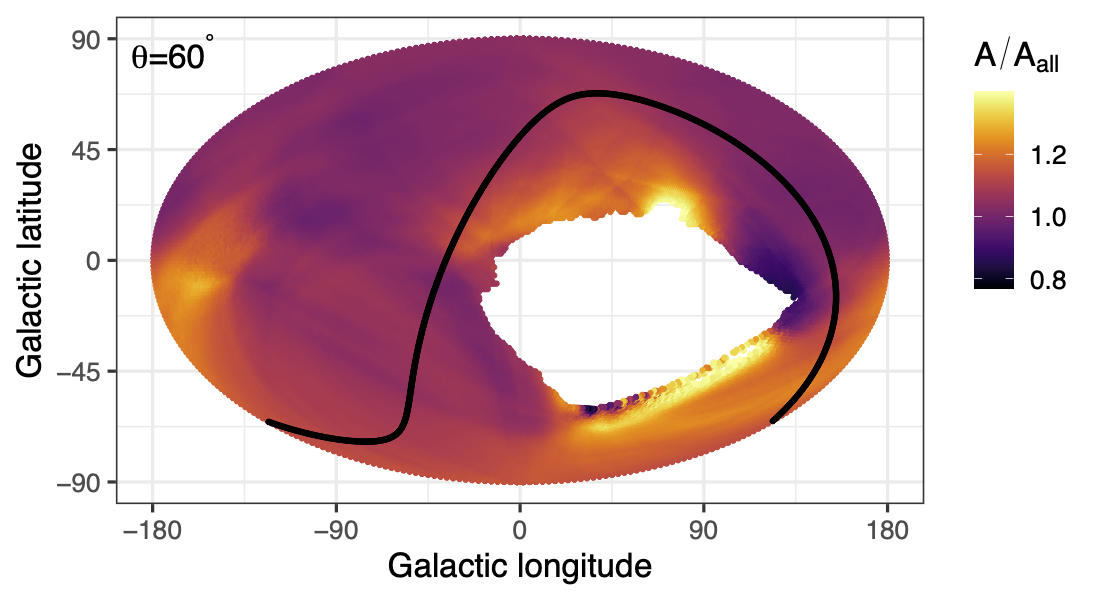} &
\includegraphics[width=8.5cm]{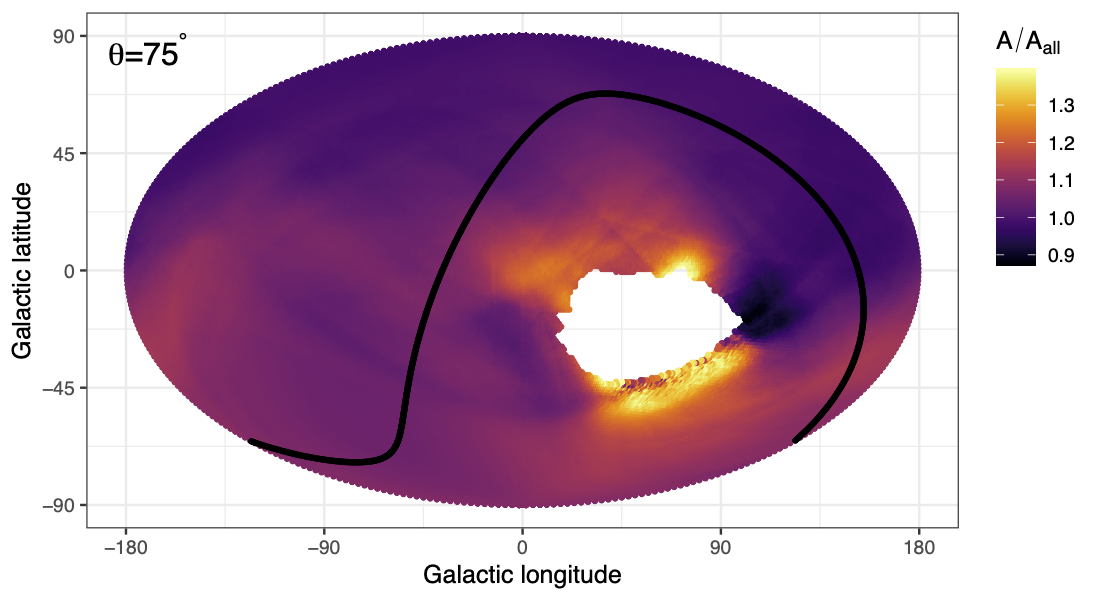} \\
(a) & (b) \\
\end{tabular}
\end{center}
\caption{Ratio of the local best fit $L^{(0.15-1)r_{500}}_{X,52} - T^{(0.15-1)r_{500}}_{X}$ normalisation ($A$) to the all sky normalisation ($A_{\rm all}$), as a function of sky position.  Maps are created, assuming at each sky position, clusters within a cone of (a) $\theta$=60$^{\circ}$ and (b) $\theta$=75$^{\circ}$, are used for the local relation.  The black line highlights the Galactic plane.} \label{fig:ltisotropysdss}
\end{figure*}

We then apply the same method to the SDSSRM-XCS$_{T_{X}}$ sample.  Note we use core excluded properties for this, in line with Mig20, who used (0.2-0.5)$r_{500}$ values.  The ratio of $A$/$A_{\rm all}$ over the sky is then determined where there are $>$30 clusters in the bin.  Figures~\ref{fig:ltisotropysdss}(a) and (b) displays the sky distribution of $A$/$A_{\rm all}$, assuming cones of $\theta$=60$^{\circ}$ and $\theta$=75$^{\circ}$ respectively.  The $\theta$=60$^{\circ}$ cone was chosen, as the dipole feature found in Mig20 is the most prominent at this scale.  The $\theta$=75$^{\circ}$ cone was chosen to increase the sky coverage.  Based upon the distribution of $A$/$A_{\rm all}$ (Fig.~\ref{fig:ltisotropysdss}), we do not observe the anisotropy feature found in Mig20 for the 60$^{\circ}$, although for the 75$^{\circ}$ we start to see hints of a decrease in $A$/$A_{\rm all}$, coincident with the position of the isotropy feature found in Mig20. However, it is not possible to yet confirm the existence of an anisotropy feature because there is a region in the Southern sky where we are not able to measure $A$/$A_{\rm all}$ because SDSS is a northern survey. The strong edge features around the empty area correspond to local regions where all clusters in the respective cones have an angular separation of $\geq$55$^{\circ}$.  Assuming Equation~\ref{equ:anisotropy_err_scale}, and $\theta_{d}$>55$^{\circ}$, the uncertainties on the measured cluster properties are divided by $\leq$0.13.  The resulting local $L_{X}-T_{X}$ relation thus becomes unconstrained.  We therefore test the use of a new error scaling method as given in \cite{2021A&A...649A.151M}.  The updated error scaling in \cite{2021A&A...649A.151M} follows the form $\cos(\theta / 90^{\circ})$, and is noted as a more conservative scaling approach.  Furthermore, we apply another update given in \cite{2021A&A...649A.151M}, where the slope of the local $L_{X}-T_{X}$ relation is left free to vary (as opposed to being fixed as in Mig20).  The results of these updates are presented in Figure~\ref{fig:ltmigkas}(b).  The edge feature around the empty area appears less scattered, however, again due to this empty feature, no anisotropy feature is observed.

In summary, while our sample size is larger than the one presented in Mig20, and we have replicated the results using Mig20 data, further data is required due to the SDSS sky coverage.  For this, the sample used here will be combined with RM clusters detected from the DES Y3 Gold catalogue \citep{2021ApJS..254...24S} to improve the sky coverage. This technique (of combining SDSS and DES RM clusters) has been successfully applied in \cite{2021arXiv210707631W} to measure the correlations between velocity dispersion, $\lambda_{\rm RM}$, $T_{\rm X}$ and $L_{\rm X}$ for RM clusters. The results shown in Figure~\ref{fig:tlambda_comp}(d) also give us confidence that SDSS and DES cluster samples can be jointly analysed.

\section{Summary}
\label{sec:conc}

In this paper, we detail the X-ray analysis of SDSS DR8 redMaPPer (SDSSRM) clusters using data products from the XMM Cluster Survey (XCS). In summary:

\begin{itemize}

\item In total, 1189 SDSSRM clusters fall within the cleaned {\em XMM}-Newton footprint. This has yielded 456 confirmed detections accompanied by X-ray luminosity ($L_{X}$) measurements. Using an updated version of the XCS Post Processing Pipeline ({\sc XCS3P}), we have extracted 381 X-ray temperature measurements ($T_{X}$) from these 456 clusters. This represents one of the largest samples of coherently derived cluster $T_{X}$ values to date. We have also shown that the reliability of derived $T_{X}$ values improves when low quality spectra are removed from joint fits.

\item We find that the SDSSRM clusters in the {\em XMM} footprint that were not detected are primarily lower richness systems (75\% at $\lambda <30$). It was possible to estimate $L_{X}$ upper limits for most 599 (of 733) of these non-detections.

\item Our analysis of the X-ray observable to richness scaling relations has demonstrated that scatter in the $T_{X}-\lambda$ relation is roughly a third of that in the $L_{X}-\lambda$ relation, and that the $L_{X}-\lambda$ scatter is intrinsic, i.e. will not be significantly reduced with larger sample sizes.

\item Our analysis of the scaling relation between $L_{X}$ and $T_{X}$ has shown that the fits are sensitive to the selection method of the sample, i.e. whether the sample is made up of clusters detected ``serendipitously'' compared to those deliberately targeted by {\em XMM}. These differences are also seen in the $L_{X}-\lambda$ relation and, to a lesser extent, in the $T_{X}-\lambda$ relation. Exclusion of the emission from the cluster core does not make a significant impact to the findings. A combination of selection biases is a likely, but as yet unproven, reason for these differences. 

\item We have used our data to probe recent claims of anisotropy in the $L_{X}-T_{X}$ relation across the sky \citep{2020A&A...636A..15M}. We find no evidence of anistropy, but stress that this may be masked in our analysis by the incomplete declination coverage of the SDSS DR8 sample.

\end{itemize}

The methods outlined in this work have further been employed in the analysis of large cluster samples, such as those constructed from the RM analysis of the Dark Energy Survey data e.g. \citep{2019MNRAS.487.2578Z,2019MNRAS.490.3341F}.  Although optically selected samples are free from X-ray selection biases, when matching to available X-ray data, future archival studies should consider only the use of serendipitously detected X-ray clusters to avoid observer biases. Furthermore, future use of the {\em XMM} Cluster Survey data will be of critical importance for upcoming cluster samples such as those constructed from the Legacy Survey of Space and Time undertaken by the Vera C. Rubin Observatory, of which currently $\approx$450 deg$^{2}$ of the LSST sky has been covered by {\em XMM}.  

\section*{Data availability}

The data underlying this work can be found at:

\noindent \href{http://users.sussex.ac.uk/pag22/SDSSRM-XCS/sdssrm-xcs-sample-data.csv}{http://users.sussex.ac.uk/pag22/SDSSRM-XCS/sdssrm-xcs-sample-data.csv}, along with a table description:

\noindent \href{http://users.sussex.ac.uk/pag22/SDSSRM-XCS/column_names.txt}{http://users.sussex.ac.uk/pag22/SDSSRM-XCS/column\_names.txt}.

\section*{Acknowledgements}

PG, KR, RW, DT and EU recognises support from the UK Science and Technology Facilities Council via grants ST/P000525/1 and ST/T000473/1 (PG, KR), ST/P006760/1 (RW, DT), ST/T506461/1 (EU) and ST/N504452/1 (SB).  MH acknowledges financial support from the National Research Foundation.  PTPV and LE were supported by Fundação para a Ciência e a Tecnologia (FCT) through research grants UIDB/04434/2020 and UIDP/04434/2020 (PTPV), and SFRH/BD/52138/2013 (LE). TJ acknowledges support by the U.S. Department of Energy, Office of Science, Office of
High Energy Physics, under Award Numbers DE-SC0010107 and A00-1465-001. We thank K. Migkas for useful discussions regarding cluster isotropy.  We thank M. Sereno for useful discussions on the use of the {\sc lira} fitting package.  We thank M. Oguri for providing the CAMIRA data used in Figure~\ref{fig:tlambda_comp}(c).




\bibliographystyle{mnras}

\bibliography{xcs-sdssrm} 


\appendix

\section{Examples of problematic {\em XMM} observations}
\label{sec:problem_xmm}

\begin{figure*}
\begin{center}
\begin{tabular}{cc}
\includegraphics[clip, trim=4cm 0.25cm 6cm 0.25cm, width=8cm]{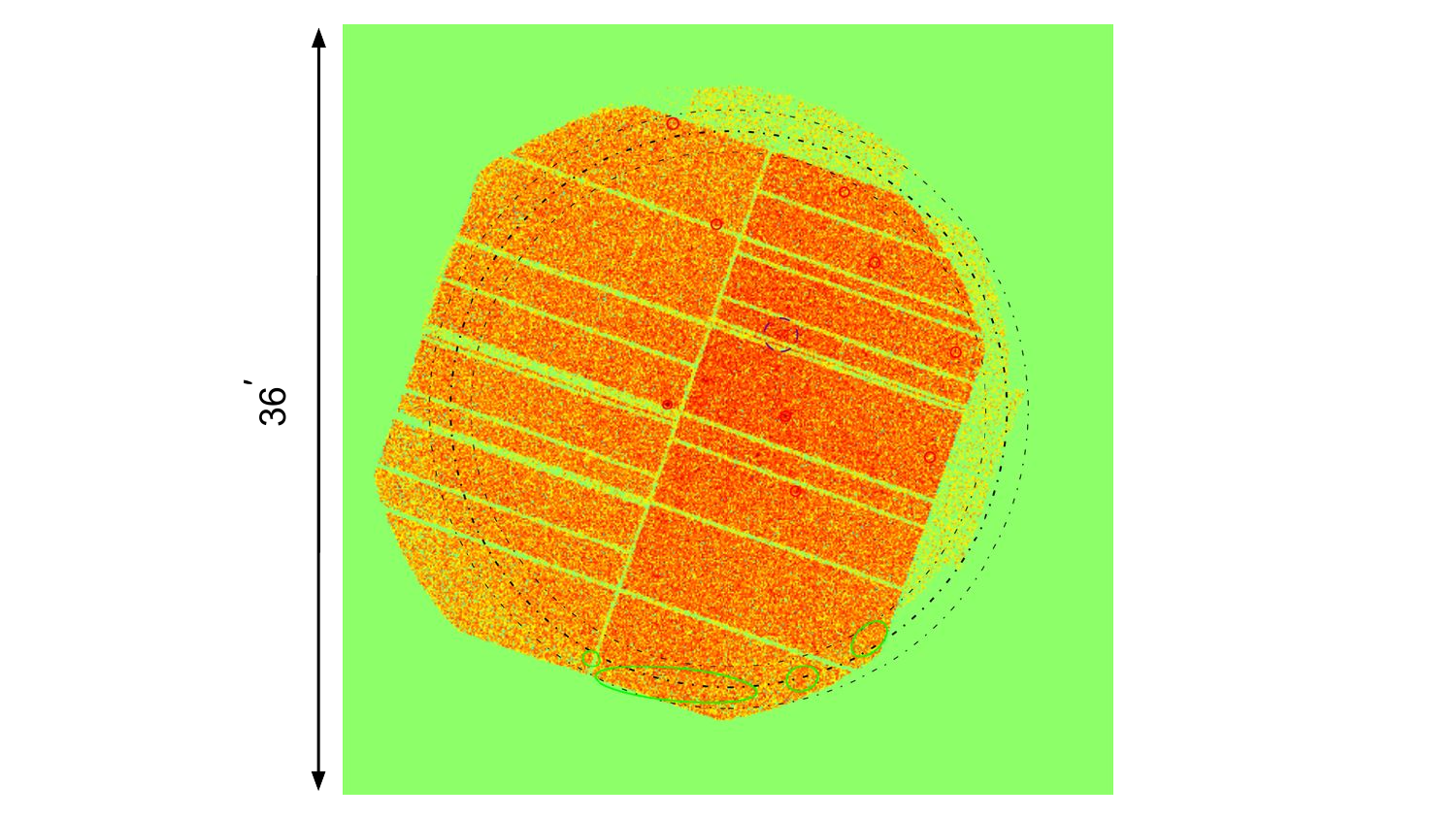} &
\includegraphics[clip, trim=4cm 0.25cm 6cm 0.25cm, width=8cm]{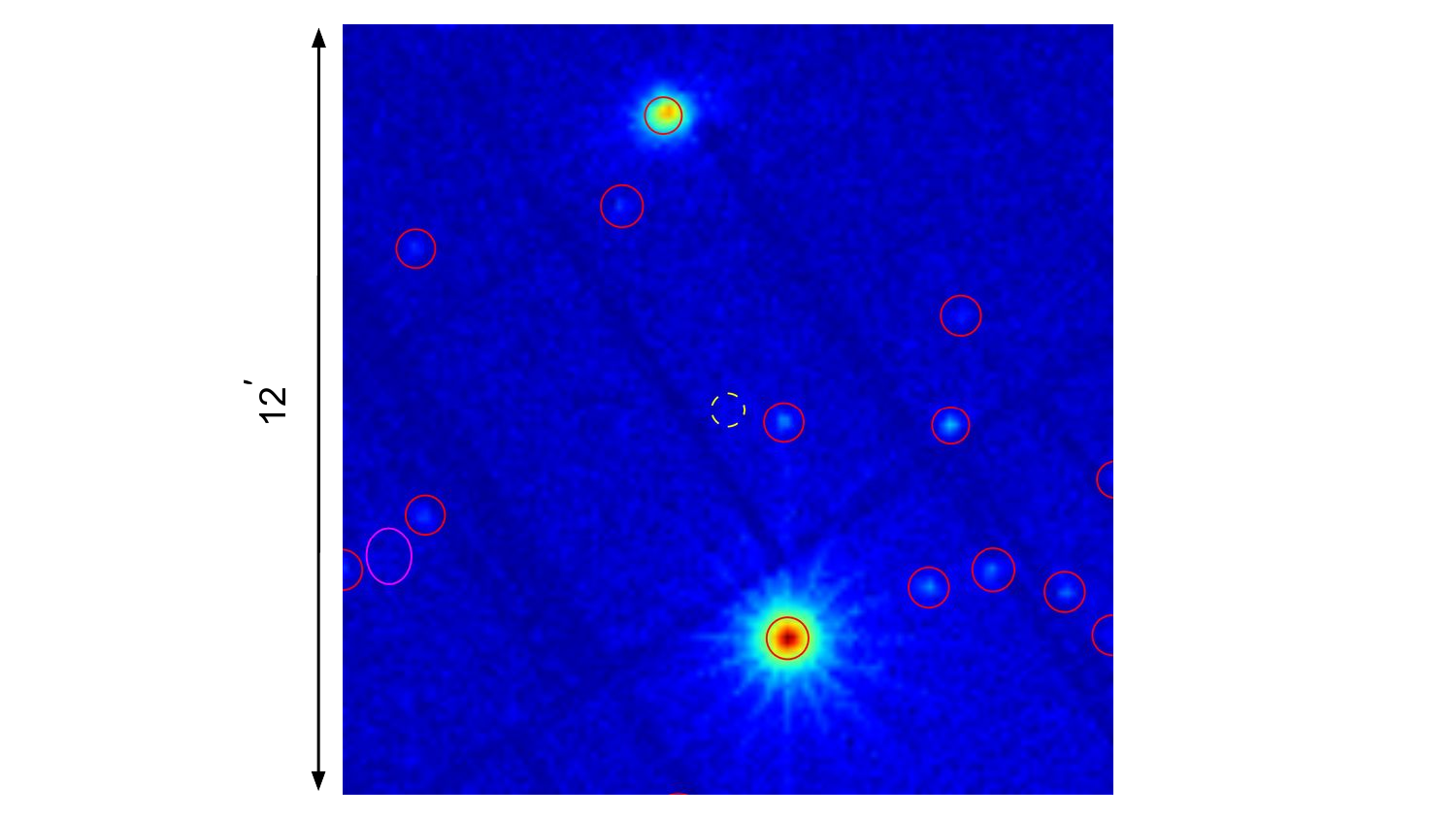} \\
(a) & (b) \\
\end{tabular}
\end{center}
\caption[]{Examples of problematic {\em XMM} observations found during the visual inspection process described in Sect~\ref{sec:sdss-xmm}. (a) Soft-band field-of-view image of the {\em XMM} observation ObsiD=0556213801.  The observation was rejected from further analysis because it is effected by periods of high background rates; (b) {\em XMM} image in the 0.5-2.0~keV band of the region surrounding SDSSRM cluster RMID=42060 (centroid indicated with the dashed yellow circle).  The cluster falls nearby to a bright point source that has created artifacts in the image (characterised by the ``spokes'') and was subsequently dropped from the cluster sample.}
\label{fig:xmmbad}
\end{figure*}

Here we show examples of SDSSRM clusters that were removed from the SDSSRM-{\em XMM} sample due to high levels of background, Fig~\ref{fig:xmmbad}(a), and strong point source contamination, Fig~\ref{fig:xmmbad}(b).  See Section~\ref{sec:sdss-xmm} for further details.

\section{Example of a cluster excluded from the SDSSRM-XCS sample after visual inspection}
\label{sec:unassociated}

\begin{figure*}
\begin{center}
\begin{tabular}{cc}
\includegraphics[clip, trim=4cm 0.25cm 6cm 0.25cm, width=8cm]{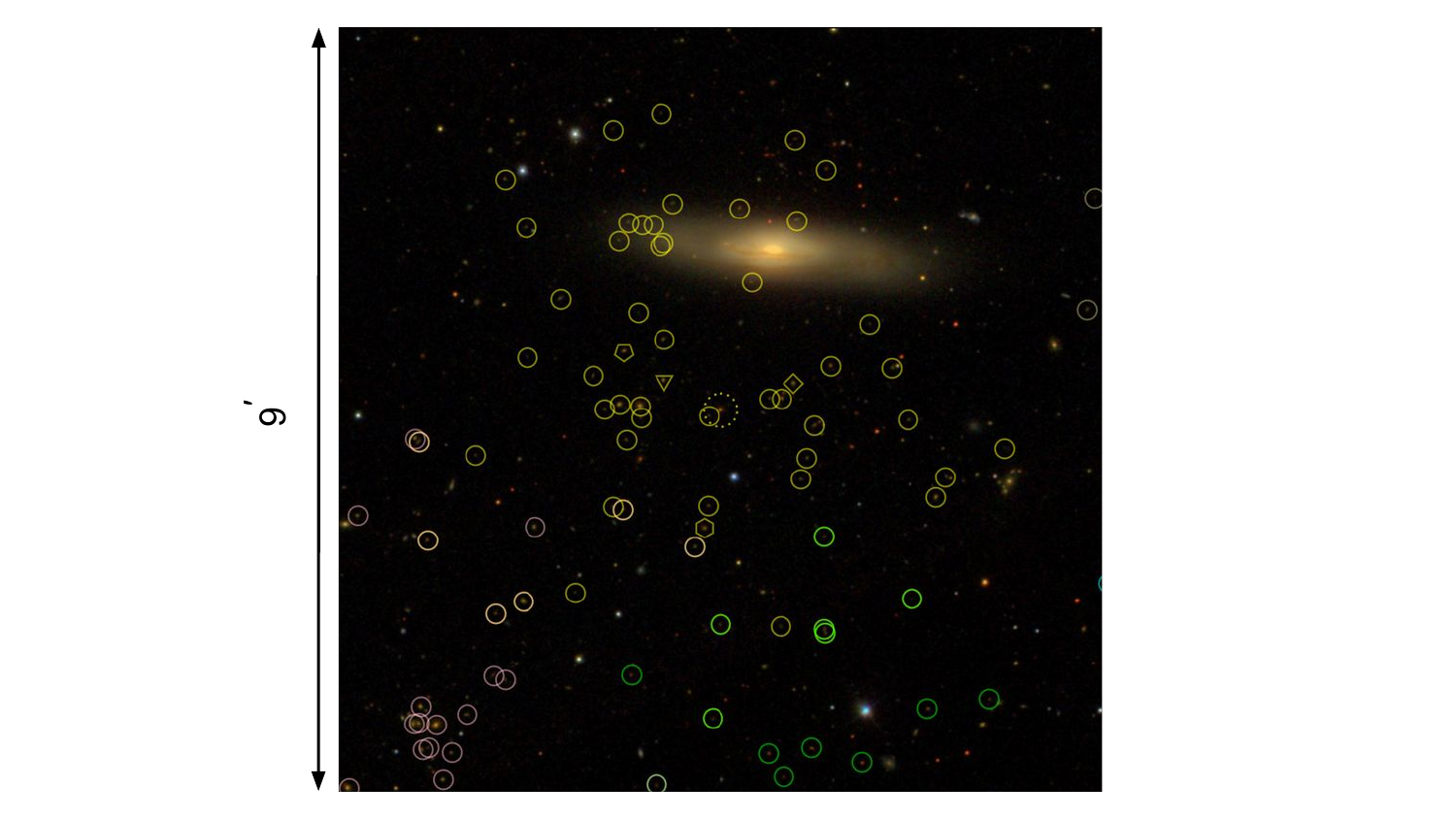} &
\includegraphics[clip, trim=4cm 0.25cm 6cm 0.25cm, width=8cm]{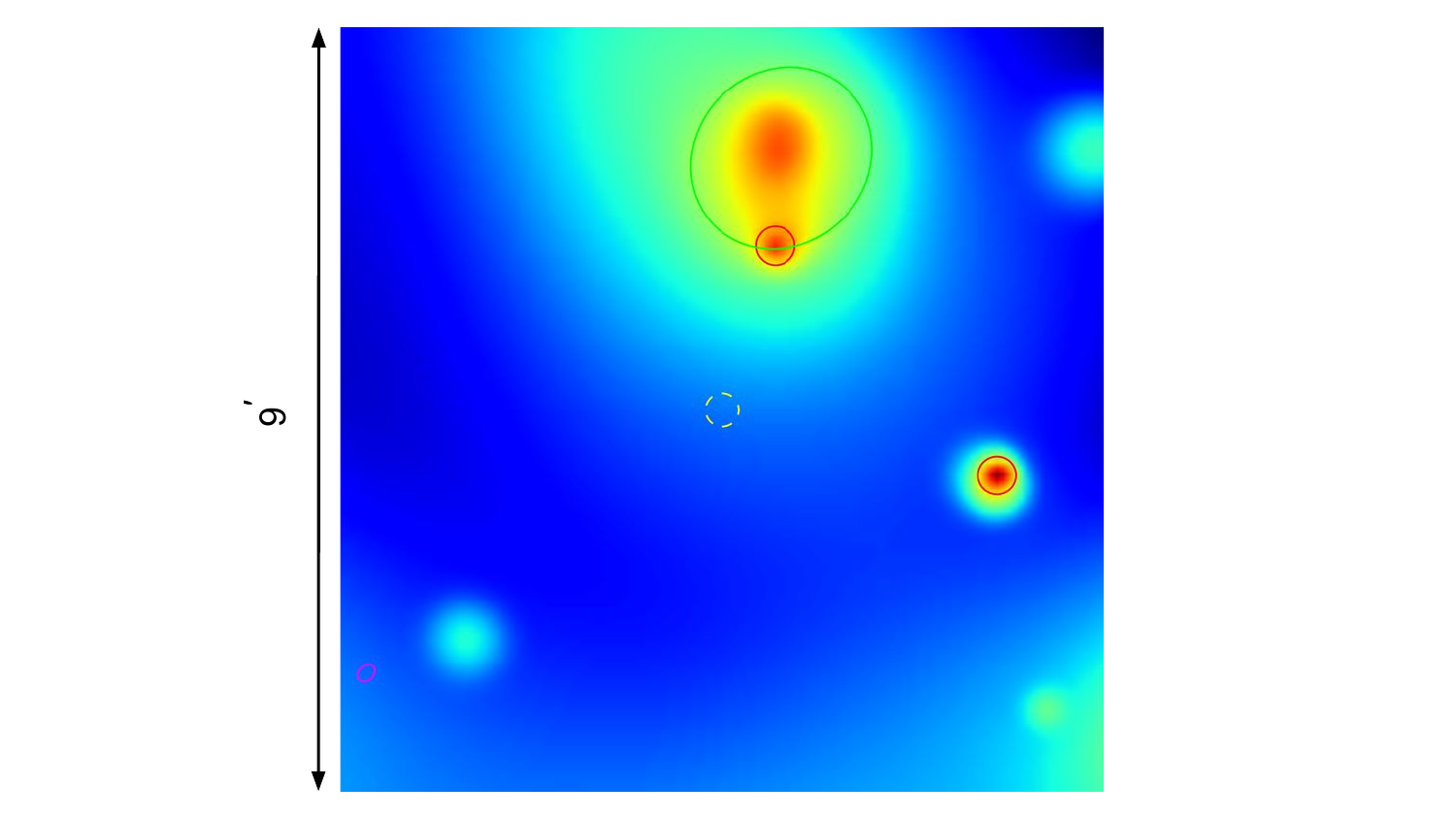} \\
(a) & (b) \\
\end{tabular}
\end{center}
\caption[]{An example of SDSSRM cluster that is not physically associated with nearest extended X-ray source. (a) SDSS optical image of the cluster SDSS RMID=55078, $z=0.39$, $\lambda=26$.  The yellow dashed circle shows the position of the RM defined central galaxy (remaining circles/shapes as defined in Fig~\ref{fig:cutouts}); (b) Corresponding {\em XMM} X-ray observation.  Green (red) outlines highlight {\sc xapa} extended (point) sources.  In this case, the extended emission is associated with an outflow from a nearby galaxy}
\label{fig:unassociated}
\end{figure*}

Here we show an example of SDSSRM-XCS clusters that were initially matched to an extended XCS source, but after visual inspection (see Sect.~\ref{sec:sdss-xcs}), the X-ray emission was found not to be associated with the RM cluster.  In Figure~\ref{fig:unassociated}, the SDSSRM-XCS cluster has been matched to an extended source where the X-ray emission comes from an outflow from a low redshift galaxy.  The extended XCS source was deemed un-associated with the SDSSRM cluster in question.

\section{Clusters effected by mispercolation}
\label{sec:misperc_data}

\begin{figure*}
\begin{center}
\begin{tabular}{cc}
\includegraphics[clip, trim=4cm 0.25cm 6cm 0.25cm, width=8.5cm]{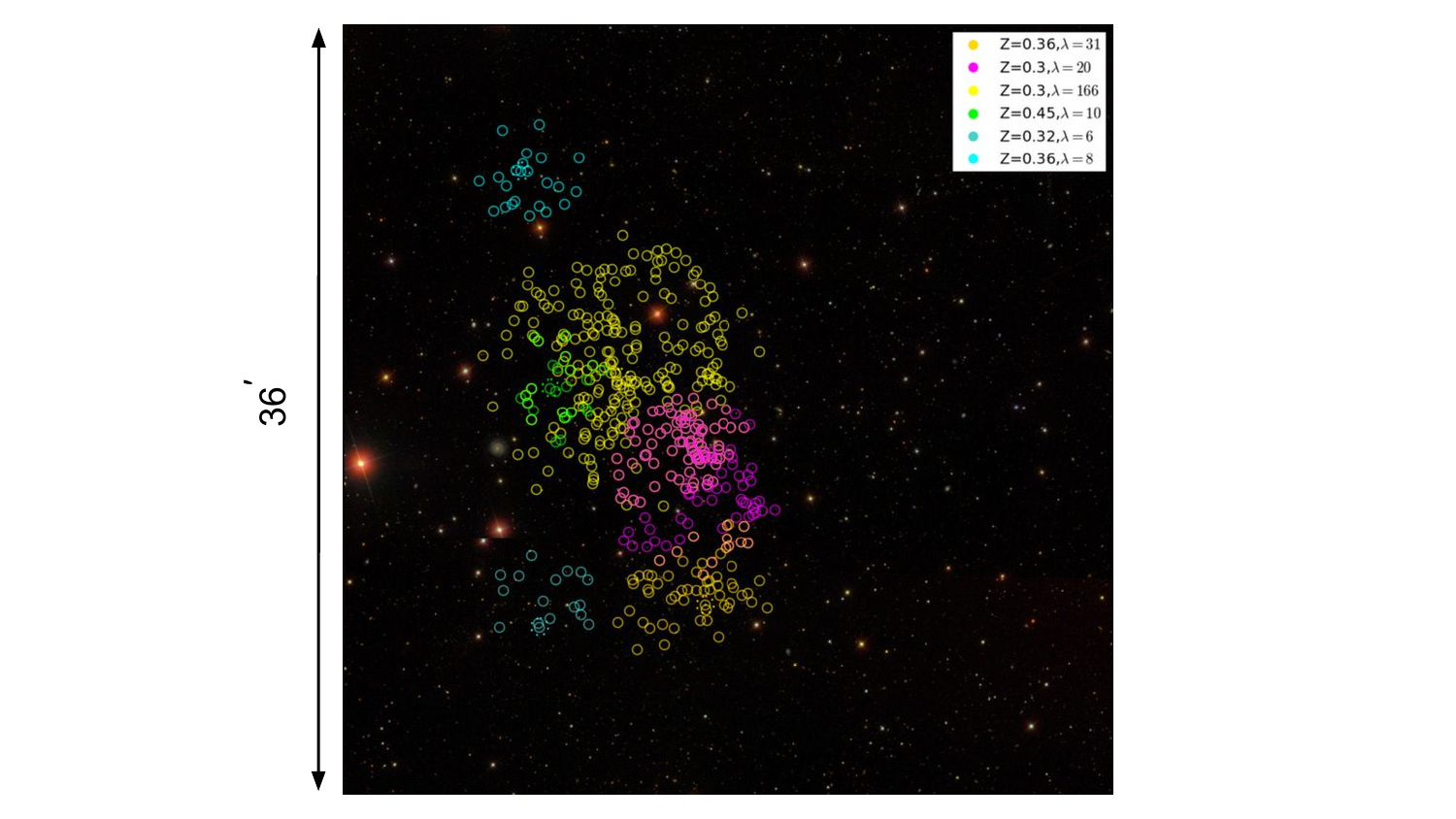} &
\includegraphics[clip, trim=4cm 0.25cm 6cm 0.25cm, width=8.5cm]{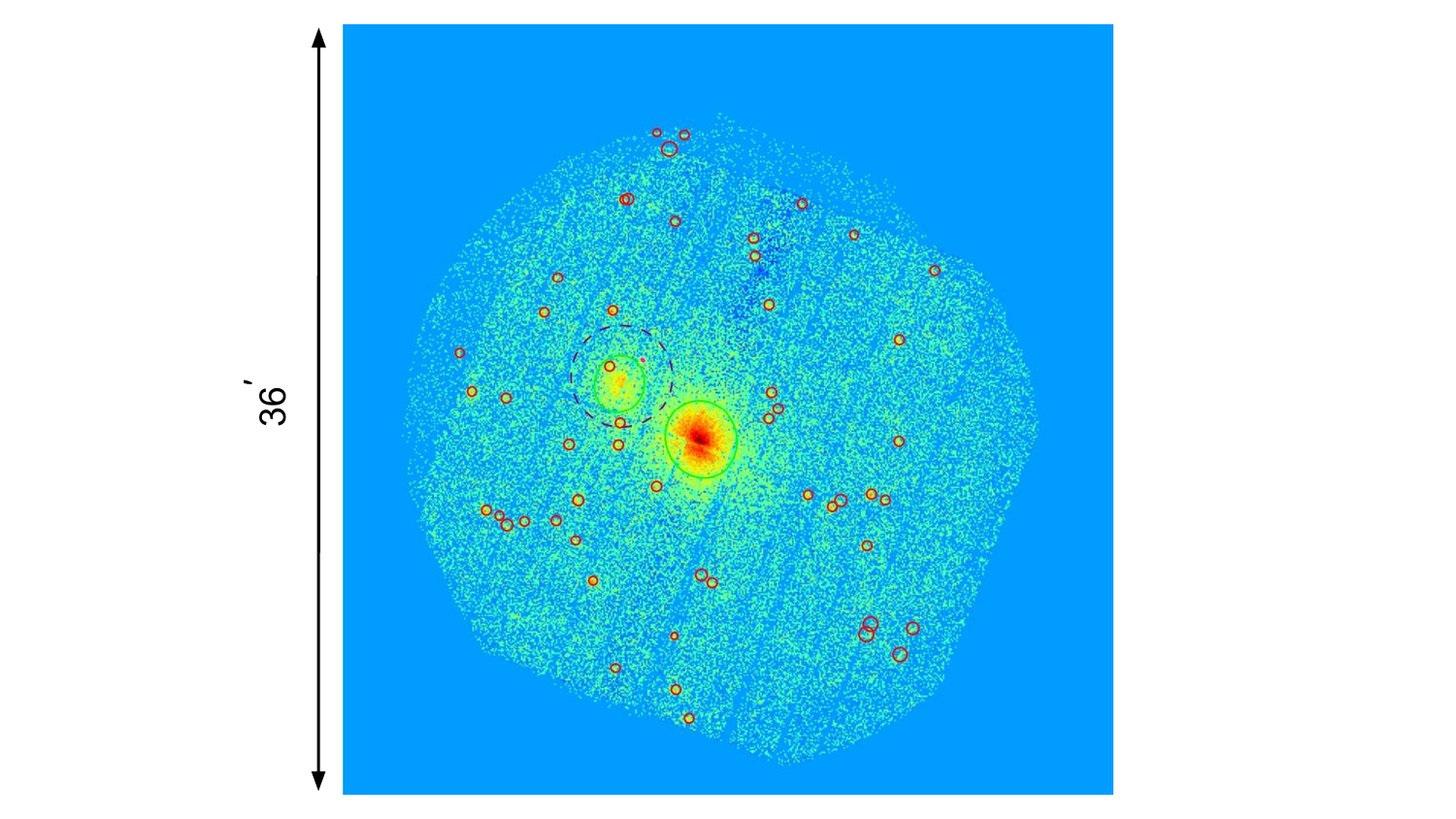} \\
(a) & (b) \\
\end{tabular}
\end{center}
\caption[]{Example of a mispercolated cluster (as described in Sect.~\ref{sec:misperc}).  Each image measures 36$^{\prime}\times$36$^{\prime}$ on a side.  (a) SDSS optical image with the yellow circles highlighting galaxies associated with the lower flux cluster ($\lambda_{\rm RM}=166$).  Pink circles highlight galaxies RM associated with the higher flux cluster ($\lambda_{\rm RM}=20$).  Other RM clusters nearby are highlighted by their respective colours (the colourbar inlay gives the redshift and richness of the highlighted clusters); (b) Corresponding {\em XMM} image.}
\label{fig:misperc}
\end{figure*}

In Section~\ref{sec:misperc}, we identified three pairs of clusters effected by mispercolation.  In Figure~\ref{fig:misperc}, an example of a mispercolated cluster is shown, and Table~\ref{tab:misperc_data} highlights the three pairs of clusters effected by mispercolation and detail manual adjustments made to their properties.

\begin{table*}
\centering
\caption[]{{\small Clusters effected by mispercolation.}\label{tab:misperc_data}}
\vspace{1mm}
\begin{tabular}{lccccc}
\hline
\hline
 RMID & $\lambda_{\rm RM}$ & z & XCS match & swap $\lambda_{\rm RM}$ & Notes\\
\hline
9 & 151 & 0.32 & XMMXCS J100213.9+203222.7 & 15 & Dropped from sample \\
12 & 15 & 0.32 & XMMXCS J100227.5+203102.1 & 151 & Retained, $\lambda_{\rm RM}$ swapped with RMID 9 \\
\hline
21 & 39 & 0.30 & XMMXCS J092021.2+303014.5 & 129 & Retained, $\lambda_{\rm RM}$ swapped with RMID 23 \\
23 & 129 & 0.29 & XMMXCS J092052.5+302803.5 & 39 & Retained, $\lambda_{\rm RM}$ swapped with RMID 21 \\
\hline
34 & 166 & 0.30 & XMMXCS J231148.8+034046.7 & 20 & Retained, $\lambda_{\rm RM}$ swapped with RMID 41 \\
41 & 20 & 0.30 & XMMXCS J231132.6+033759.9 & 166 & Retained, $\lambda_{\rm RM}$ swapped with RMID 34 \\
\hline
\end{tabular}
\end{table*}

\section{Additional scaling relation fits}
\label{sec:targets_exc_core}

\begin{table*}
\begin{center}
\caption[]{{\small Best-fit parameters of the scaling relations studied in this work when considering core excluded and $r_{2500}$ cluster properties.  In each case, parameters are given for the SDSSRM-XCS$_{T_{X},vol}$ cluster sample, and the targeted and serendipitous sub-samples (as defined in Sect.~\ref{sec:targets}).  Best-fit parameters are given for the $L_{X}-T_{X}$, $T_{X}-\lambda_{\rm RM}$, and $L_{X}-\lambda_{\rm RM}$ relations, given by equations~\ref{equ:lt},~\ref{equ:l-lambda} and~\ref{equ:t-lambda} respectively (see Sects.~\ref{sec:ltrelation} and~\ref{sec:xray_richness}).}\label{tab:exccorebestfit}}
\vspace{1mm}
\begin{tabular}{ccccc}
\hline
\hline
 Relation & Normalisation & Slope & Scatter & Figure \\
(sample) & & & & \\
\hline
\multicolumn{5}{c}{Core-excluded relations} \\
\hline
$L_{X,52}^{(0.15-1)r500}-T_{X}^{(0.15-1)r500}$ & $A_{LT}$ & $B_{LT}$ & $\sigma_{LT}$ & \\ 
SDSSRM-XCS$_{T_{X},vol}$ & 0.74$\pm$0.03 & 2.46$\pm$0.10 & 0.51$\pm$0.04 & -- \\
Targets & 0.73$\pm$0.05 & 2.58$\pm$0.16 & 0.53$\pm$0.04 & \ref{fig:lx_kt_targ_serend}(b) \\
Serendipitous & 0.54$\pm$0.07 & 1.84$\pm$0.21 & 0.43$\pm$0.06 & \ref{fig:lx_kt_targ_serend}(b) \\
\hline
$L_{X,52}^{(0.15-1)r500}-\lambda_{\rm RM}$ & $A_{L\lambda}$ & $B_{L\lambda}$ & $\sigma_{L\lambda}$ & \\ 
SDSSRM-XCS$_{T_{X},vol}$ & 0.79$\pm$0.06 & 1.49$\pm$0.12 & 0.88$\pm$0.06 & -- \\
Targets & 1.06$\pm$0.10 & 1.13$\pm$0.16 & 0.88$\pm$0.07 & \ref{fig:x-ray_lambda_targ_serend}(c) \\
Serendipitous & 0.42$\pm$0.07 & 1.15$\pm$0.25 & 0.66$\pm$0.08 & \ref{fig:x-ray_lambda_targ_serend}(c) \\
\hline
$T_{X}^{(0.15-1)r500}-\lambda_{\rm RM}$ & $A_{T\lambda}$ & $B_{T\lambda}$ & $\sigma_{T\lambda}$ & \\ 
SDSSRM-XCS$_{T_{X},vol}$ & 1.04$\pm$0.03 & 0.58$\pm$0.05 & 0.32$\pm$0.02 & -- \\
Targets & 1.17$\pm$0.04 & 0.43$\pm$0.05 & 0.26$\pm$0.02 & \ref{fig:x-ray_lambda_targ_serend}(d) \\
Serendipitous & 0.80$\pm$0.07 & 0.46$\pm$0.13 & 0.34$\pm$0.04 & \ref{fig:x-ray_lambda_targ_serend}(d)\\
\hline
\multicolumn{5}{c}{r$_{2500}$ relations} \\
\hline
$L_{X,52}^{r2500}-T_{X}^{r2500}$ & $A_{LT}$ & $B_{LT}$ & $\sigma_{LT}$ & \\ 
SDSSRM-XCS$_{T_{X},vol}$ & 0.57$\pm$0.04 & 2.89$\pm$0.13 & 0.71$\pm$0.05 & -- \\
Targets & 0.68$\pm$0.06 & 2.69$\pm$0.19 & 0.71$\pm$0.06 & -- \\
Serendipitous & 0.44$\pm$0.07 & 2.56$\pm$0.33 & 0.62$\pm$0.08 & -- \\
\hline
$L_{X,52}^{r2500}-\lambda_{\rm RM}$ & $A_{L\lambda}$ & $B_{L\lambda}$ & $\sigma_{L\lambda}$ & \\ 
SDSSRM-XCS$_{T_{X},vol}$ & 0.57$\pm$0.06 & 1.69$\pm$0.15 & 1.14$\pm$0.07 & -- \\
Targets & 0.88$\pm$0.11 & 1.15$\pm$0.20 & 1.13$\pm$0.09 & -- \\
Serendipitous & 0.43$\pm$0.07 & 1.60$\pm$0.24 & 0.66$\pm$0.03 & -- \\
\hline
$T_{X}^{r2500}-\lambda_{\rm RM}$ & $A_{T\lambda}$ & $B_{T\lambda}$ & $\sigma_{T\lambda}$ & \\ 
SDSSRM-XCS$_{T_{X},vol}$ & 1.01$\pm$0.03 & 0.59$\pm$0.04 & 0.30$\pm$0.02 & \ref{fig:tlambda_comp}(d) \\
Targets & 1.10$\pm$0.04 & 0.49$\pm$0.05 & 0.29$\pm$0.02 & -- \\
Serendipitous & 0.85$\pm$0.06 & 0.42$\pm$0.11 & 0.27$\pm$0.04 & -- \\
\hline
\end{tabular}
\end{center}
\end{table*}

\begin{figure*}
\begin{center}
\begin{tabular}{cc}
\includegraphics[width=7.5cm]{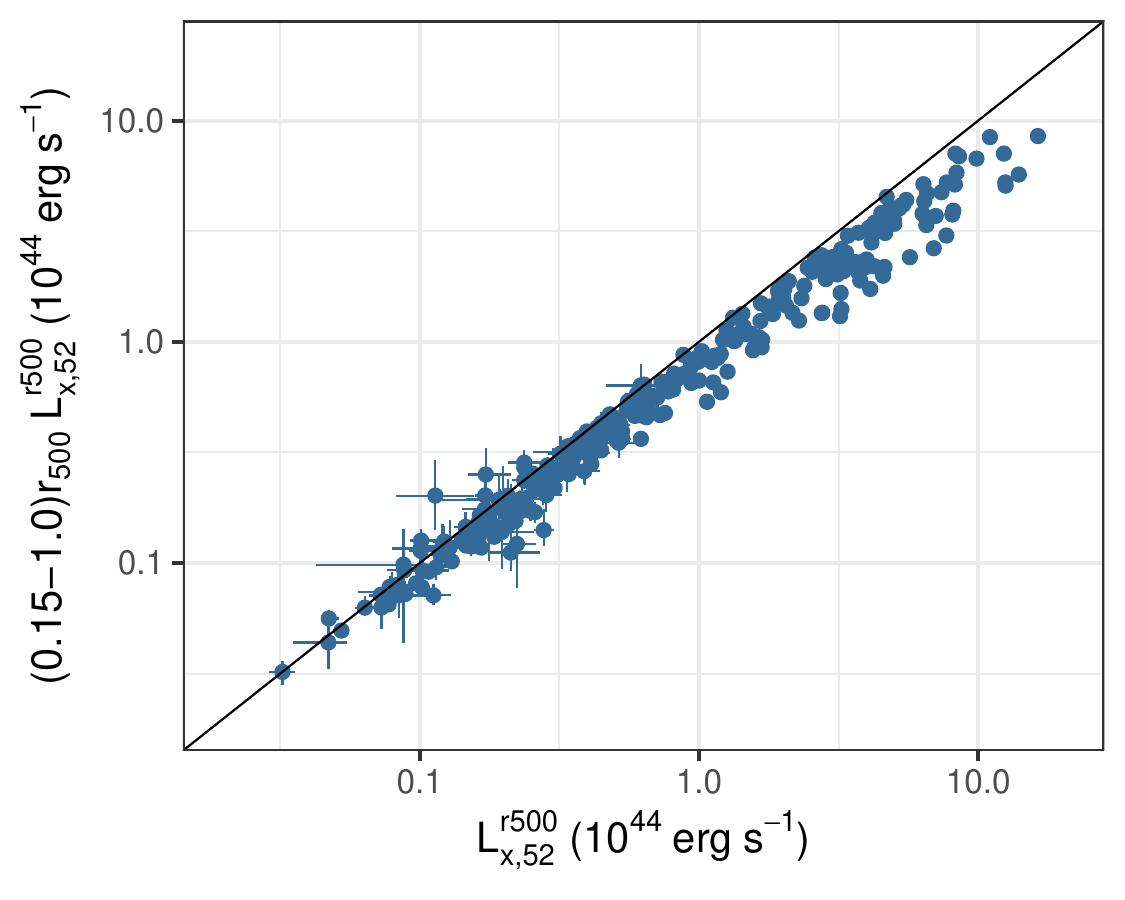} &
\includegraphics[width=7.5cm]{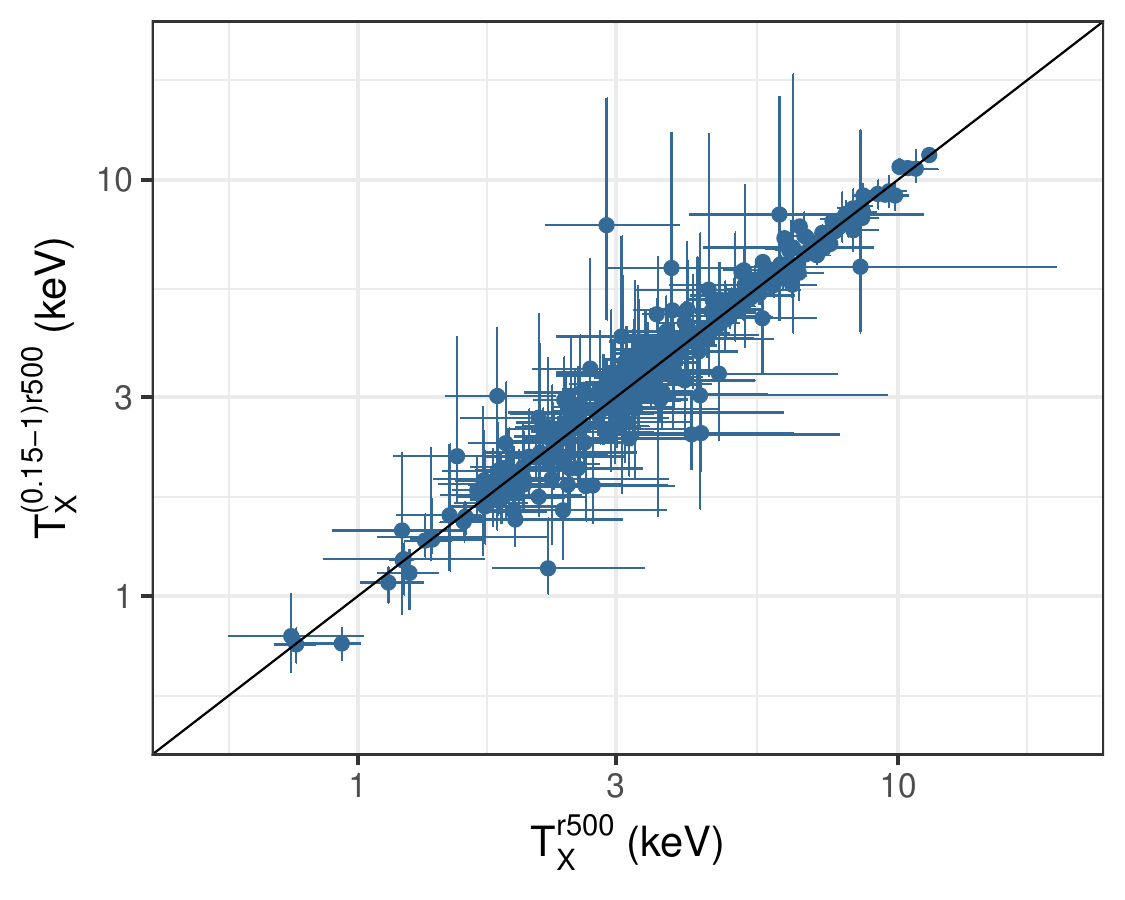} \\
(a) & (b) \\
\includegraphics[width=7.5cm]{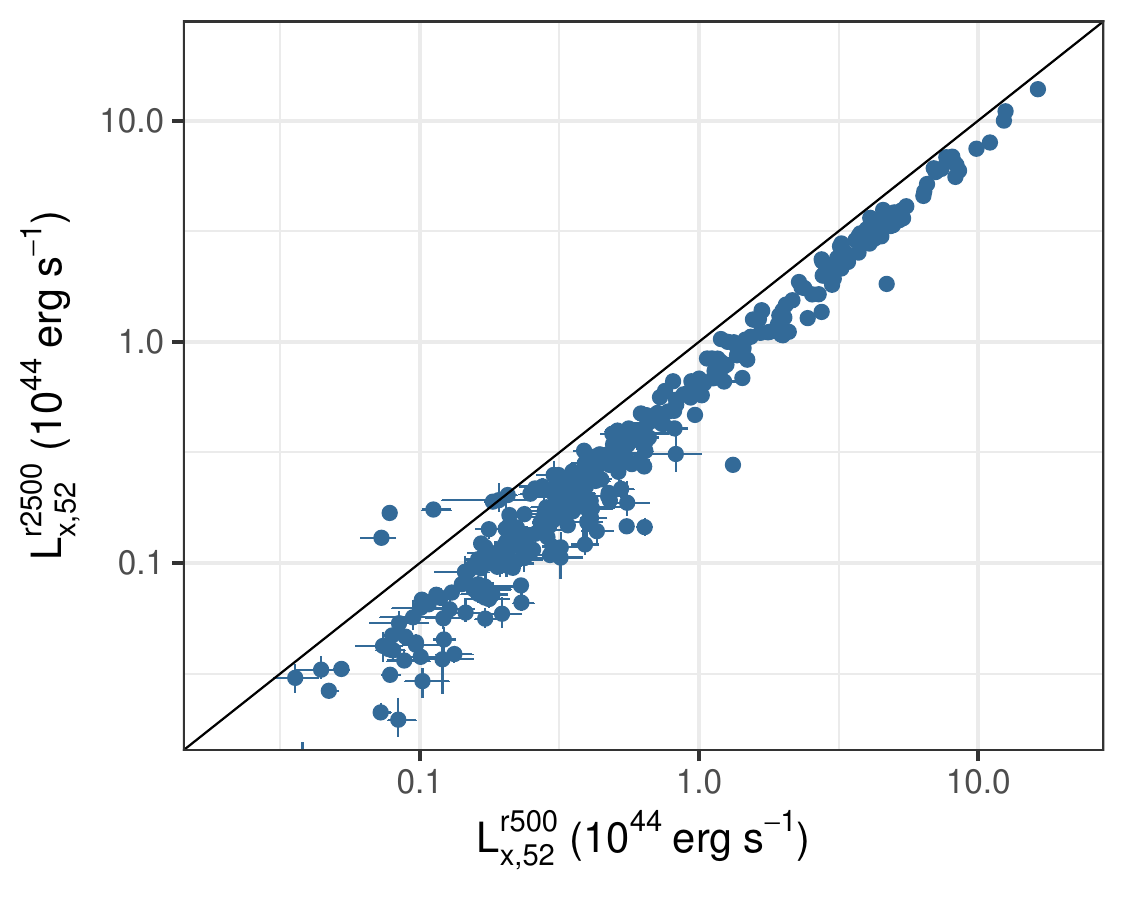} &
\includegraphics[width=7.5cm]{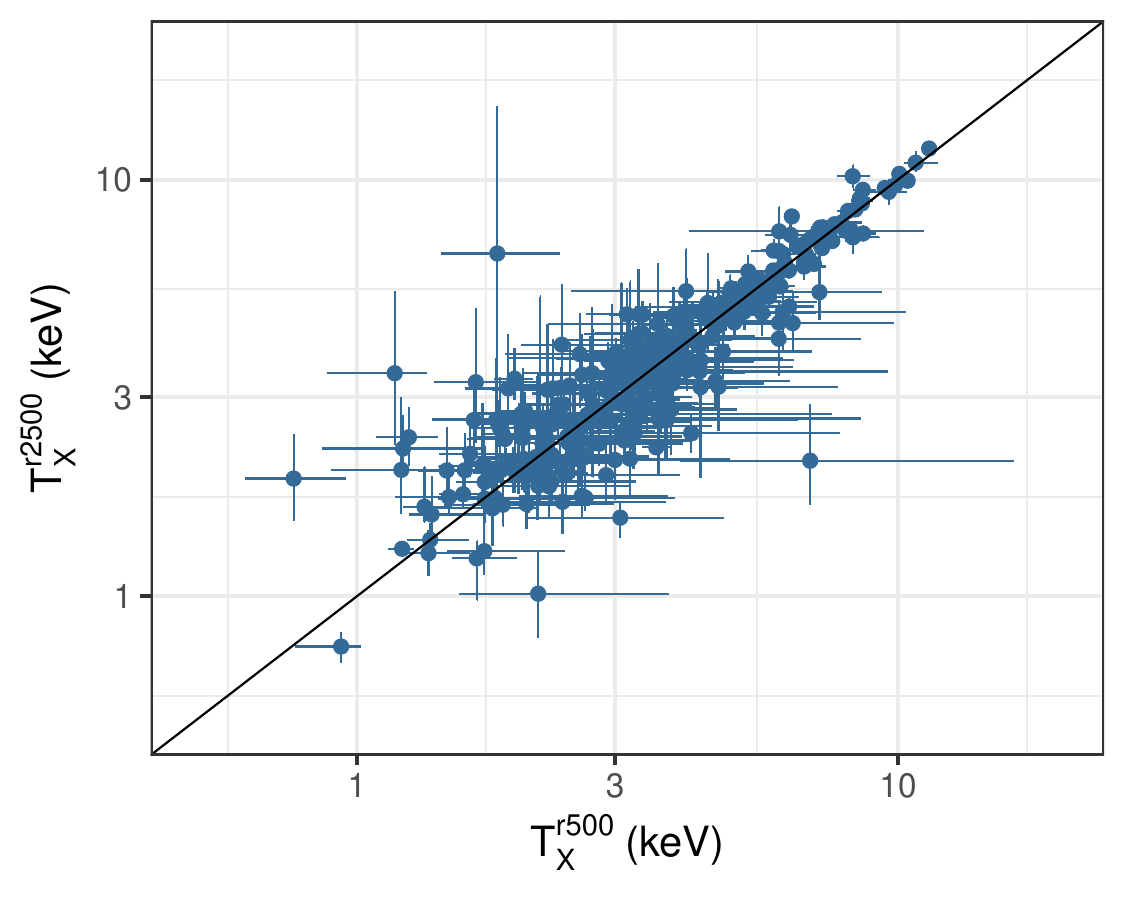} \\
(c) & (d)
\end{tabular}
\end{center}
\caption[]{Comparison of cluster properties measured for our fiducial analysis (i.e. core-included, $r_{500}$).  Plots (a) and (b) compare the core excluded (0.15-1.0$r_{500}$) luminosities and temperatures respectively.  Plots (c) and (d) compare $r_{2500}$ luminosities and temperatures respectively.  In each plot, the 1:1 relation is highlighted by the solid black line.}
\label{fig:lumin_temp_comps}
\end{figure*}

In Table~\ref{tab:exccorebestfit}, we present results of our cluster scaling relation analyses using other apertures: core-excluded  (0.15-1)$r_{500}$, and $r_{2500}$.  The X-ray properties derived from these apertures are compared to those from our fiducial $r_{\rm 500c}$ analysis in Figure~\ref{fig:lumin_temp_comps}.

\section{Replicating the observed $L_{X}-T_{X}$ anisotropy}
\label{sec:migkascom}

\begin{figure*}
\begin{center}
\begin{tabular}{cc}
\includegraphics[width=8.5cm]{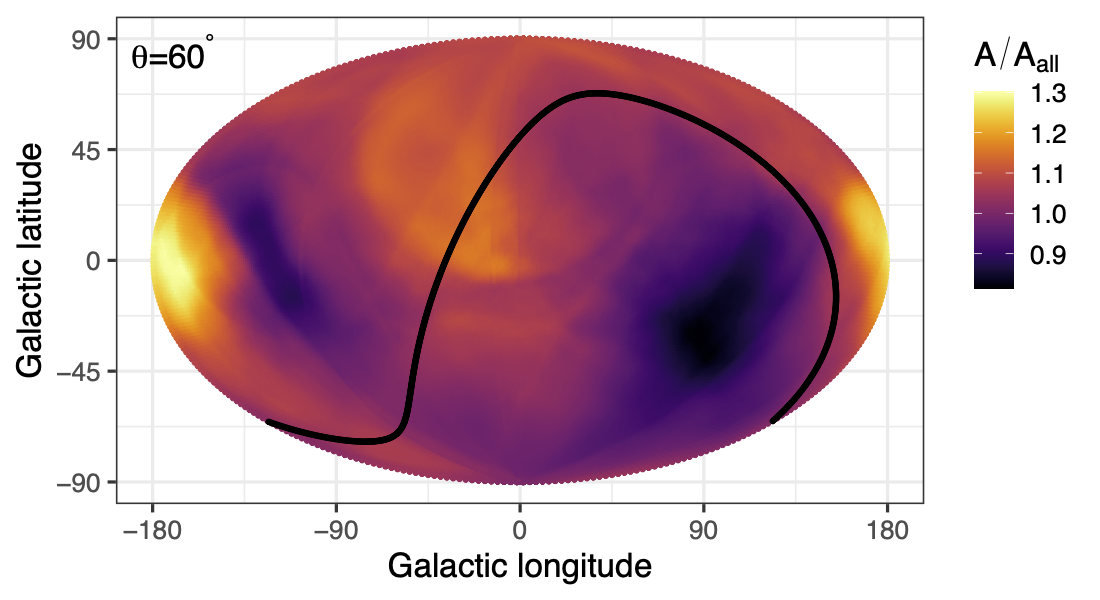} &
\includegraphics[width=8.5cm]{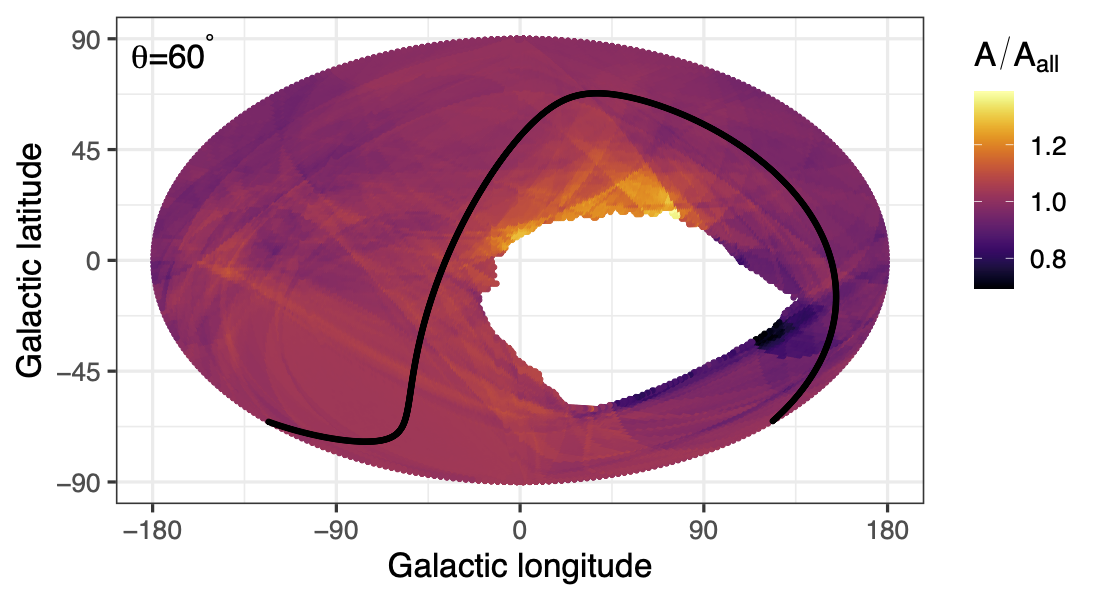} \\
(a) & (b) \\
\end{tabular}
\end{center}
\caption{Ratio of the local best fit $L_{X}-T_{X}$ normalisation ($A$) to the all sky normalisation ($A_{\rm all}$), as a function of sky position. (a) The data used to create the map was taken from Mig20, hence replicating the results presented therein (see Mig20 Fig 8). (b) The sky map was created using the SDSSRM-XCS$_{T_{X}}$ sample and following the updated method of Mig21, see Section~\ref{sec:ltisotropy}.  In each case, the maps are generated using cones of $\theta$=60$^{\circ}$ at each sky position and the black solid line represents the Galactic plane.}
\label{fig:ltmigkas}
\end{figure*}

In Section~\ref{sec:ltisotropy}, we show the results of our investigation into the possible anisotropic behaviour of the $L_{X}-T_{X}$ relation using the SDSS-XCS$_{T_{X}}$ cluster sample.  While we conclude that the SDSS-XCS$_{T_{X}}$ sample does not have the required sky coverage to probe such effects, here, we show that the method (adopted from Mig20) indeed replicates the results shown in Mig20.  Cluster data was obtained from Mig20 and using the replicated method (see Sect.~\ref{sec:ltisotropy}), the results shown in Figure~\ref{fig:ltmigkas}.   

\bsp	
\label{lastpage}
\end{document}